\documentclass[aps,prd,a4paper,preprintnumbers,amsmath,amssymb,superscriptaddress,twocolumn,floatfix,showpacs,showkeys,nofootinbib]{revtex4}

\usepackage{latexsym}
\usepackage{epsfig}
\usepackage{amssymb}
\newcommand{\beq}{\begin{eqnarray}}
\newcommand{\eeq}{\end{eqnarray}}
\newcommand{\be}{\begin{eqnarray}}
\newcommand{\ee}{\end{eqnarray}}

\newcommand{\ve}{\varepsilon}

\begin{document}

\title{Gauss-Bonnet Quintessence: Background Evolution, Large Scale Structure and Cosmological Constraints}
\date{\today}
\author{Tomi Koivisto\footnote{tomikoiv@pcu.helsinki.fi}}
\affiliation{ Helsinki Institute of Physics,FIN-00014 Helsinki, Finland}
\author{David F. Mota\footnote{d.mota@thphys.uni-heidelberg.de}}
\affiliation{  Institute for Theoretical Physics, University of Heidelberg, 69120 Heidelberg,Germany}
\affiliation{ Institute of Theoretical Astrophysics, University of Oslo, Box 1029, 0315 Oslo, Norway}
\affiliation{Perimeter Institute, Waterloo, Ontario N2L 2Y5, Canada}

\begin{abstract}

%We investigate a class of dark energy models in which a scalar field is
%coupled to the quadratic curvature invariants. Such
%couplings are present in the one-loop corrected string effective actions and appear
%generically in theories with extra dimensions. Hence, from the high
%energy theoretical point of view, these models are much better
%motivated than perhaps most of the modified gravity or usual (minimally coupled) quintessence 
%models in the literature. We show that from the cosmological and observational 
%perspective, the Gauss-Bonnet quintessence might also be an interesting
%model of dark energy, because it can
%1) Be derived from an action principle, so one can make definite
%predictions, and it is simple enough to calculate these in practice;
%2) Provide a mechanism to viably onset the late time acceleration after a
%scaling matter era; 3) Alleviate the coincidence problem; 4) Cross the phantom
%divide at the present and avoid Big Rip; 5) Be compatible with the CMB and LSS power
%spectra; 6) Present specific features which can be tightly constrained by both
%local and astrophysical experiments, such as Solar system, supernovae Ia, cosmic microwave
%background radiation, large scale structure and Big Bang nucleosynthesis;
%7) Possibly provide an exit from the acceleration in the future; and
%8) Do all this with just two extra parameters (compared to the concordance model) and
%without introducing unnatural scales into the Lagrangian.

We investigate a string-inspired dark energy scenario featuring a scalar field with a
coupling to the Gauss-Bonnet invariant. Such coupling can trigger the onset of
late dark energy domination after a scaling matter era. 
%and thus possibly alleviate the coincidence problem.
The universe may then cross the phantom divide and perhaps also exit from the acceleration.
We discuss extensively the cosmological and astrophysical implications of the coupled scalar field.
Data from the Solar system, supernovae Ia, cosmic microwave background radiation, large scale structure and
big bang nucleosynthesis is used to constrain the parameters of the model.
A good Newtonian limit may require to fix the coupling. 
With all the data combined, there appears to be some tension with the nucleosynthesis bound,
and the baryon oscillation scale seems to strongly disfavor the model. 
These possible problems might be overcome in more elaborate models. 
In addition, the validity of these constraints in the present context
is not strictly established. Evolution of fluctuations in the scalar field and their impact to clustering 
of matter is studied in detail and more model-independently. Small scale limit is 
derived for the perturbations and their stability is addressed. A divergence is found and discussed.
The general equations for scalar perturbations are also presented and solved
numerically, confirming that the Gauss-Bonnet coupling can be compatible
with the observed spectrum of cosmic microwave background radiation as well as the
matter power spectrum inferred from large scale surveys.

\end{abstract}

\keywords{Cosmology: Theory} 
\pacs{98.80.-k,98.80.Jk}
\maketitle

\section{Introduction}

General relativity predicts singularities. Therefore, in spite of its 
being a highly successful as a classical theory of gravitation, its 
modification seems inevitable at high energy scales. The low energy action 
could then also feature
%These modifications could then manifest in
corrections to the Einstein-Hilbert term from additional fields and
curvature invariants. Among the possible corrections, a particular
combination of the quadratic Riemann invariants, the Gauss-Bonnet term,
is of special interest. It appears in extensions of gravitational actions,
whether motivated by the form of most general of a scalar-tensor theory,
uniqueness of the Lagrangian in higher dimensions or the
leading order corrections from string theory. An interesting possibility
then arises that the modifications could change the way the universe itself
gravitates at large scales.
%This article is an investigation of such possibility.

Indeed, the current cosmological observations cannot be explained
by the standard model of particle physics and general relativity.
Precise cosmological experiments have confirmed the standard Big Bang
scenario of a flat universe undergoing an inflation in its earliest
stages, where the perturbations are generated that eventually form into
structure in matter. Most of this matter must be non-baryonic, dark
matter. Even more curiously, the universe has presently entered into
another period of acceleration. Such a result is inferred from
observations of extra-galactic supernovae of type Ia (SNeIa) \cite{Riess:2004nr,Astier:2005qq}
and is independently supported by the cosmic microwave background radiation (CMBR)
\cite{Spergel:2006hy} and large scale structure \cite{sloan} data. It seems that some
dark energy, with its negative pressure that speeds up the universal expansion, dominates the density
of the universe \cite{Copeland:2006wr}. This concordance model agrees very well
with the astrophysical data, but features inflation, dark matter and dark
energy as phenomenological ingredients of undisclosed nature.
Perhaps dark energy is the most enigmatic of the latter two, as
at least there are reports of tentative discoveries of dark
matter \cite{bertone}.

The problem of dark energy reduces to the questions 1) what is the
culprit (perhaps in fundamental physics) for the acceleration, and 2) why
did the speed-up begin just at the present stage of the cosmological evolution.
In recent years dark energy and its cosmological and astrophysical
signatures has been addressed in many papers considering both
modifications of the energy-momentum tensor, with the inclusion of
scalar fields \cite{Caldwella,brookfield2,bertolami,carsten}
or imperfect fluids \cite{tomicard,koivisto} and of gravitational physics at large
scales \cite{Capozziello02,Nojiri:2006ri,Straumann:2006tv,Borowiec:2006qr}. The
latter approach has proven to be perhaps surprisingly difficult. Extending the
action with curvature invariants rather generically results in fourth-order gravity, which
is problematical because of the implied instabilities\cite{Woodard:2006nt}. An $f(R)$ extension,
although not generically, can be stable\cite{Faraoni:2006sy}. However, though some $f(R)$ -type 
modifications might be compatible with the acceleration following a standard matter dominated 
phase\cite{Nojiri:2006gh,Nojiri:2006su}, the simplest modifications as proposed this far seem not 
to be able to generate a viable cosmology\cite{Amendola:2006kh}. On the other hand, within the Palatini 
formulation, the corresponding extensions result in second order field equations, but it turns out that the 
observationally allowed modifications are then practically indistinguishable from a cosmological constant 
\cite{tomimod2,morad,tomimod3,Li:2006ag}. 

Scalar-tensor theories of gravity are interesting alternatives to the concordance
model and seem to have potential to provide a linkage between the acceleration and fundamental 
physics \cite{st1,st2,Albrecht:2001xt,Kainulainen:2004vk}. They have a
desirable feature which is their quasi-linearity: the property that the highest
derivatives of the metric appear in the field equations only linearly, so as to make
the theory ghost free. Interestingly, there is a particular combination of the curvature
squared terms with such behavior, known as the Gauss-Bonnet (GB) integrand.
It is constructed from the metric as
\be
R^2_{GB} \equiv R^{\mu\nu\rho\sigma}R_{\mu\nu\rho\sigma} -
4R^{\mu\nu}R_{\mu\nu} + R^2.
\ee
As mentioned in the beginning, this term appears frequently in attempts at a quantum
gravity, especially in stringy set-ups. In fact, all versions of string theory (except Type II)
in 10 dimensions ($D=10$) include this term as the leading order
$\alpha^\prime$ correction\cite{Callan,gross}. In $D=4$ the Gauss-Bonnet
term is a topological invariant. It's appearance alone in the action can then be
neglected as a total divergence (though even then this topological term can have interesting
role in view of the boundary terms and regularization of the $D=4$ action in 
asymptotically AdS spacetimes\cite{Aros:1999id,Olea:2005gb}). However, if coupled, its presence
may lead to contributions to the field equations also in $D=4$.
The low energy string action typically features scalar fields with such couplings.
Namely the moduli, associated to the internal geometry of the hidden dimensions, become
non-minimally coupled to the curvature terms. Hence, in a general $D=4$ scalar-tensor theory 
one might couple the scalar field non-minimally not only to the Ricci curvature $R$ but to the
Gauss-Bonnet invariant $R_{GB}^2$ as well, and in $D>4$ it is necessary to
include this term in order to preserve the uniqueness of the gravitational action.

Such a coupling could thus be useful in modeling both the early inflation and
the late acceleration\cite{Carter:2005fu,Neupane:2006dp,Tsujikawa:2006is,Carter:2005sd,Neupane:2006ip}.
Here we will concentrate on the post-inflationary epochs in such a universe, and more specifically,
within the scenario we recently studied\cite{Koivisto:2006xf}. There the scalar field lives in an exponential
potential, and the coupling can likewise be exponential. This kind of model
was originally proposed by Nojiri, Odintsov and Sasaki\cite{Nojiri:2005vv}. More recently
it was shown that if the coupling grows steeper than the potential decays with the field,
acceleration can occur with a canonic scalar, after a transition from
a scaling era\cite{Koivisto:2006xf}. Then the interaction
with the Gauss-Bonnet curvature term causes the universe to enter from a scaling
matter era to an accelerating era. Existence of scaling solutions with
more general parameterizations has then been taken under investigation\cite{Tsujikawa:2006ph},
noting that requiring exact scaling is otherwise possible only with a tachyon field which
is non-minimally coupled with matter. Other dark energy solutions in a broad variety of
second order string cosmologies, taking into account both coupled and uncoupled
matter, have been explored also previously \cite{Sami:2005zc,Calcagni:2005im},
%The cosmological dynamics in all the above
%references were explored mainly in view of the asymptotic attractor solutions.
higher order terms have been also incorporated and a method to reconstruct the
coupling and the potential has been found\cite{Nojiri:2006je}. If the scalar field is
non-dynamical, the model is equivalent with a modified Gauss-Bonnet 
gravity, featuring a function of $R^2_{GB}$ the action \cite{Cognola:2006eg,Nojiri:2005am}.

The coupling mechanism can also momentarily push the universe to a phantom
era, until the scalar field potential begins to dominate. Therefore
it is possible to get a very good accordance with the SNeIa data.
Dark energy models other than of the Gauss-Bonnet form but featuring transition
to phantom expansion and having possible origin in string theory have been considered also 
\cite{Aref'eva:2005fu,Aref'eva:2006et}. In addition, one should clearly distinguish the models investigated here 
from those featuring a coupling to the Ricci scalar \cite{Baccigalupi:2000je}, since
for them the so-called R-boost \cite{Pettorino:2005pv} plays a role in the early 
universe, while the potential has to be shallow in order to
have accelerating behaviour at the present. On the other hand, one can
couple the self-tuning scaling field to matter \cite{Amendola:1999er,manera,brookfield1}, 
and thereby indeed set the acceleration ongoing and thereby perhaps alleviate the   
coincidence problem \cite{Bean:2000zm,Huey:2004qv}.
However, implications for the large-scale structure formation seem to make this
possibility problematical \cite{Koivisto:2005nr,nunes}. Here we will not find the
same, since modifications to the growth of matter perturbations are
much less drastic as the scalar field remains minimally
coupled to the matter sector. 
%The impact of the Gauss-Bonnet interaction
%does not enhance the clustering of the quintessence field, and the small
%scale evolution of matter inhomogeneities stays scale-invariant as in
%the $\Lambda$CDM case regardless of the coupling.

In this paper, besides investigating the cosmological evolution of such models,
we also derive quantitative constraints on the scenario of
Gauss-Bonnet dark energy mentioned above. Our aim is to present in more detail
this scenario and to generally uncover in more depth what are
the possible effects of the Gauss-Bonnet combination on dark energy cosmologies,
in order to assess whether they could be compatible with present days observations, and to
estimate the amount of fine-tuning that this might require. To this end the
the model in Ref.\cite{Koivisto:2006xf} is subjected to further constraints. In addition to tightening
the limits for this particular parameterization, we will discuss in more
detail and generality the phenomenology of Gauss-Bonnet dark energy
at cosmological, astrophysical and Solar system scales during the cosmological
evolution. Especially, we consider the impact of the coupling to the CMBR and
large scale structure. In view of the remarks above, this is a crucial aspect
of a dark energy model with any non-minimal couplings or modifications of gravity.
Since the class of models studied here features both and these can
lead to completely different predictions for the expansion rate of the
Universe than without the $R^2_{GB}$ term, it is rather non-trivial to
find that the perturbation evolution in these models is in some sense only modestly altered. Namely,
the shape of the matter power spectrum is retained during the presence of the Gauss-Bonnet
gravity and the effects one expects to see in the large multipoles of the CMBR spectrum
are typically small. Still, there are distinguishable features peculiar to these models that
can lead to essential constraints for the model parameters.

The article is organized as follows. The cosmological dynamics for these models is
described in Section \ref{cs} for both the background expansion and for linear perturbations.
In Section \ref{constraints} we discuss the implications and derive constraints of the Gauss-Bonnet
coupling to nucleosynthesis, Solar system, large scale structure, CMBR
and SNeIa. Section \ref{conclusions} contains our conclusions.
Some technical details are confined to the Appendixes.

\section{Cosmological Dynamics}
\label{cs}

Before writing down the cosmological equations, let us briefly discuss the
model and its motivation. A peculiar property of the string effective action is the
presence of scalar fields and couplings which are field dependent, and 
thus in principle space-time dependent. The scalar fields are moduli associated to
geometrical properties of compactified extra dimensions. Thus one could consider 
multiple scalar fields, representing the dilaton and various other moduli, but 
here we stick to one field $\phi$. The reason is that the dilaton couples to the Ricci curvature, 
and this would lead to variations of the Newton's constant.
When going from the string frame to the Einstein frame, the coupling to $R$ is transformed 
into a non-minimal interaction with matter. This in turn would lead to violations of the equivalence principle.
Thus, in general, also the matter Lagrangian $L_m$ could be non-minimally coupled to the scalar field. 
There are tight constraints from observations to such effects \cite{uzan} (see however, \cite{doug1,doug2}). However, it is (usually)
only the dilaton field which in heterotic string theory acquires the geometrical coupling to $R$
and thus enters in the conformal factor and the matter Lagrangian in the Einstein frame\cite{Antoniadis2}.
If the conformal factor is not (nearly) trivial, it can evolve so, for instance 
due to the so called least coupling principle \cite{Damour:1994zq}.  
In the present article we will make the usual assumption in cosmology, that possible non-minimal couplings to the 
matter sector are negligible, since our specific purpose here is to study the novel features resulting 
from the Gauss-Bonnet curvature interaction. Thus we identify $\phi$ as a run-away 
modulus without direct matter couplings. Then the equivalence principle is automatically satisfied,
though gravitational dynamics are modified due to the modulus-dependent loop corrections.

Thus we can write the action for the system to consider as 
\be &S& =  \label{action} \\
&\int& d^4 x \sqrt{-g}\left[ \frac{R}{2\kappa^2} - 
\frac{\gamma}{2}(\nabla\phi)^2 - V(\phi) - f(\phi)R^2_{GB} + L_m\right],\nonumber
\ee
here $\kappa=(8\pi G)^{-1/2}$, $G$ being the Newton's constant. The 
$\gamma$ could be a function of $\phi$ as well, and there could be 
additional kinetic terms. In the following we will mostly consider canonic scalar 
field and thus set $\gamma=1$ (which can be achieved by a redefinition of the field
for any constant $\gamma>0$). $V(\phi)$ is the field potential which could result 
from nonperturbative effects. In four dimensions the GB term makes no 
contribution if $f(\phi)=\text{const}$. However, it is natural to consider $f(\phi)$ to be
dynamical. This follows, for example, from the one-loop corrected string effective 
action \cite{Antoniadis2,Antoniadis3}, where the function 
$f(\phi)=\sigma-\hat{\delta} \xi(\phi)$: the coupling $\sigma$ may be related to string 
coupling $g_s$ via $\sigma\sim 1/g_s^2$, and the numerical coefficient $\hat{\delta}$ typically 
depends on the massless spectrum of every particular model~ \cite{Antoniadis}. 
It thus seems that the dilaton might couple already at the string tree-level next to leading order 
expansion in the inverse string tension $\alpha'$, and the modulus functions $\xi(\phi)$ are non-trivial 
at the one-loop order.

%Then the where the equivalence principle is well preserved and the Newton's constant $G$ is (almost) 
%time-independent. 

In the numerical examples we will adopt an exponential form for the potential and the coupling, 
\be \label{exps}
V(\phi) = V_0 e^{-\lambda\kappa \phi/\sqrt{2}}, \quad
f = f_0 e^{\alpha\kappa \phi/\sqrt{2}}.
\ee
On one hand, the nonperturbative effects from instantons or gaugino condensation typically result 
in an exponential potential. On the other hand, there is also phenomenological motivation for this, since, 
besides its being a simple choice, an exponential potentials allows to consider 
scaling solutions in cosmology. 
An exponential field-dependence for coupling is a reasonable assumption in
supergravity actions. For massless dilaton one in fact has $f(\phi) = \sum C_n e^{(n-1)\phi}$
\cite{Gasperini:2002bn}. For another instance, a known example from heterotic 
string theory \cite{Antoniadis} yields for the modulus 
coupling, in terms of the Dedekind 
function $\eta$, the form $f(\phi) \sim 
\log{[2e^{\alpha\kappa\phi}\eta^4(ie^{\kappa\alpha\phi})]}$, which behaves like 
Eq.(\ref{exps}) to a good approximation. 
Again, it is a simple choice for $f$, introducing but one extra 
parameter $\alpha$. Such minimalism is practical when considering a 
phenomenological model of dark energy, though from a particle physics 
point of view one might expect that corrections to the Eqs.(\ref{exps}) 
come into play, in particular when the Gauss-Bonnet term begins 
participate in cosmological dynamics in the late time in our model.

In the presence of the coupling $f$, the field equations can be written as
\be \label{fe}
G_{\mu\nu} = \kappa^2\left[ T^m_{\mu\nu} + T^\phi_{\mu\nu} + T^f_{\mu\nu}\right]
\ee
where the two first terms in the right hand side are the energy-momentum tensors
for matter and the scalar field, respectively. The curvature corrections resulting
from taking into account the Gauss-Bonnet contribution involve only terms
proportional to derivatives of $f(\phi)$. A partial integration of the 
coupled term in action (\ref{action}) gives a vanishing boundary term but in addition,
because of the field dependent interaction, a contribution involving the integral of 
$R^2_{GB}$ which is of the first order by construction. Explicitly, one obtains
\be \label{tf}
T^{f}_{\mu\nu} & = & -8\Big[f_{; \alpha\beta}R_{\mu\phantom{\alpha\beta}\nu}^{\phantom{\mu}\alpha\beta}
+\Box{f}R_{\mu\nu} 
 -  2f_{;\alpha(\mu}R^\alpha_{\phantom{\alpha}\nu)}
\nonumber \\
& + & \frac{1}{2}f_{;\mu\nu}R\Big] 
- 4\left[2f_{;\alpha\beta}R^{\alpha\beta}-\Box{f}R\right]g_{\mu\nu}.
\ee
The equation of motion for the scalar field reads
\be \label{kl}
\gamma \Box \phi - V'(\phi) - f'(\phi) R^2_{GB} = 0.
\ee
Thus the field lives in an effective potential given by $V(\phi) + f(\phi)R^2_{GB}$. 
Matter, since minimally coupled, is conserved as usually \cite{tomimod1}, $\nabla^\mu T^m_{\mu\nu} = 0$.
The general covariant expression for curvature corrections Eq.(\ref{tf}) is a rather 
complicated combination of the derivatives of the coupling and their contractions
with the Riemann and Ricci objects, but due to the geometric properties of these corrections, 
they vanish in constant curvature spacetimes, and also otherwise can assume rather 
tractable forms, as we will see in the following.  

\subsection{Background}

We consider a flat, homogeneous and isotropic background universe with 
scale factor $a(t)$ in the Friedmann-Robertson-Walker (FRW) metric
\be \label{metric_bg}
ds^2 = -dt^2 + a^2(t)\delta_{ij}dx^idx^j.
\ee
The action (\ref{action}) then yields the Friedmann equation
as the $t$-$t$ component of Eq.(\ref{fe}),  
\be \label{friedmann}
\frac{3}{\kappa^2}H^2 = \frac{\gamma}{2}\dot{\phi}^2+V(\phi) + \rho_m + 
24H^3f'(\phi)\dot{\phi},
\ee
where an overdot denotes derivative with respect to the cosmic time
$t$, $H\equiv \dot{a}/a$ is the Hubble rate and
$\rho_m$ represents the matter component.
The Klein-Gordon equation Eq.(\ref{kl}) reads
\be \label{klein-gordon} 
\gamma (\ddot{\phi}+3H\dot{\phi})+V'(\phi)+f'(\phi)R^2_{GB}=0,
\ee
where the Gauss-Bonnet invariant is $R^2_{GB}=24H^2(\dot{H}+H^2)$. 
It will be convenient to work with the  
dimensionless variables defined as
\beq \label{variables}
\Omega_m &\equiv& \frac{\kappa^2\rho_m}{3H^2}, \quad
x \equiv \frac{\kappa}{\sqrt{2}}\frac{\dot{\phi}}{H}, \quad
y \equiv \kappa^2\frac{V(\phi)}{H^2}, \nonumber \\
\mu  &\equiv& 8\kappa^2\dot{\phi} H f'(\phi), \quad
\epsilon \equiv \frac{\dot{H}}{H^2}.
\eeq 
In addition it is useful to define, in analogy with $\Omega_m$, the relative contributions from the scalar
field and the Gauss-Bonnet correction as
\be
\Omega_\phi \equiv \frac{\kappa^2\rho_\phi}{3H^2} = \frac{\gamma x^2+y}{3}, \quad
\Omega_f \equiv 1 - \Omega_m-\Omega_\phi = \mu. 
\ee
Then
\be
w_{eff} = 
w_m\Omega_m + w_\phi\Omega_\phi + w_f\Omega_f = -\frac{2}{3}\epsilon - 1
\ee
is the total effective equation of state in the sense that the Universe
expands as if dominated by a fluid with this relation between its pressure and
energy. One notes that the Gauss-Bonnet term can be written as
$R^2_{GB} = -12H^4(3w_{eff}+1)$, and is thus negative if and only if the scale factor
is decelerating. 

With the aim to understand the behaviour of the background cosmology for the model 
we perform a dynamical system analysis. In terms of the dimensionless variables (\ref{variables}),
the complete dynamical system is then given by \cite{Neupane:2006dp}
\be \label{system}
0 & = & -3 + \gamma x^2 + y + 3\mu + 3\Omega_m, \nonumber \\
0 & = & 2\epsilon + 3 + \gamma x^2 - y - \mu' - (\epsilon+2)\mu + 3w_m\Omega_m, 
\nonumber \\
0 & = & 2\gamma\left[xx'+x^2(\epsilon+3)\right] + y' + 
2y\epsilon+3\mu(\epsilon+1), 
\nonumber \\
0 & = & 2xx'+y' + 3\mu' + 3\Omega_m', \nonumber \\
\Omega_m' & = & -3(1+w_m)\Omega_m -2\epsilon\Omega_m, \nonumber \\
\mu' & = & (x'/x +2\epsilon + \alpha x)\mu, \nonumber \\
y' & = & -(\lambda x + 2\epsilon)y,
\ee 
where prime means a derivative with respect to $\log(a)$ and $w_m$ is the 
equation of state of the background. The first five equations hold generally
for an action like (\ref{action}), whereas the two last equations encode
the information about the specific model (\ref{exps}). For simplicity, we 
set from now on $\gamma=1$, so the scalar field kinetic term is canonical. 

We find several fixed points for the system (\ref{system}), 
characterized by $x'=y'=\mu'=0$:

\begin{itemize}

\item{A}: $(x,y,\mu)=(0,0,0)$. This is an unstable point which would correspond to
domination of the background fluid with $w_{eff}=w_m$ and completely vanishing
contribution of the scalar field.

\item{B}: $(x,y,\mu)=(\pm\sqrt{3},0,0)$. In this case, the kinetic energy of the scalar field 
dominates. Then $w_{eff}=1$. The kination phase is not stable, since the kinetic 
energy will always redshift away faster than the other contributions to 
the energy density. This solution with $\dot{\phi}>0$ is a 
saddle point (it has one positive eigenvalue, implying that the solution attracts from some direction while 
repelling from some other), whenever $\alpha<2\sqrt{3}$ or $\lambda,\alpha \ge 2\sqrt{3}$.

\item{C}: $(x,y,\mu)=(0,3,0)$. In this fixed point, the potential of the scalar field
dominates. Thus this is a de Sitter solution with $w_{eff} = -1$. In the absence of the
coupling, all other contributions to the energy density have larger effective equations of 
state, and the solution is stable. With the coupling, stability condition is simply 
$\alpha \ge \lambda$.

\item{D}: $(x,y,\mu)=(\lambda/2,3-\lambda^2/4,0)$.  This fixed point 
corresponds to scalar field domination with $w_{eff} = \lambda^2/6 -1$.
It does not exist when $\lambda>2\sqrt{3}$ and is thus irrelevant
to us here. 
When $\alpha=\lambda$, the Gauss-Bonnet term can reduce 
$w_{eff}$ of this solution\cite{Tsujikawa:2006ph}.

\item{E}: 
$(x,y,\mu)=(\frac{3}{\lambda}(1+w_m),\frac{9}{\lambda^2}(1-w_m^2),0)$.  
This is the well-known scaling solution, where a scalar 
field with exponential potential 
mimics exactly the background equation of state $w_m$.
This fixed point is a stable spiral when $\lambda > \alpha, 
\sqrt{6}(1+w_m)$. It is, however, a saddle point when 
$\lambda,\alpha<\sqrt{6}(1+w)$ or $\alpha \ge \lambda,\sqrt{6}(1+w_m)$.
Also this solution exists in a modified form in the special case 
$\alpha=\lambda$ \cite{Tsujikawa:2006ph}.

\item{F}:
$(x,y,\mu)=(\frac{3}{\alpha}(1+w_m),
0,\frac{18}{\alpha^2}\frac{(1-w_m)(1+w_m)^2}{1+3w_m})$. This is an 
a new scaling solution, in which field has run to large values and the
the potential can be neglected. Then 
$\Omega_m$ is nonzero and the effective equation of state in this case is again just $w_m$.
The condition for the existence
of this fixed point is that either $w_m>7/3$ or $\alpha^2 = 3(1+w_m)^2(7-3w_m)/(1+3w_m)$.
Then it might be a saddle point, but it is generically unstable. Thus this
possibility of the kinetic energy together with the Gauss-Bonnet contribution scaling like 
matter seems not to be useful for cosmological applications, as it will 
not be reached by dynamical means. Even if one sets
the field to this solution as an initial condition during radiation domination,
the Gauss-Bonnet contribution will drop away when transition to matter domination 
occurs. 

\item{G}: There exists also a solution where only the Gauss-Bonnet term 
and the kinetic term of the scalar field survive while $y=\Omega_m=0$ (see Appendix 
\ref{eqs}). This solution is complicated and of no cosmological 
interest to us here. 

\end{itemize}

From this we can see that the standard tracking behaviour of exponential 
quintessence is available whenever the coupling term is negligible. This 
tracker solution has been shown to exist for extremely wide range of 
initial conditions. Unfortunately, while in the tracking regime, the scalar 
field equation of state equals exactly the background $w$, and thus 
this solution cannot account for the accelerating universe. 

\begin{figure}
\begin{center}
\includegraphics[width=0.4\textwidth]{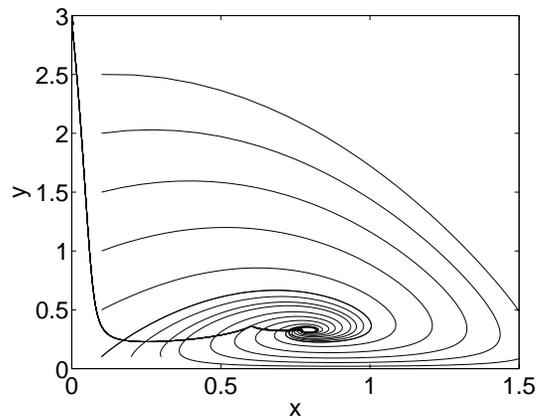}
\caption{\label{pp} A phase portrait of the model. The saddle 
point $E$ attracts the field to the scaling regime but later always passes it, 
along the same track, to the fixed point $C$ which corresponds to an 
accelerated expansion. The requirement for the fixed point $C$ to be 
reached and be stable is $\alpha \ge \lambda$. For the acceleration to 
begin only after a sufficiently long scaling matter era, the scale $f_0$ 
has to bet set suitably.}
\end{center}
\end{figure}

If, however, the coupling becomes significant at late times, the 
situation will change. Since for the fixed point E we have $H^2 \sim 
\rho \sim a^{-3(1+w_m)}$, the last term in the Friedmann 
equation (\ref{friedmann}) scales like 
$\rho_f \equiv H^3f'(\phi)\dot{\phi} \sim a^{-3(1+w_m)(2-\alpha/\lambda)}$. 
This 
follows from the tracking behaviour of the scalar field; since $\phi' = 
3(1+w_m)/\lambda$, we have that $\phi = \phi_0 + 
3(1+w_m)\log(a)/\lambda$, 
which implies $f'(\phi) \sim a^{3\alpha(1+w_m)/\lambda}$. Thus we find that
the effective energy density due to the presence Gauss-Bonnet term, $\rho_f$, 
dilutes slower than the energy density due to ordinary matter, $\rho_m$, 
if and only if $\alpha>\lambda$. We have confirmed this simple 
result by numerically integrating our system (\ref{system}) (see 
Appendix \ref{eqs} for details). Then we found that as the coupling 
begins to affect the evolution of the field, 
it will always be passed to the fixed point $C$ from the saddle point $E$ as
depicted in FIG.\ref{pp}. Hence the universe is approaching a 
de Sitter phase and an acceleration occurs as observations indicate. 

\begin{figure}
\begin{center}
\includegraphics[width=0.4\textwidth]{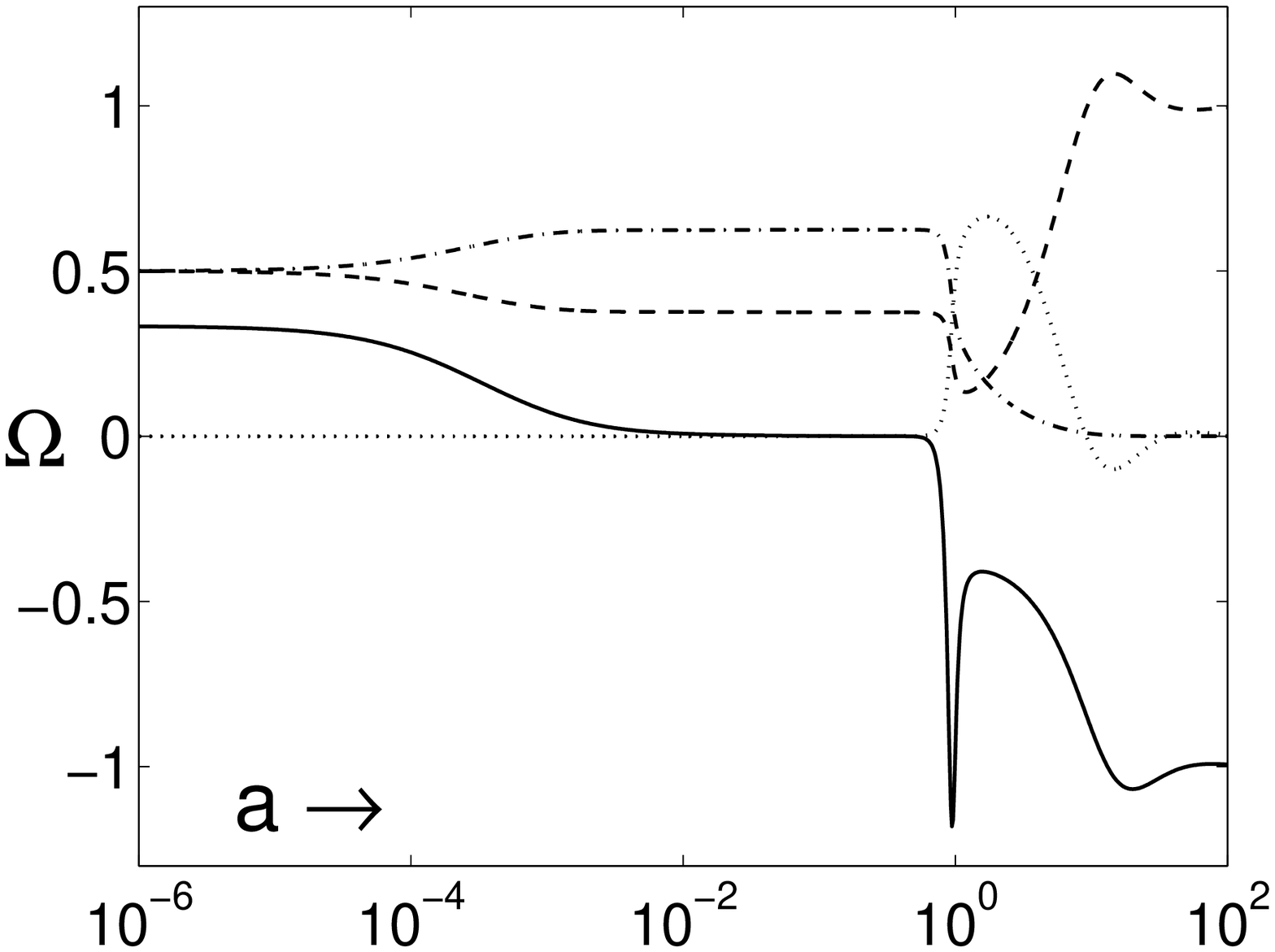}
\includegraphics[width=0.4\textwidth]{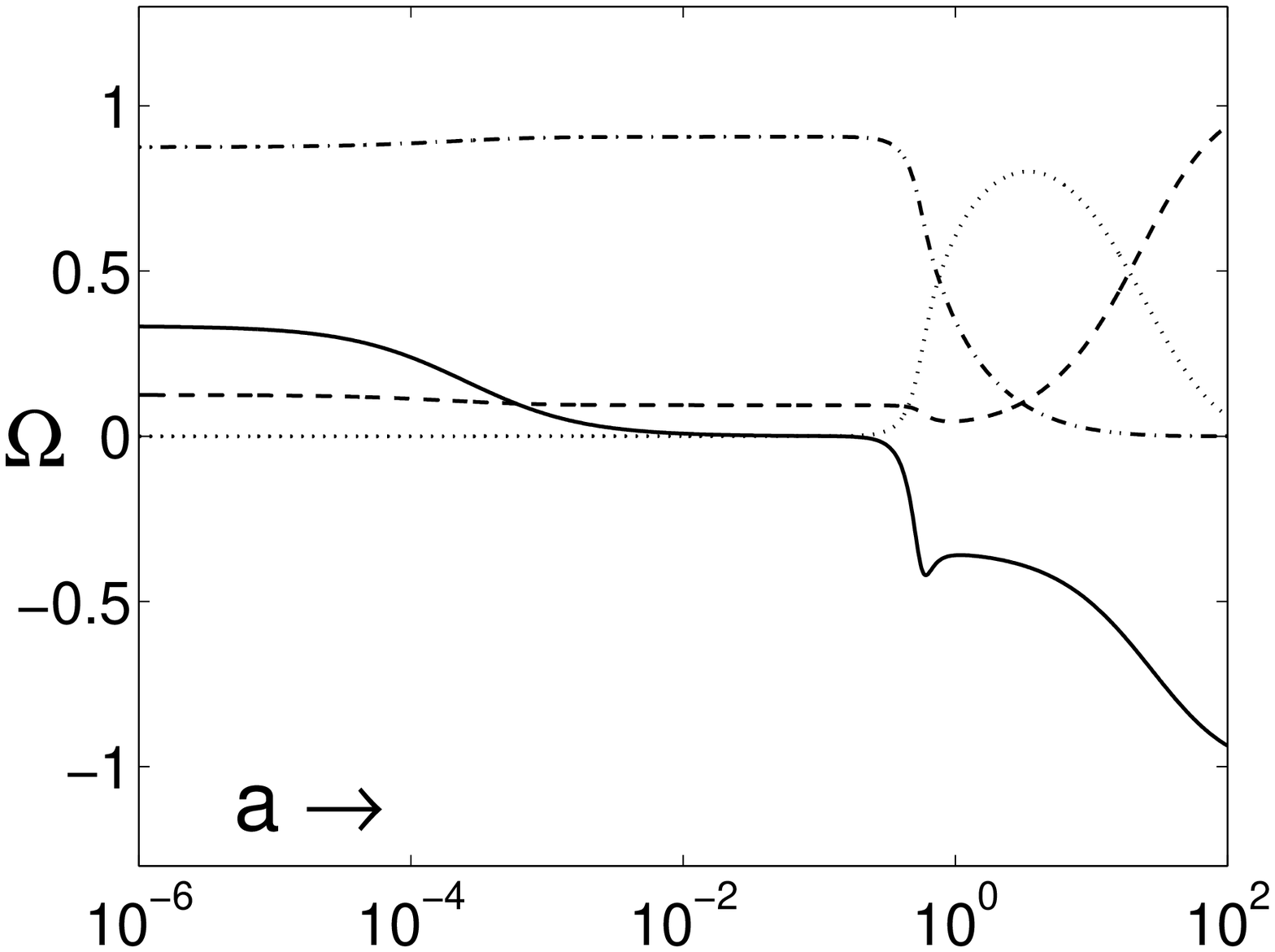}
\caption{\label{omegas} The fractional energy densities
for matter, $\Omega_m$ (dash-dotted line), the scalar field
$\Omega_\phi$ (dashed line) and Gauss-Bonnet
term, $\Omega_f$ (dotted line). The solid line is the total
equation of state $w_{eff}$. The upper panel is for $\lambda=4$
and the lower panel for $\lambda=8$. For both cases $\Omega_m^0=0.35$
and $\alpha=20$. The transient phantom era in the upper plot
is caused by the dynamics of the coupling.}
\end{center}
\end{figure}

Furthermore, during the transient epoch between matter and scalar field
domination, the universe can enter into a transient phantom stage (with 
$w_{eff}<1$), which is possible due to the $\rho_f$ term in the 
Friedmann equation. Even if $w_{eff}>-1$, a result 
$w_{eff}<-\Omega_{\phi}$ would imply effective phantom dark
energy when $\Omega_{\phi}$ is interpreted as due to an uncoupled dark energy
component. We plot two typical examples of such evolution in 
FIG. \ref{omegas} with $\lambda = 4$ or $\lambda=8$  and $\alpha=20$. 
Note that $\Omega_m + \Omega_\phi>1$ is possible, since the 
(effective) Gauss-Bonnet energy density $\rho_f$ can be negative when 
the field momentarily rolls backwards. This can occur since, as noted previously,
when the universe begins to accelerate the Gauss-Bonnet term flips its
sign, and might overturn the slope of the effective potential. 

Having now found the background expansion in the model, we can check
what parameter values are needed to produce it. We see from Fig.(\ref{omegas}) that 
all the variables $x$, $y$ and $\mu$ are roughly of the order of
one today. Then we have that 
\be \label{est_f}
\mu_0 & = & 8\kappa^2f'(\phi_0)\dot{\phi_0}H_0 = 8\kappa^4 H_0^2x_0\alpha f_0 e^{\alpha\phi_0/\sqrt{2}}
\nonumber  \\
& \sim & 10^{-120} e^{\alpha\kappa\phi_0/\sqrt{2}} \sim 1,
\ee
if we set $f_0$ of order one as expected in string theory 
and recall that $(\kappa^2H)^2 \approx 10^{122}$.
Note that though $f(\phi_0)$ has to be very large,
%, as already noted in Ref.\cite{Esposito-Farese:2003ze,Esposito-Farese:2004cc}, 
because of the exponential amplification of $f(\phi)$ we can here have 
avoid introducing huge numbers into the Lagrangian (\ref{action}).
Then it is seen that the present value of the scalar field is about $\kappa\phi_0 
\sim 390/\alpha$. It seems that typically our models the field has run to perhaps
a couple of dozen Planck masses. Then the scale of the potential can be deduced from 
\be \label{est_p}
y = \kappa^2V(\phi_0)/H_0^2 \sim \kappa^4 V_0 e^{280(1-\lambda/\alpha)} \sim 1.
\ee
In the case $\lambda=\alpha$ the potential scale $V_0$ 
would be of the order of $M_P^4$, where $M_P = 1/\kappa$ is the Planck mass. The
mass of the field at present, $m_\phi \equiv \sqrt {-V_0''(\phi_0)}$, on the other
hand, will turn out to be very small, as usual with quintessence fields.
The potential energy is exponentially suppressed, while the coupling is exponentially enhanced 
as a function of the field; this allows to consider natural magnitudes for both
simultaneously. However, for the same reason these estimates are also very
sensitively dependent on the particular values of the parameters. Since in the 
models we consider here $\alpha>\lambda$, the potential scale is in fact usually
much less than $M_P^4$. According to the very rough estimates (\ref{est_f}) and 
(\ref{est_p}), if $f_0  \sim 1$, then $\alpha$ could be allowed to be only
slightly less than $\lambda$, for the potential scale to be of the order of, say 
$V_0^{1/4} \sim (m_{3/2}M_P)^{1/2} \sim 10^{10}$-$10^{11}$ GeV, which corresponds
to the effective potential for low energy supersymmetry breaking, with the
soft supersymmetry mass scale of about $m_{3/2}=10^3$ GeV. Nevertheless, this 
suggests that the scales of the model could be obtained from more fundamental physics. 
%though not
%perhaps quite generically within the simplest parameterization (\ref{exps}).
  
%If the dimensionless constant
%$f_0$ is of order one, we have that  
%We plot $w_{eff}$ for several values of $\alpha$ in FIG. \ref{eos_a}. 
%The situation is analogous for different values $\lambda$: the 
%evolution is steeper for larger $\alpha$. In this particular plot, the 
%equation of state today is less than $-1/3$, but it might not be 
%negative enough right now for good accordance with the SNeIa data. We 
%have, however, set the value of $\Omega_m$ today to $0.27$ for all the 
%models plotted here. It is thus clear from the plots that if we would allow 
%smaller - or larger, since there is a dip in $w_{eff}$ before $a=1$ - 
%matter densities at the present, it would be possible to get more negative 
%values for $w_{eff}$ today. 
%
%\begin{figure}
%\begin{center}
%\includegraphics[width=0.4\textwidth]{eos_a.eps}
%\caption{\label{eos_a} The effective equation of state of the
%universe for $\lambda=4$ and $\alpha=5,10,15,20,25,30$. The
%curves are steeper for steeper couplings (larger $\alpha$).
%All the curves will asymptote to $-1$, so no singularity awaits
%in the future. 
%\end{center}
%\end{figure}

\subsection{Linear Perturbations}
\label{perturbations}

The line-element Eq.(\ref{metric_bg}) generalizes in the perturbed FRW
spacetime to 
 \be \label{metric}
  ds^2 & = & -\left(1+2\varphi_A\right)dt^2 - 2a(t)\beta_{,i}dt dx^i \nonumber \\
       & + & a^2(t)\left[ \delta_{ij}\left(1+2\varphi\right) + \psi_{|ij}\right]dx^idx^j.
 \ee
We consider variables in the Fourier
space. The transformation is simple since at linear order each
$k$-mode evolves independently. We do not consider vector perturbations since their evolution
is unmodified\cite{Cartier:2001is}, and we will only briefly comment on tensor perturbations.
We characterize the scalar perturbations in a general gauge by the
four variables $\varphi_A,\beta,\varphi,\psi$. Some of these degrees of freedom are
due to arbitrariness in separating the background from the perturbations. One can
deal with these gauge degrees of freedom by noting that the homogeneity and isotropy of the background
space implies invariance of all physical quantities under purely
spatial gauge transformations\cite{Bardeen:1980kt}. 
Therefore one can trade $\beta$ and
$\psi$ to the shear perturbation 
 \be \label{gready}
 \chi \equiv a\beta+a^2\dot{\psi} \,.
 \ee
Since both $\beta$ and $\psi$ vary under spatial gauge
transformation, they appear only through the spatially invariant
linear combination $\chi$ in all relevant equations. In addition, one
can define the perturbed expansion scalar 
 \be \label{kappa}
 \kappa \equiv 3(H\varphi_A -
 \dot{\varphi})+\frac{k^2}{a^2}\chi\,.
 \ee
Using these variables in the so called gauge-ready formalism is useful in studying 
generalized gravity\cite{Cartier:2001is,Hwang:2005hb,tomimod2}. The general equations for scalar 
perturbations are given in Appendix \ref{peqs}. From here on we set $(8\pi G)^{-1/2}$ to unity to avoid confusion 
since it was also denoted by $\kappa$. 

\subsubsection{Small scales}

Because the Gauss-Bonnet interaction becomes here dynamically important at late times,
let us for now neglect the contribution from radiation and consider the 
matter dominated and subsequent epochs, when $w_m=c_m^2=\Pi_m=0$. 
Dropping the subscripts $m$, the continuity equations
(\ref{dd}) and (\ref{dv}) then read
 \be \label{dd_s}
 \dot{\delta} = -\frac{k^2}{a}v +\kappa-3H\varphi_A,
\quad
 \dot{v} = -Hv + \frac{1}{a}\varphi_A.
 \ee
We choose to work in a synchronous gauge, defined by $\varphi_A=0$. One can then see 
from Eq.(\ref{dd_s}) that it is convenient and legitimate to set $v=0$ to fix the 
remaining gauge mode. Then Eq.(\ref{dd_s}) tells that $\kappa = \dot{\delta}$. Thus already 
three variables are eliminated from the system. The momentum constraint (\ref{mc})
then allows to relate the evolution of metric potential $\varphi$ to the fluctuations
in the scalar field by 
  \be \label{mc_s}
 (\frac{1}{4}-2H\dot{f})\dot{\varphi} =  [H^2\dot{f}-Hf'-\frac{1}{8}\dot{\phi}]\delta\phi
   + H^2f'\dot{\delta\phi}.
  \ee
Note that in the conventional synchronous gauge notation \cite{Ma:1995ey} 
$\kappa = -\dot{h}/2$ and $\eta = -\varphi$, from which follows that the metric shear
$\chi = -a^2(\dot{h}/2+3\dot{\eta})/k^2$. The evolution of this $\chi$ is governed by
 \be
  \label{propagation_s}
  (1-8H\dot{f})\dot{\chi}  + 
  [H-8(H\ddot{f}+\dot{H}\dot{f}
  + H^2\dot{f})]\chi 
 \nonumber \\
   -  (1-8\ddot{f})\varphi = 
    - 8(\dot{H}+H^2)f'\delta\phi.
   \ee

\begin{figure}
\begin{center}
\includegraphics[width=0.4\textwidth]{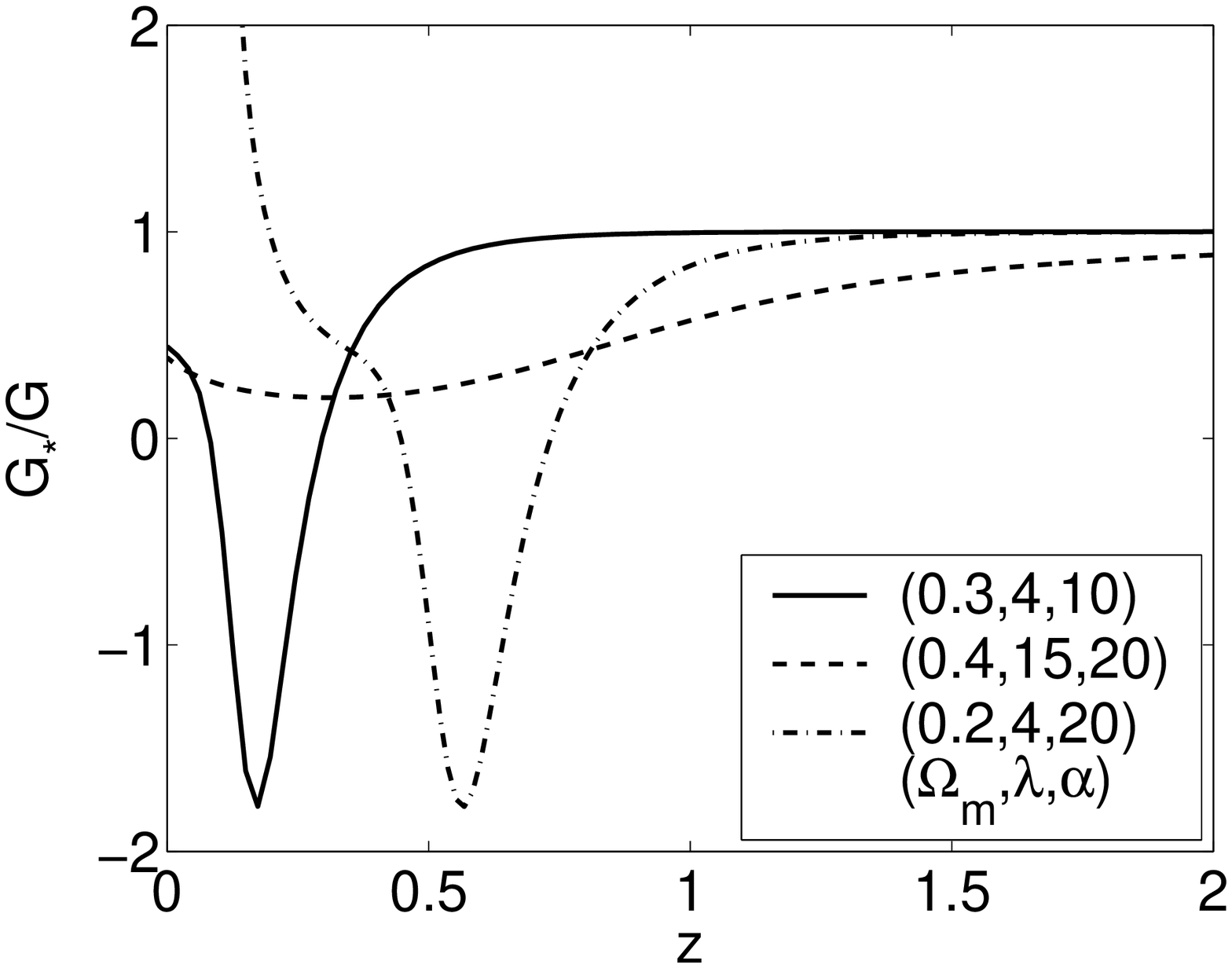}
\includegraphics[width=0.4\textwidth]{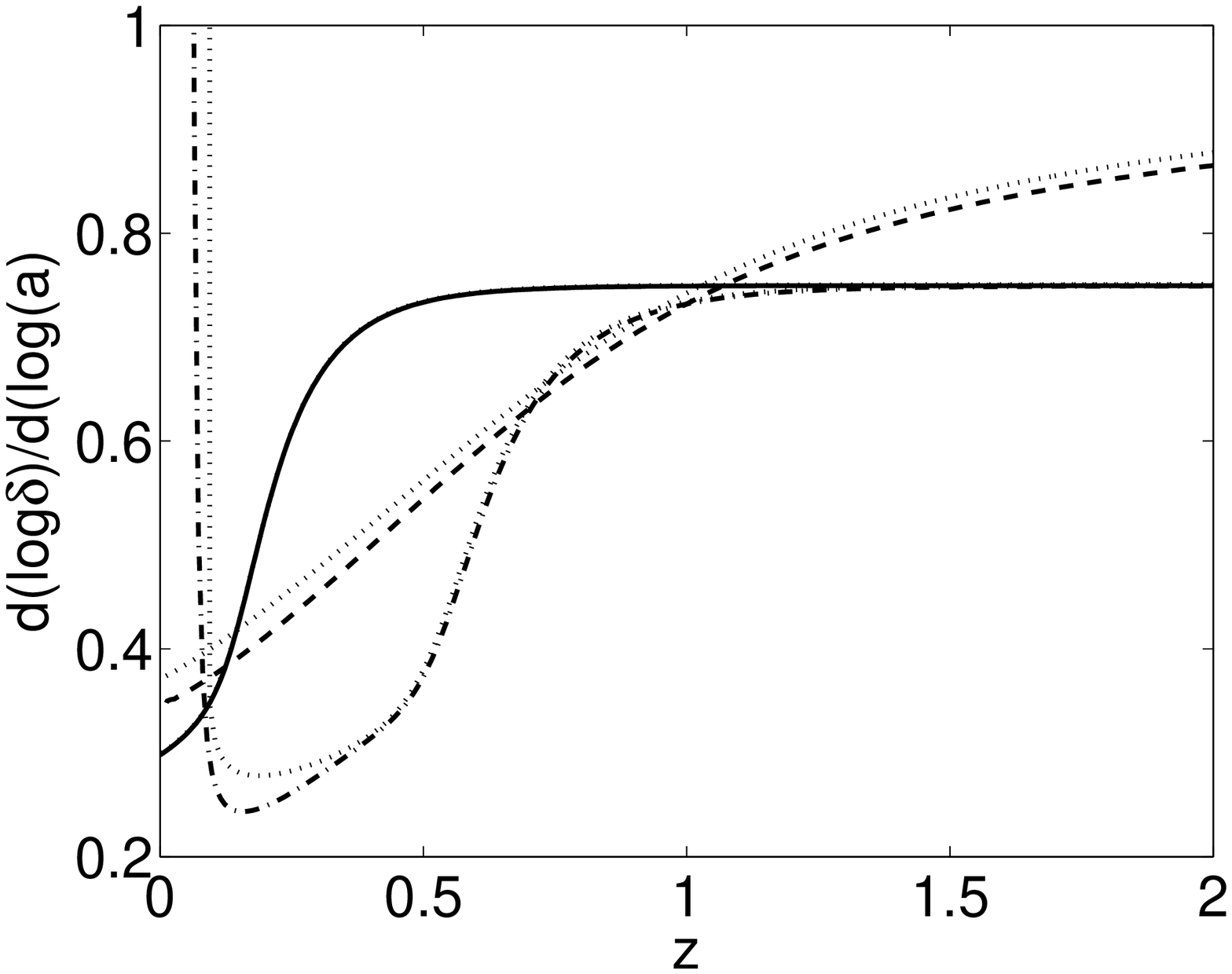}
\caption{\label{geff} Evolution of effective gravitational constant
$G_*$ (top panel) and of the dimensionless growth rate $d\log{\delta_m}/(d\log{a})$
(lower panel) as functions of redshift. We see that a divergence is possible for some
parameter combinations. The thick lines are numerical solutions
of the full linearized equations, and the thin dotted lines in the 
lower panel are solutions to the approximative equation (\ref{d_evol}). 
The agreement is excellent in most of
the parameter space, though in the two extreme cases depicted here
(dashed line corresponding to very slow transition, the dash-dotted line corresponding
to instability) deviation is visible.}
\end{center}
\end{figure}

Now we turn to consider small scales in order to find approximations for the
perturbation evolution. 
Then carefully neglecting the terms that
are subdominant at the large-$k$ limit yields the energy constraint
(\ref{admenergy}) in the form
 \be \label{admenergy_s}
  \frac{a^2\rho\delta}{2k^2} =  (1-8H\dot{f})\varphi
   -(H-12H^2\dot{f})\chi- 4H^2\delta f'\delta\phi.
 \ee
Equations (\ref{mc_s}) and (\ref{propagation_s}) imply that as usually, here the metric fluctuations
$\varphi$ and $\chi$ (or $\eta$) are small at large $k$, since their only sources
are the fluctuations of the scalar field. However, from Eq.(\ref{dd_s}) we
do not expect this to be the case for $\kappa$ (or $h$). The large part
of the spatially gauge-invariant variable $\kappa$ is separated, by the
definition (\ref{kappa}), into the gradient of the metric shear $\chi$, which
we have kept in Eq.(\ref{admenergy_s}). We don't either drop the scalar field
fluctuations or their derivatives, since their gradients can influence the
matter inhomogeneities. Since the field is very light, $m_\phi^2 \ll 
k^2$ is easily satisfied at the scales we consider, and the potential 
term could be dropped from the perturbed Klein-Gordon equation.
Finally, we have the small-scale limit
of the Raychaudhuri equation (\ref{ray}), 
 \be\label{ray_s}
\frac{a^2\rho\delta}{2k^2} & = & -(1-8H\dot{f})\dot{\chi}
+ (8H\ddot{f}- 4H^2\dot{f}+8\dot{H}\dot{f})\chi 
\nonumber \\
& - & 8(\ddot{f}-H\dot{f})\varphi -4(H^2+2\dot{H})f'\delta\phi.
 \ee
With some algebra, it is now possible to deduce the evolution equation
for the matter overdensity $\delta$. One arrives at\footnote{Assuming the Birkhoff
theorem (or Jebsen-Birkhoff theorem\cite{morad}) and deriving the evolution equation for spherical overdensities at
subhorizon scales as suggested Ref.\cite{Lue:2003ky} for some different modified gravity models than the present,
would give $G_*^{J-B} = 2[2\epsilon(1+\epsilon)+\epsilon']/(3\Omega_m)$. The reason for discrepancy is 
that the Jebsen-Birkhoff theorem is not respected in the model.} 
\be \label{d_evol}
\ddot{\delta} + 2H\dot{\delta} = 4\pi G_*\rho \delta.
\ee
The effective gravitational constant $G_*$ seen by the matter inhomogeneities
depends rather non-trivially on the evolution of the background quantities, and in terms of our
dimensionless variables defined in Eq.(\ref{variables}) it can be written as
\be \label{g_eff}
G_* = 4\frac{-x^4 + \mu^2(1+\epsilon)^2 +
        x^2[2(1 + \epsilon)(\mu-1) + y]}{x^2[4 + \mu(5\mu -8)] - \mu^2[6(1 + \epsilon)(\mu-1) + y]}
\ee
We plot $G_*/G$ in the upper panel of Fig. \ref{geff} for selection of parameter values. 
It is clear that at high redshifts, when the scalar field and so coupling term is negligible, all the models reduce to the standard value of G.

When the coupling can be neglected ($f', f'' \approx 0$),
$G_*$ reduces to unity. This is understandable since standard uncoupled
quintessence does not cluster at small scales, and the evolution equation
for linear growth stays unmodified. On the other hand, in the
slow-roll limit where $\dot{\phi}$ and $\ddot{\phi}$ can be neglected
we have
\be
G_* =\frac{1 + 32f'^2(-2H^2 + \rho)^2}{1 + 32H^2f'^2(3H^2 - 2\rho)}.
\ee
In the case that the coupling is subdominant, so that $H^2 f',H^2f'' \sim \ve$,
one has
\be
G_*  & = & 1 - 8\left[\ddot{\phi}f' + \dot{\phi}(\dot{\phi}f'' - 2f'H)\right]
\\
 & + & 32f'\Big[\dot{\phi}^4f' - 4\ddot{\phi}\dot{\phi}f'H
\nonumber \\ \nonumber
 & - & 4\dot{\phi}^3f''H + f'(H^2 - \rho)^2 + 2\dot{\phi}^2f'(2H^2 + \rho)\Big],
\ee
where the first and second square brackets respectively embrace corrections
of the order $\mathcal{O}(\ve)$ and $\mathcal{O}(\ve^2)$.
%
%
%\end{figure}
%\begin{figure}
%\begin{center}
%\includegraphics[width=0.4\textwidth]{growth.eps}
%\caption{\label{growth} The growth rate of matter
%perturbations $F=(\log({\delta}))'$ when $\Omega^0_m=0.4$ and  
%$\lambda=5,6,7,8,9,10,11$ and $\alpha = 5\lambda$.}
%\end{center}
%\end{figure}
 
In any case, we can here make the important conclusion from Eq.(\ref{d_evol}) 
that the shape of the matter power spectrum is the same as for $\Lambda$CDM cosmology. The 
superhorizon scales, where our approximation breaks down and the density perturbation becomes gauge dependent, 
are not effeciently probed by the present large scale structure surveys. As we have shown that at 
subhorizon scales the growth rate of structure is the same at all scales, which is the also case when 
the acceleration is driven by vacuum energy, we cannot distinguish between these 
two very different scenarios by comparing the shape of matter power spectrum (assuming
of course that the primordial spectrum is the same in both of the cases). Therefore the primary 
constraints arising from the matter power spectrum could be deduced from the overall normalization.
Note that these conclusions are model independent, the only assumption being that the scalar field
is not very massive.
We did not make specific assumptions on the coupling or the potential or assume that
$\Omega_m$ dominates in order to deduce these results. The constraints from large scale structure
are still of course relevant, since it is possible that for some models the (scale-independent)
growth rate is heavily modified and the normalization of the power spectrum can be used to exclude such 
cases. On the other hand, the perturbations at large scales will affect the integrated Sachs-Wolfe effect
of the CMBR, which might constrain models although the cosmic variance weakens the  
significance of the low CMBR multipoles.

\subsubsection{Numerical solutions}

To investigate the cosmological effects of the Gauss-Bonnet coupling in more detail, 
we have numerically integrated the full perturbations 
equations\footnote{In numerical calculations we use a modified version of the 
public CAMB code\cite{Lewis:1999bs}, see http://camb.info/. Our numerical 
implementation is discussed in the Appendixes.} and computed the full 
matter power and CMBR spectra for the example model presented in the 
previous subsection. Then we have
to fix several other parameters besides the $\Omega_m^0$, $\lambda$ and $\alpha$. Since
these parameters govern the background expansion and thus fix the distance
to the last scattering surface, they determine the locations of the 
peaks in the CMBR spectrum, as will be seen in next section. However, the relative
peak heights depend on the amount of baryonic matter. To get results that are (roughly) compatible with
the WMAP observations we fix $\Omega^0_b/\Omega^0_c = 1.716$ for all the cases shown in this section. For 
simplicity, we set the scalar spectral index $n_S$ to one and the optical depth $\tau$ to zero and keep
the present Hubble constant $h$ fixed to $0.7$. The normalization we use for each model is
such that the first peak in the CMBR spectrum matches with WMAP observations. For comparison, we 
plot also the concordance $\Lambda$CDM model with its parameters set according to the
best-fit obtained by combining the WMAP and SDSS data \citep{Spergel:2006hy}. 

Firstly we can confirm the considerations based on appropriate approximations and to 
compute the linear growth rate found in the previous subsection. 
The dimensionless quantity $F = (\log({\delta}))'$ can be introduced to characterize 
the growth rate\cite{morad}. We have checked that its value is now independent of scale in 
the matter dominated era, when $k \gtrsim 0.02 h^{-1}$. Note that there \cite{Ferreira:1997hj}
\be \label{phicdm}
F_{MD} = 1 + \frac{5}{4}\left(1 - \sqrt{1-\frac{24}{25}\Omega_\phi}\right),
\ee
which is the small scale solution to Eq.(\ref{d_evol}) when the Gauss-Bonnet terms can be neglected. 
The subsequent evolution of this quantity is plotted for various parameter choices in the lower pannel of FIG. \ref{geff}, 
showing that this evolution is indeed reproduced to high accuracy by the simple second order 
differential equation Eq.(\ref{d_evol}). Discrepancy with the numerical solutions to the 
full linearized equations appear only for some extreme cases. 

\begin{figure}
\begin{center}
\includegraphics[width=0.4\textwidth]{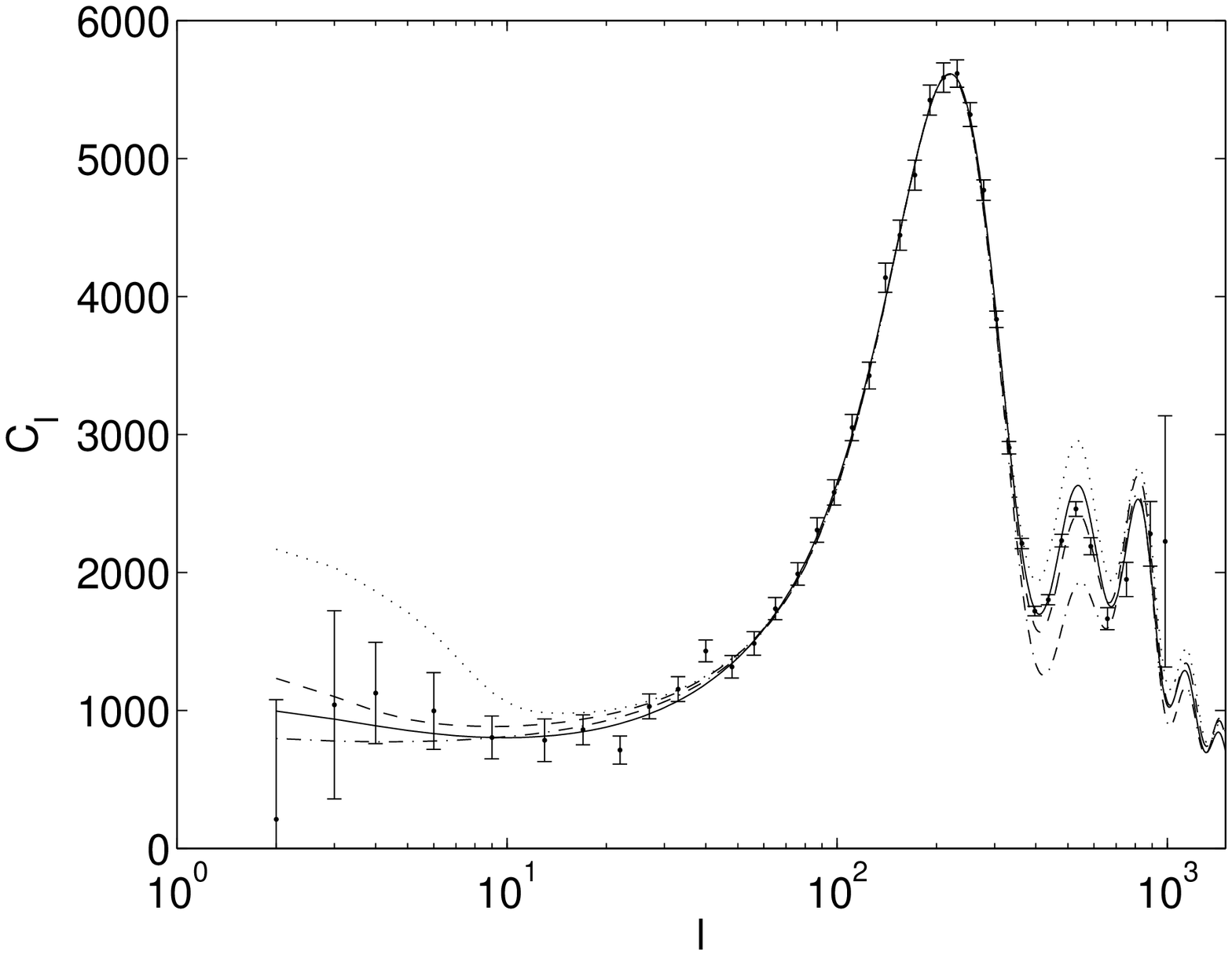}
\includegraphics[width=0.4\textwidth]{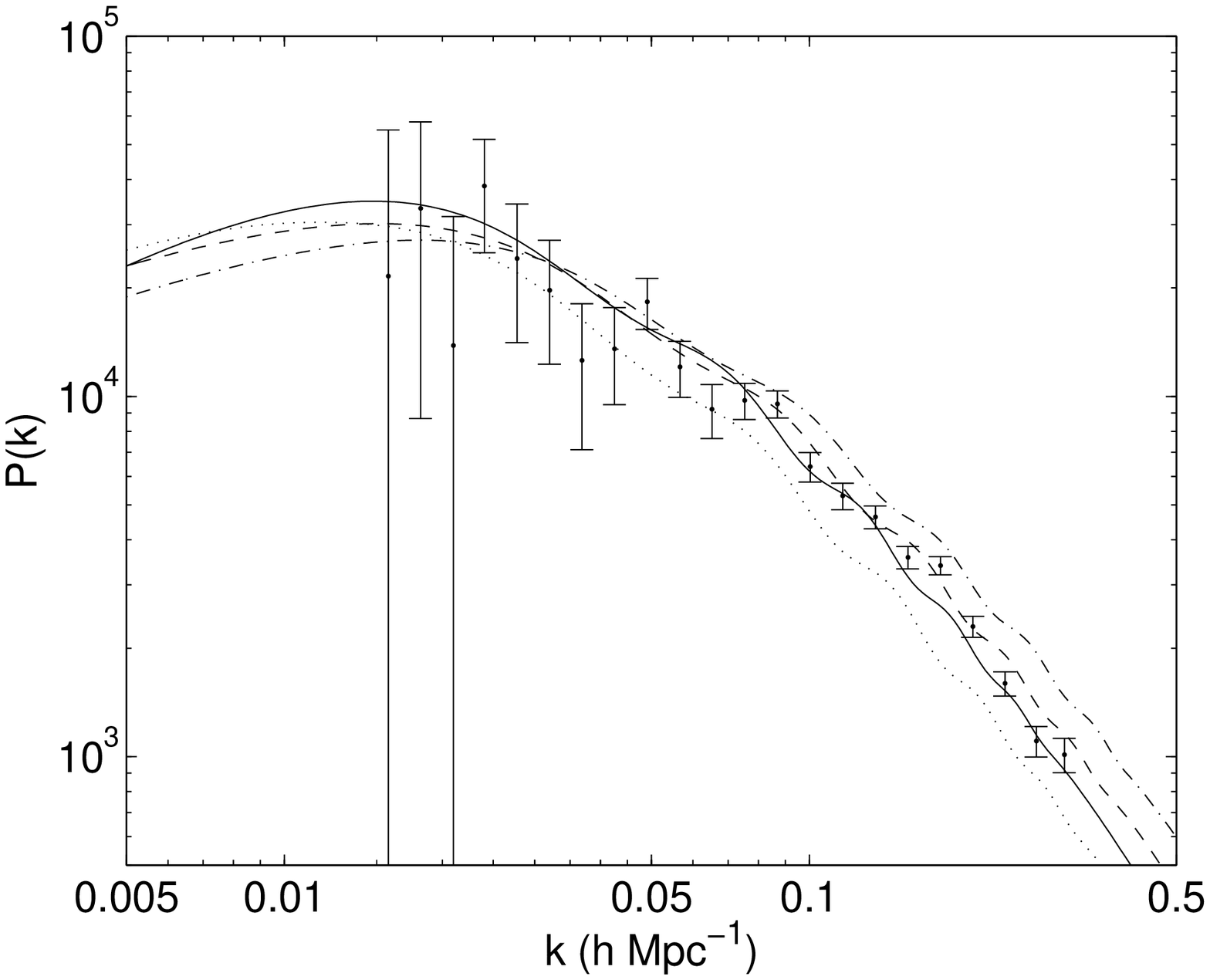}
\caption{\label{figs_1} The effect of matter density on the CMBR and matter power spectra.
Here $\lambda=6$ and $\alpha=20$. Dotted lines are for $\Omega^0_m=0.3$, dashed line for $\Omega^0_m=0.4$),
and dash-dotted for $\Omega^0_m=0.5$. The solid line is $\Lambda$CDM model.}
\end{center}
\end{figure}

\begin{figure}
\begin{center}
\includegraphics[width=0.4\textwidth]{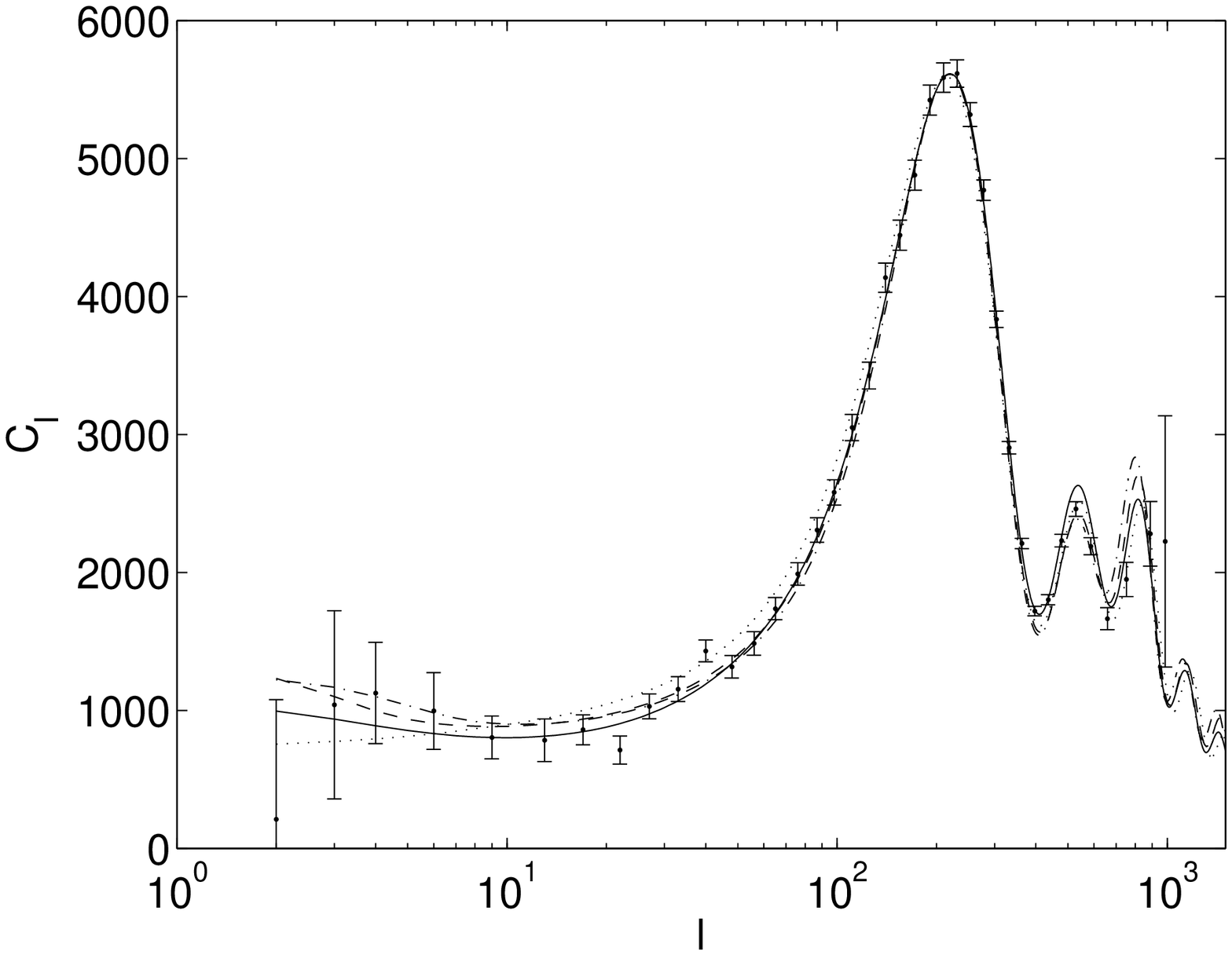}
\includegraphics[width=0.4\textwidth]{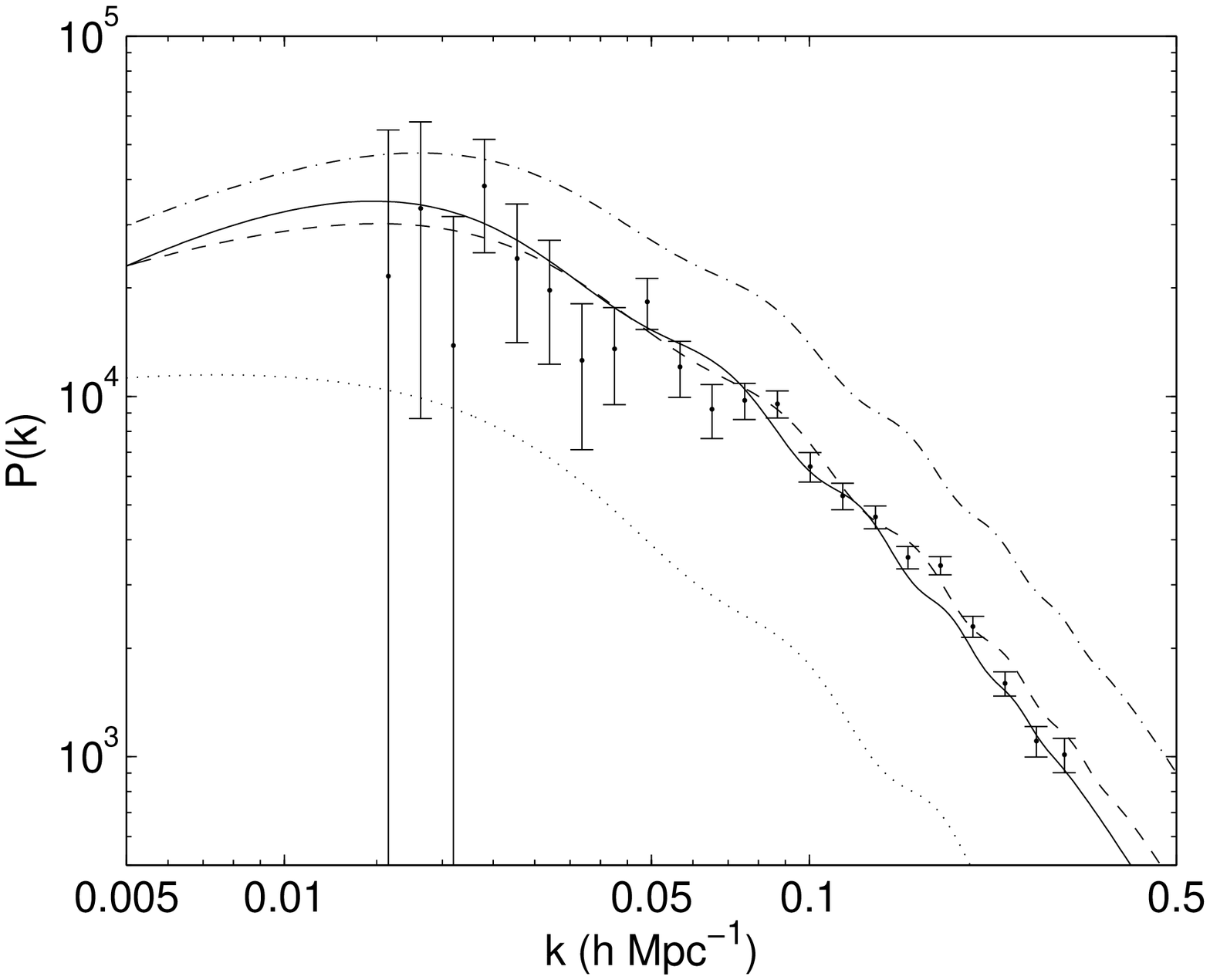}
\caption{\label{figs_2} The effect of the potential slope on the CMBR and matter power spectra.
Here $\Omega_m^0=0.4$ and $\alpha=20$. Dotted lines are for $\lambda=4.5$, dashed line for $\lambda=6.0$),
and dash-dotted for $\lambda=8.0$. The solid line is $\Lambda$CDM model.}
\end{center}
\end{figure}

\begin{figure}
\begin{center}
\includegraphics[width=0.4\textwidth]{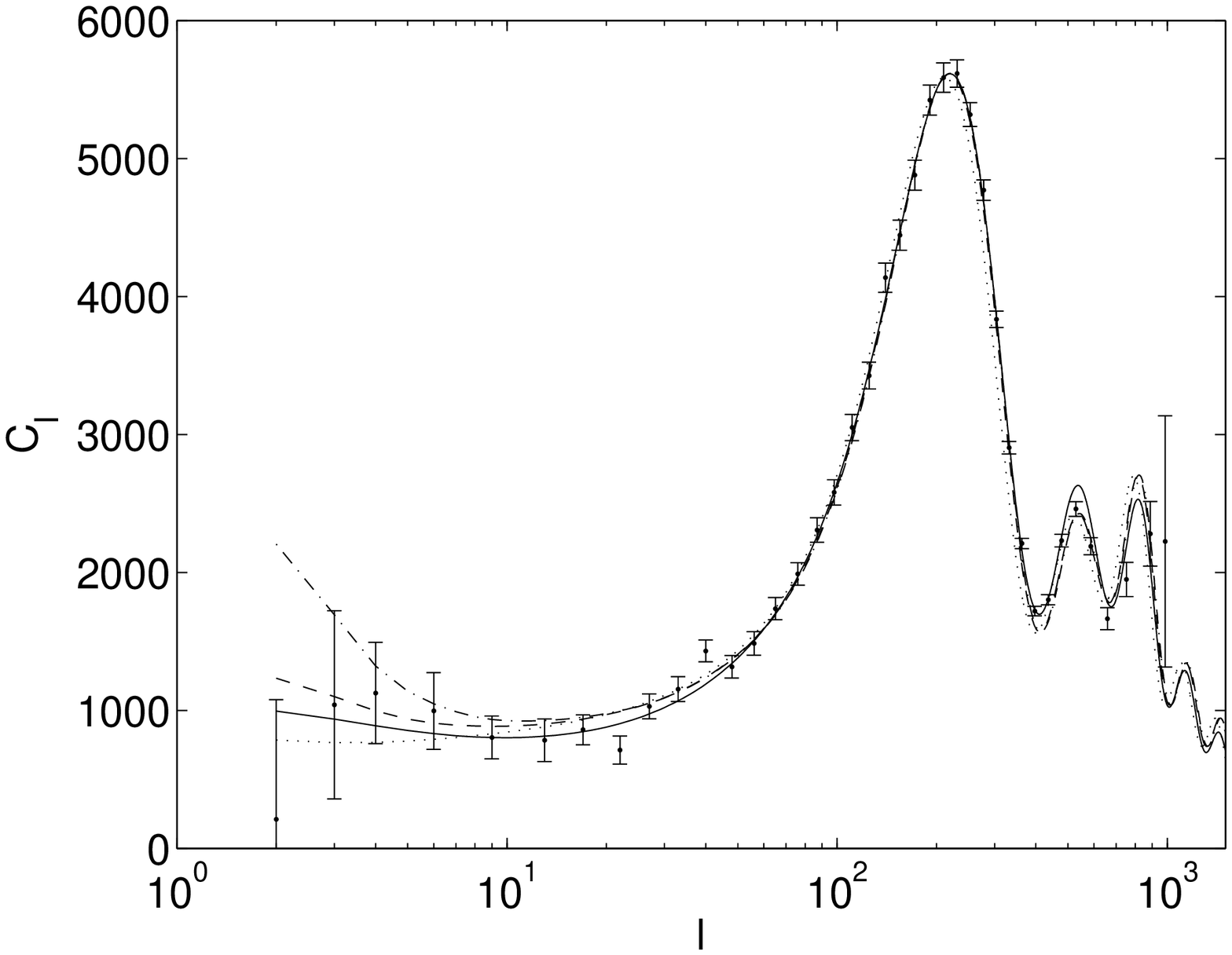}
\includegraphics[width=0.4\textwidth]{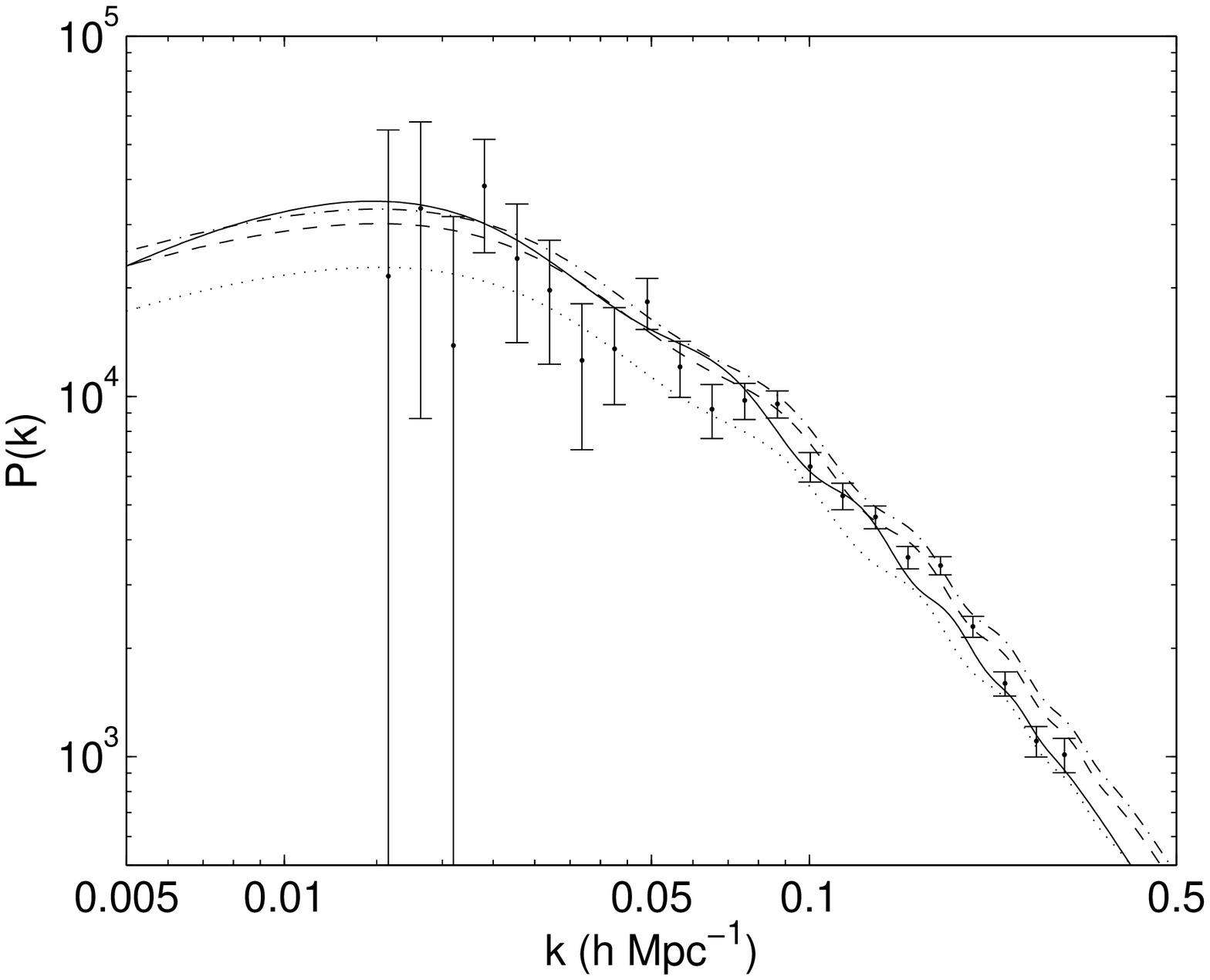}
\caption{\label{figs_3} The effect of the coupling slope on the CMBR and matter power spectra.
Here $\Omega_m^0=0.4$ and $\lambda=6.0$. Dotted lines are for $\alpha=10$, dashed line for $\alpha=20$),
and dash-dotted for $\alpha=30$. The solid line is $\Lambda$CDM model.}
\end{center}
\end{figure}

The effects on the CMBR and matter power spectra of varying the model parameters is shown in FIGS. (\ref{figs_1})-(\ref{figs_3}). Note 
that though we include the error bars in the figure, they only roughly indicate how the observations constrain
the models, since likelihoods should be computed using the window functions that depend on each
model. The CMBR error bars are from the three year WMAP data \citep{Spergel:2006hy} and the
error bars for the matter power spectra are provided by SDSS \cite{sloan}. A complete likelihood analysis
taking into account all these data would require exploring a vast parameter space, which is 
beyond the scope of the present study. Here we are rather interested in study how the model qualitatively 
differs from the concordance cosmology when the inhomogeneous evolution is 
considered. In the next Section we will  see that in fact 
it is enough to impose tight constraints on the model parameters considering only the background expansion.

FIG. (\ref{figs_1}) seems to indicate that the model favours matter densities about $\Omega^0_m \sim 0.4$. It seems 
clear that we cannot do without dark matter within this framework. 
%Another interesting point is that
%though for the concordance model the observations constrain $\Omega^0_m$ tightly to less than
%$0.3$, this is no longer the case in alternative cosmology featuring a modified gravitational
%sector. 
Low matter densities result in large relative contribution from the Gauss-Bonnet
term (this means very negative effective equation of state, as shown before), and leads to,
in addition to a modification of the peak structure in CMB, a large ISW effect. Requiring
viable normalization for both the CMBR and the matter power spectrum can also 
significantly constrain the allowed matter density.
In FIGs. (\ref{figs_2}) and (\ref{figs_3}) we show how the cosmological predictions
are changed when the slope of the potential or of the coupling are varied. The main imprint from different
potential slopes $\lambda$ seems to be in the normalization. For low values of $\lambda$, there is significant
contribution of the scalar field during the scaling matter era. This slows
down the rate of growth of matter inhomogeneities, see Eq.(\ref{phicdm}). Hence the fact that 
there is less structure nowadays than for larger $\lambda$ is not a consequence of the Gauss-Bonnet 
modification, but rather an effect of the presence of the scalar field
in the earlier scaling era. Finally, in FIG. (\ref{figs_3}) we see that the strenght of the 
coupling $\alpha$ might be more difficult to deduce from these data. With steep coupling slopes, the
scalar field domination takes place more rapidly and with more negative $w_{eff}$, which can somewhat amplify
the ISW effect. The contrary happens for smaller $\alpha$.

\subsubsection{Stability of perturbations and a possible end of acceleration}
\label{stability}

The effective gravitational constant Eq.(\ref{g_eff}) can in principle diverge . 
Indeed, we have found that in the model studied here $G_*$ typically diverges in the 
future, see FIG. \ref{geff}. 
For low matter densities or large $\alpha$ this can happen even before $a=1$, as happens 
in an example plotted in FIG. (\ref{geff}). Such a case would clearly be 
ruled out. However, in the model parameter combinations that would lead 
to this divergence of perturbations
before the present day, though may fit the SNeIa or other individual data, are not usually compatible 
with the combined constraints (these will be presented in the next section). 

Matter perturbations will at least for a while grow explosively, but this 
is different from the Big Rip singularity in phantom models where the background 
energy density will approach infinity in finite time. Note also that the perturbation
singularity does not correspond to the crossing of phantom divide, where $1+w_{eff}$  
changes its sign. In some phantom fluid dark energy models crossing this divide might seem 
problematical, since fluid perturbation equations (\ref{dd}) and (\ref{dv})
involve terms proportional to $1/(1+w)$. However, here the possibility of 
crossing the phantom divide has no direct relation to the  
divergence of perturbations linear perturbations. 

As the singularity of the linearized system is approached in future in these models, the linear 
approximation certainly breaks down at some point. 
The perturbative FRW description is no more valid as $\delta_m \rightarrow \infty$.
It seems perhaps plausible that then some kind of matter domination is restored, 
due to the energy creation from the Gauss-Bonnet interaction. 
%One justification for
%such a speculation is that the divergence of energy density perturbation $\delta_m$ 
%at the classical level, we are now discussing, seems to coincide with the 
%appearance of a ghost at quantum field theoretical level (which is shown shortly), to which 
%one can associate, among other curiosities, massive particle overproduction. 
It might be that some singularity really would occur, but this could only be assessed by considering the
field equations beyond the linear perturbation theory. Only in the case that nonlinear effects would somehow 
stabilize the growth of overdensities, would the de Sitter phase actually be reached as the background 
calculations imply. Otherwise, it is possible that the acceleration is transient. This would be favorable from a 
theoretical point of view, since the S-matrix formulation in present versions of string theory seems not to be 
consistent with an eternally accelerating universe\cite{Hellerman:2001yi}. 

Previously, instabilities have been found to occur in Gauss-Bonnet pre-big bang 
cosmology\cite{Kawai:1998ab,Kawai:1999pw}. Such have been also recently studied in the context of 
ghost conditions in Gauss-Bonnet cosmologies\cite{DeFelice:2006pg,Calcagni:2006ye}.
Though as well known, when expanded about de Sitter spacetimes, 
ghost modes do not appear in Gauss-Bonnet gravity, it has been pointed out that this does not necessarily hold
in the FRW background since such is characterized by non-constant curvature.
Quantum field theoretical consistency requires absence of ghosts, which means that a function $T$
multiplying the time derivatives of a variable to quantized should be positive. If a function $S$
multiplying the spatial gradient is negative, a related instability might occur when 
the sound speed $c^2_{S} = S/T$ is not positive definite. In fact, when 
an imaginary propagation speed appears in an action for a canonical field, one would 
expect already the solutions of its classical evolution equation to 
exhibit divergent behaviour. This is what we found here. It can be shown 
that the canonical action for the 
potential $\Phi$ in the case $\rho_m=0$ features 
an effective propagation speed \cite{Hwang:2005hb}
\be \label{sound}
c_S^2 = \frac{-x^2[4+\mu(5\mu-8)]+\mu^2[6(1+\epsilon)(\mu-1)+y]}{(\mu-1)[3\mu^2-4(\mu-1)x^2]}.
\ee 
When this exceeds one, the speed is superluminal and causality could be violated; when it is negative, the
stability might be lost. Since Eq.(\ref{sound}) corresponds to propagation of the scalar field in vacuum, the 
same $c_S^2$ would be found for any other gauge-invariant variable as there is only one scalar degree of freedom. 
However, in dark energy cosmologies featuring the Gauss-Bonnet term one should take into account both matter and 
the corrections to Einstein gravity. This does not complicate the analysis of the tensor mode, since it
decouples from other fluids (as long as they are isotropic), and one can just plug the background solution (for which
matter has been taken into account) into an expression for the tensor $c_T^2$ (which is the same
as without matter). However, scalar modes are non-trivially coupled. In the presence of matter,
Eq.(\ref{sound}) does not describe the evolution of $\Phi$. It might still describe the evolution
of $\delta\phi$ in some suitable gauge at a limit where the impact of matter perturbations on the field 
fluctuations can be neglected, but a priori it is not clear that such a limit of cosmological equations exists. 
For the models considered here, one notes by comparing Eqs.(\ref{sound}) and (\ref{g_eff}), that the 
linearized matter perturbations diverge precisely in the points where the 
propagation speed squared $c_S^2$ 
changes its sign. Our interpretation is that the fluctuation of the field $\phi$ becomes unstable when 
Eq.(\ref{sound}) dictates, and since it interacts (via gravitational potentials) with the 
fluctuation in $\rho_m$, also the latter signals instability at the same instant. 
Therefore the stability of the linear scalar modes is now indeed determined by a canonical 
action for any gauge-invariant scalar perturbation variable in the vacuum.   

\section{Constraints}
\label{constraints}

\subsection{Cosmological and Astrophysical Constraints}

Armed with all the equations describing the cosmological dynamics, we can now derive the
constraints arising from astrophysical and cosmological observations. In this section we
will consider kinematic tests related to the background expansion of the Universe: 
the SNeIa luminosity distance-redshift relationship, the CMBR shift 
parameter and the baryon oscillation length scale.

\subsubsection{Supernovae Ia}
\label{supernova}

The luminosity distance in a flat space is defined as
\be \label{dlz}
D_L(z) = (1+z)\int_0^z \frac{dz'}{H(z')}.
\ee
The distance modulus probed by SNeIa observations is then 
given by 
\be \label{distmod}
\mathcal{M} = m-M = 5\log_{10}\left(\frac{D_L(z)}{10\text{pc}}\right).
\ee
We use the ''Gold'' sample of 157 SNeIa \cite{Riess:2004nr}. There 
the observed magnitude $m$, its error and the redshift is given for each 
supernova. The absolute magnitude $M$ is unknown, and its effect is 
degenerated with $H_0$ entering the formula (\ref{distmod}) from 
Eq.(\ref{dlz}). We 
marginalize over $H_0$ (or equivalently, $M$), by integrating over the 
likelihood as $\chi^2 = \int \chi^2(H_0)e^{-\chi^2(H_0)/2}dH_0/\int 
e^{-\chi^2(H_0)/2}dH_0$. We marginalize similarly 
over the model parameter $\alpha$.

For comparison, we perform the likelihood calculations also with  
the Supernova Legacy Survey (SNLS) data \cite{Astier:2005qq} using then 
a bit different method. For each of the 115 supernovae in the SNLS set 
(labeled here with the index $i$), the distance modulus has been given as 
a function of the stretch factor 
$s_i$, color factor $c_i$ and the apparent magnitude $m_i^*$ as
\be
\mathcal{M}_i(M,a_1,a_2) = m_i^*-M+a_1(s_i-1)-a_2c_i.
\ee
We will treat the parameters $M$, $a_1$ and $a_2$ like cosmological 
parameters and find their values which maximize the likelihood. For 
individual supernovae, we keep the parameters $m_i^*$, $s_i$ and $c_i$ in 
their best-fit values. The uncertainty in $\mathcal{M}_i$ would in principle 
depend on $a_1$ and $a_2$, but should be kept fixed while optimizing 
these global parameters\cite{Astier:2005qq}. Thus we compute
\be
\chi^2 = \sum_{i=1}^{115} \frac{\mathcal{M}_i(M,a_1,a_2) - 
5\log_{10}\left[\frac{D_L(z_i,\alpha,\lambda,\Omega_m)}{10\text{pc}}\right]}
                   {\sigma^2_i(\mathcal{M}_B) + \sigma_{int}^2}.
\ee
The quantities with lower index $_i$ we get directly from the data, for the parameters
$M$, $a_1$, $a_2$ and $\beta$ we use their best-fit values, and for the 
intrinsic 
dispersion we use $\sigma_{int}=0.15$ for the nearby and $\sigma_{int}=0.12$ for the
distant supernovae. Similar procedure was used in Ref.\cite{Dick:2006ev}. We have checked
that modifications of the scheme change the results only very slightly. The SNLS data set can 
give little bit tighter contours than GOLD set for the model at hand. The contours are
shown in dotted lines in FIG. \ref{contours1}.

In principle when considering the supernovae observations in extended 
gravity theories, one should take into account the possible effects of 
modified gravity on the evolution of the supernovae\cite{Nesseris:2006jc}. Here 
we however assume that the SNeIa can be treated as standard candles. If 
variations in the scalar field $\phi$ can be neglected at the scales relevant 
to supernova evolution, it is determined by general relativity alone, 
though corrections from $R^2_{GB}$ might be crucial at cosmological scales and 
have to be taken into account in the luminosity-distance relation (\ref{dlz}). 
This assumption needs to be strictly justified; we return to a related
discussion in Section \ref{sm}.

\subsubsection{CMBR Peak Locations}

In addition, we compute the 
CMBR (cosmic microwave background radiation) shift 
parameter \cite{Odman:2002wj} given by
\be
\mathcal{R} = \sqrt{\Omega_m}H_0 \int_0^{z_{dec}} \frac{dz}{H(z)},
\ee
where $z_{dec}$ is the redshift at decoupling. 
This number $\mathcal{R}$ captures the correspondence between
the angular diameter distance to last scattering and
the relation of the angular scale of the acoustic peaks to
the physical scale of the sound horizon. 
Its value is expected to be very weakly model-dependent,
and it can be inferred rather accurately from the latest data\cite{Wang:2006ts}.  
We use here the parameter values
$\mathcal{R} = 1.70 \pm 0.03$ and $z_{dec} = 1088^{+1}_{-2}$. The resulting contours
are plotted with dashed lines in FIG. \ref{contours2}.

\subsubsection{Baryon Oscillations}

The imprint of primordial baryon-photon oscillations in the matter power 
spectrum is related to the dimensionless quantity $A$ in the following way
\be \label{bao}
\mathcal{A} = 
\frac{\sqrt{\Omega_m}H_0}{H^{\frac{1}{3}}(z_1)}\left(\frac{n_S}{0.98}\right)^{1.2}
\left[\frac{1}{z_1}\int^{z_1}_0\frac{dz}{H(z)}\right]^{\frac{2}{3}}.
\ee
The physical length scale associated with the oscillations is set by the
sound horizon at recombination, which can be estimated from the
CMBR data  \citep{Spergel:2006hy}. Measuring the apparent size of the
oscillations in a galaxy survey allows one to determine
the angular diameter distance at the survey redshift.
Together with the angular size of the CMBR sound horizon, the
baryon oscillation size can then be a powerful probe of the properties 
and evolution of the universe. From the large scale structure data one 
can infer that  \cite{Eisenstein:2005su} $A=0.469 \pm 0.017$ and $z_1 = 
0.35$, assuming the $\Lambda$CDM model. For the scalar spectral index we 
use the best-fit WMAP value 0.95 \cite{Spergel:2006hy}. The resulting 
constraints are plotted with solid lines in FIG. \ref{contours2}. 

\subsubsection{Combined constraints}

\begin{figure}
\begin{center}
\includegraphics[width=0.4\textwidth]{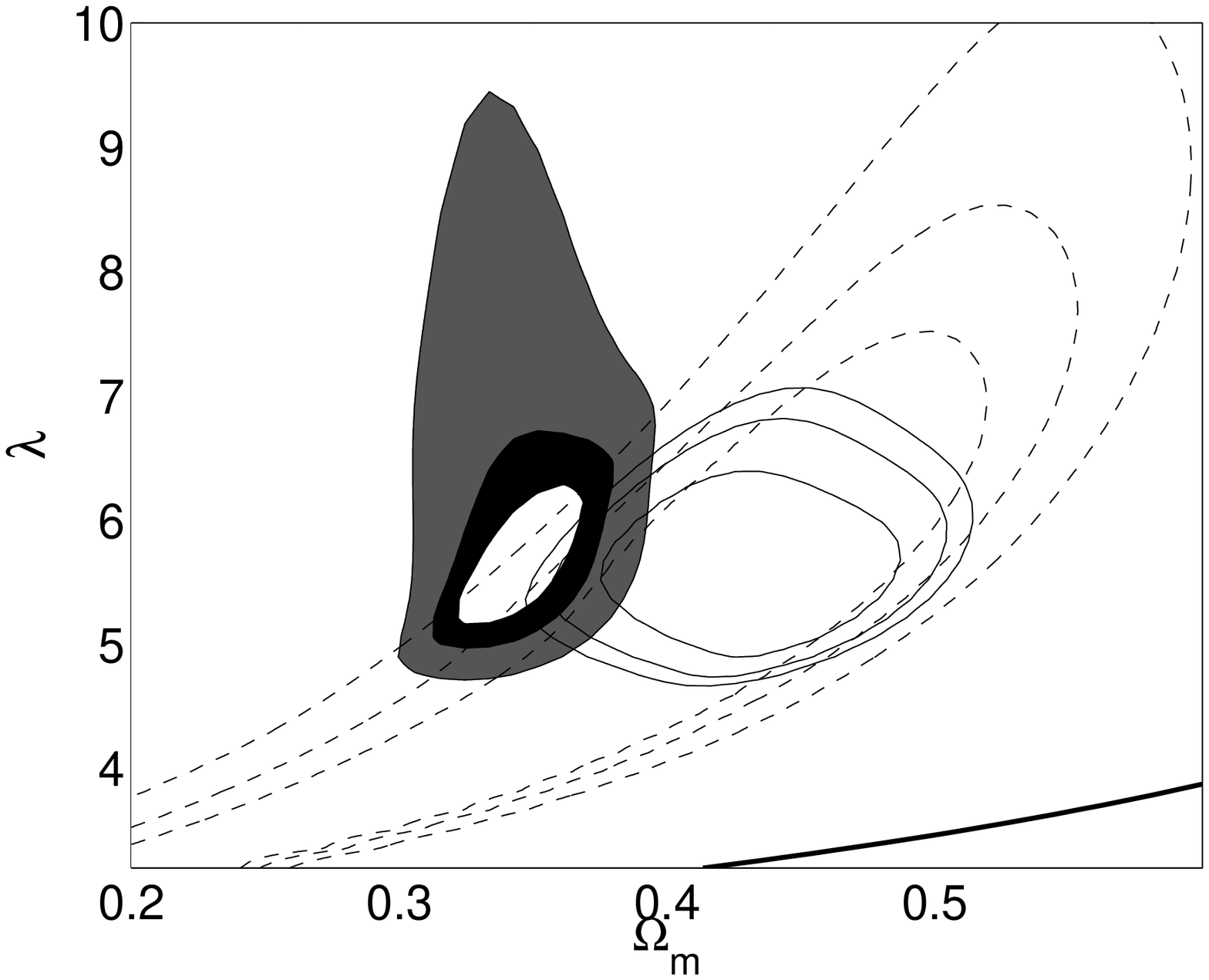}
\includegraphics[width=0.4\textwidth]{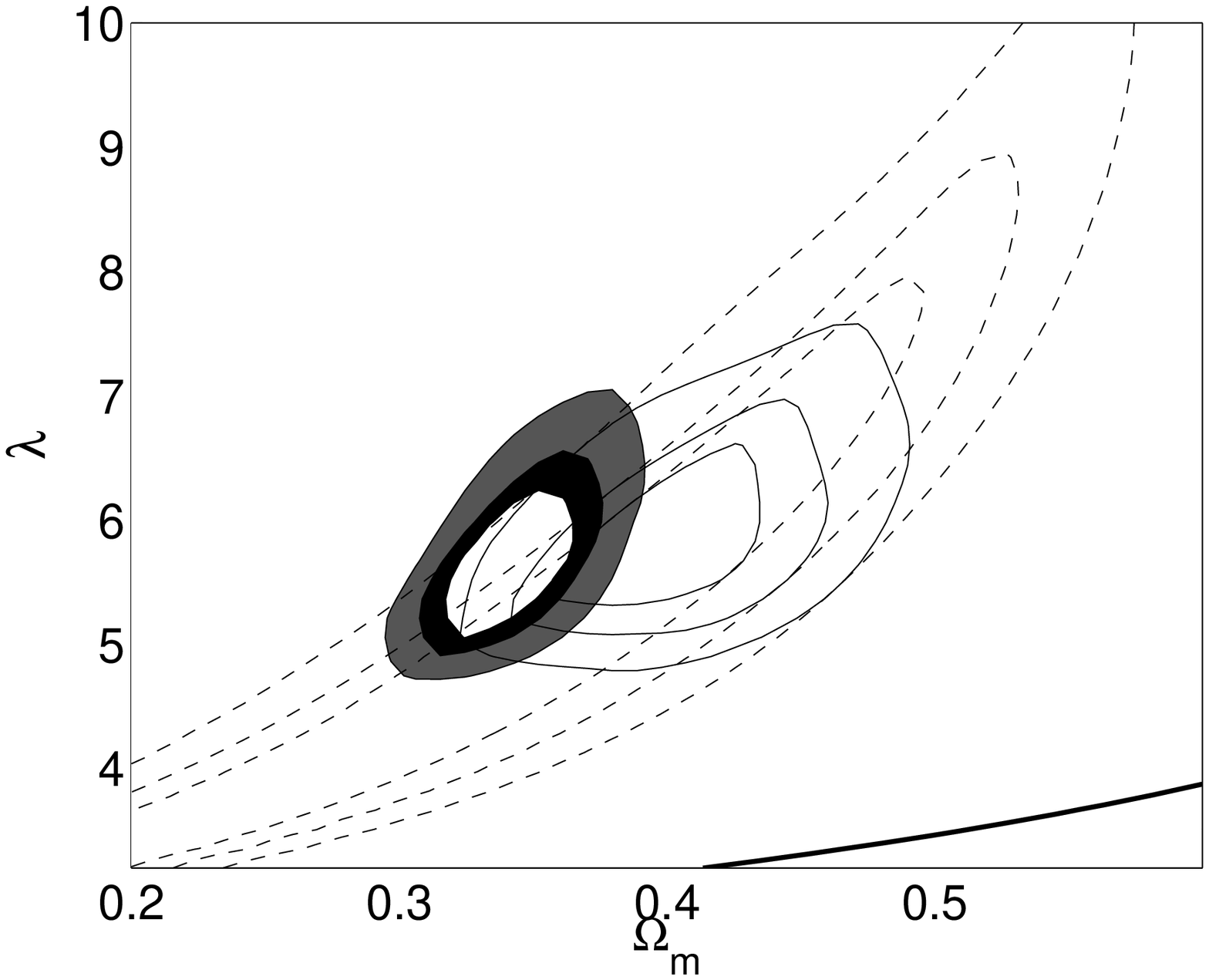}
\caption{\label{contours1} The confidence limits corresponding to $\Delta\chi^2=2.3$,
$4.61$ and $9.21$ for the model in the $\Omega_m$ - $\lambda$ plane. Dashed lines are constraints from 
the SNeIa data, solid lines from the combined SNeIa and CMBR shift 
parameter data, and shaded regions include in addition the baryon oscillation scale. Below the thick
line the scaling solution E is unstable regardless of $\alpha$. In the upper panel the SNeIa data
is the Gold data set, and $\alpha$ (together with $H_0$) is integrated over $1.5\lambda < \alpha <10\lambda$.
In the lower panel the SNeIa data is the SNLS data set and $\alpha$ (together with $M$, $a_1$ and
$a_2$) is set to minimize the $\chi^2$.}
\end{center}
\end{figure}

\begin{figure}
\begin{center}
\includegraphics[width=0.23\textwidth]{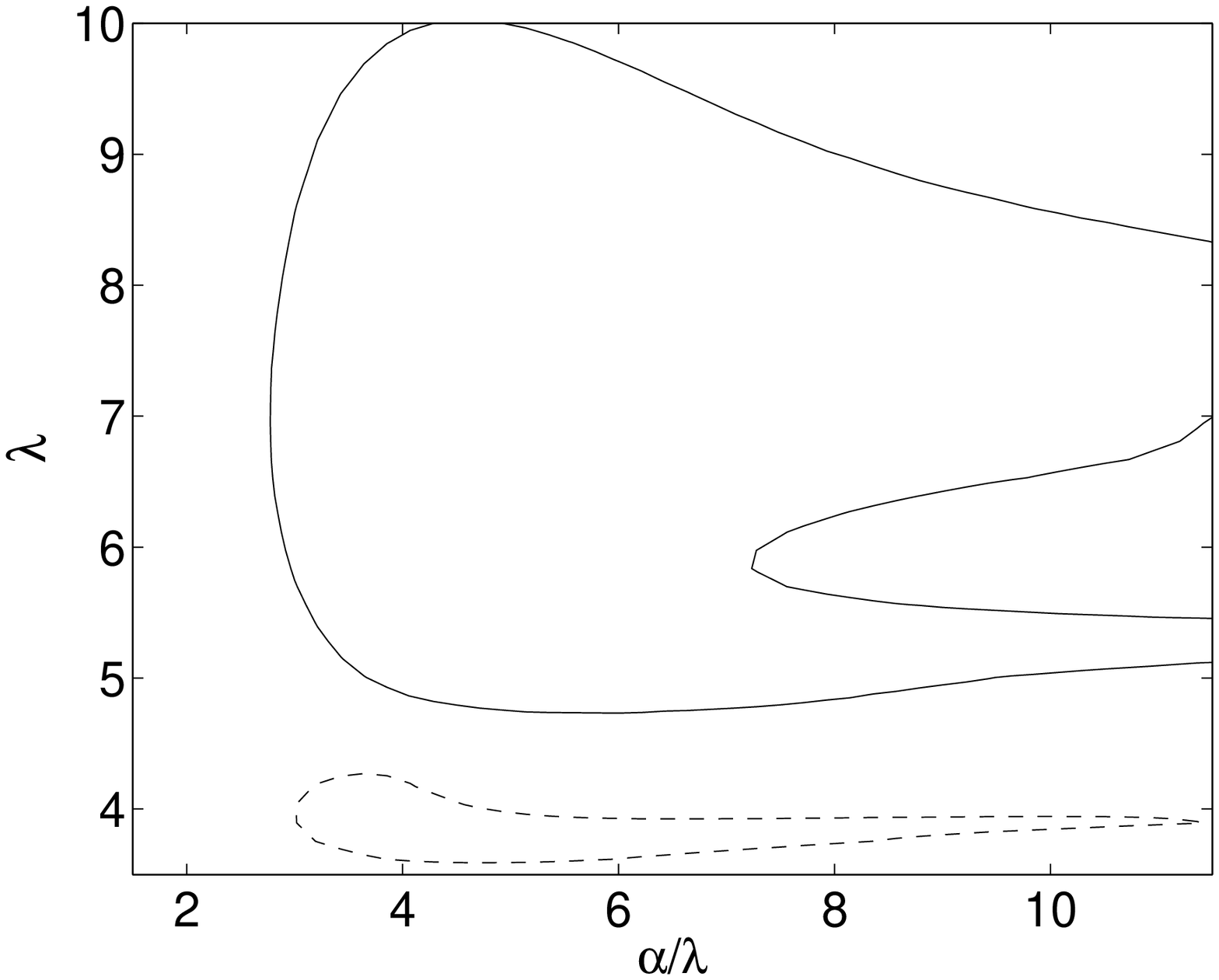}
\includegraphics[width=0.23\textwidth]{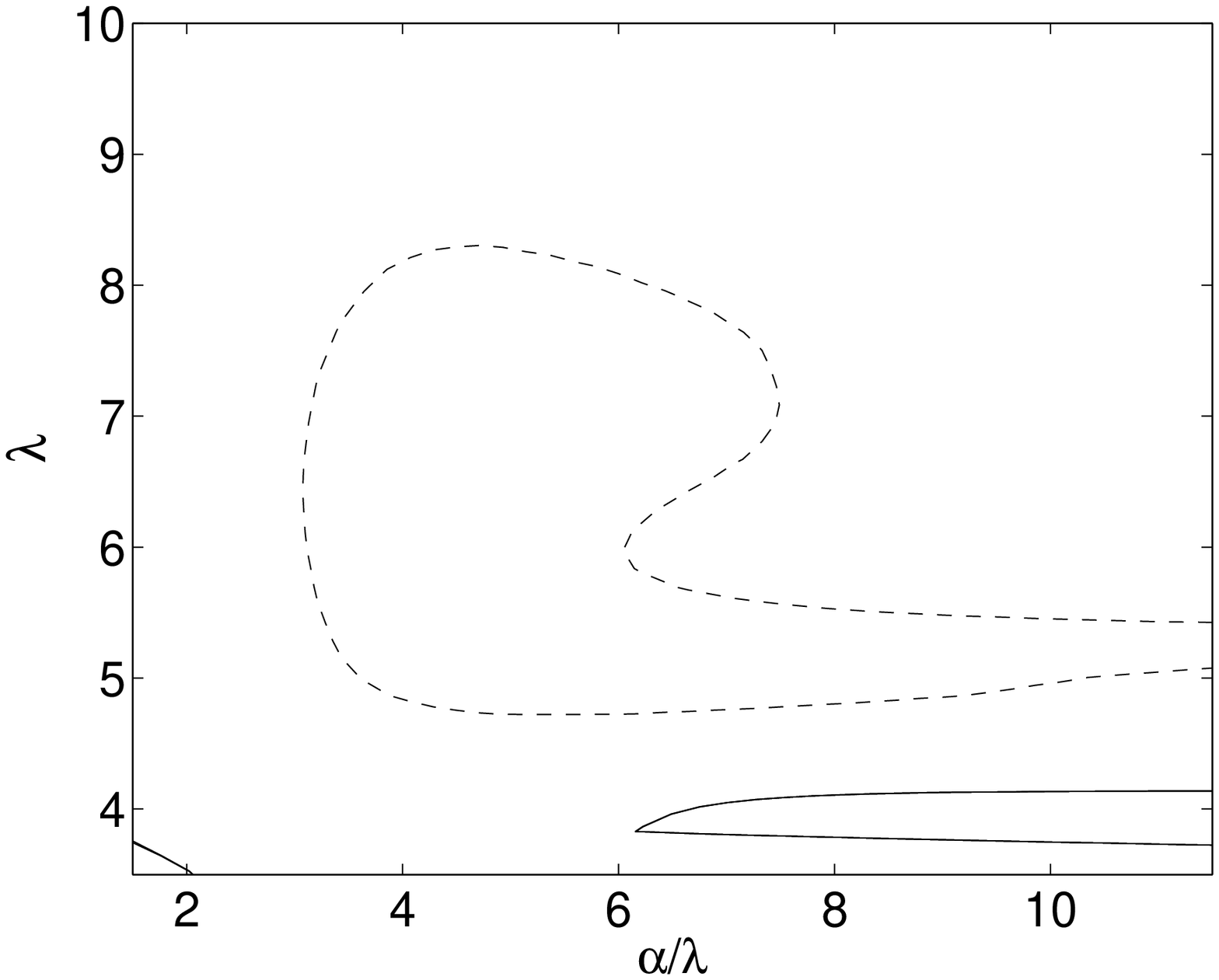}
\includegraphics[width=0.23\textwidth]{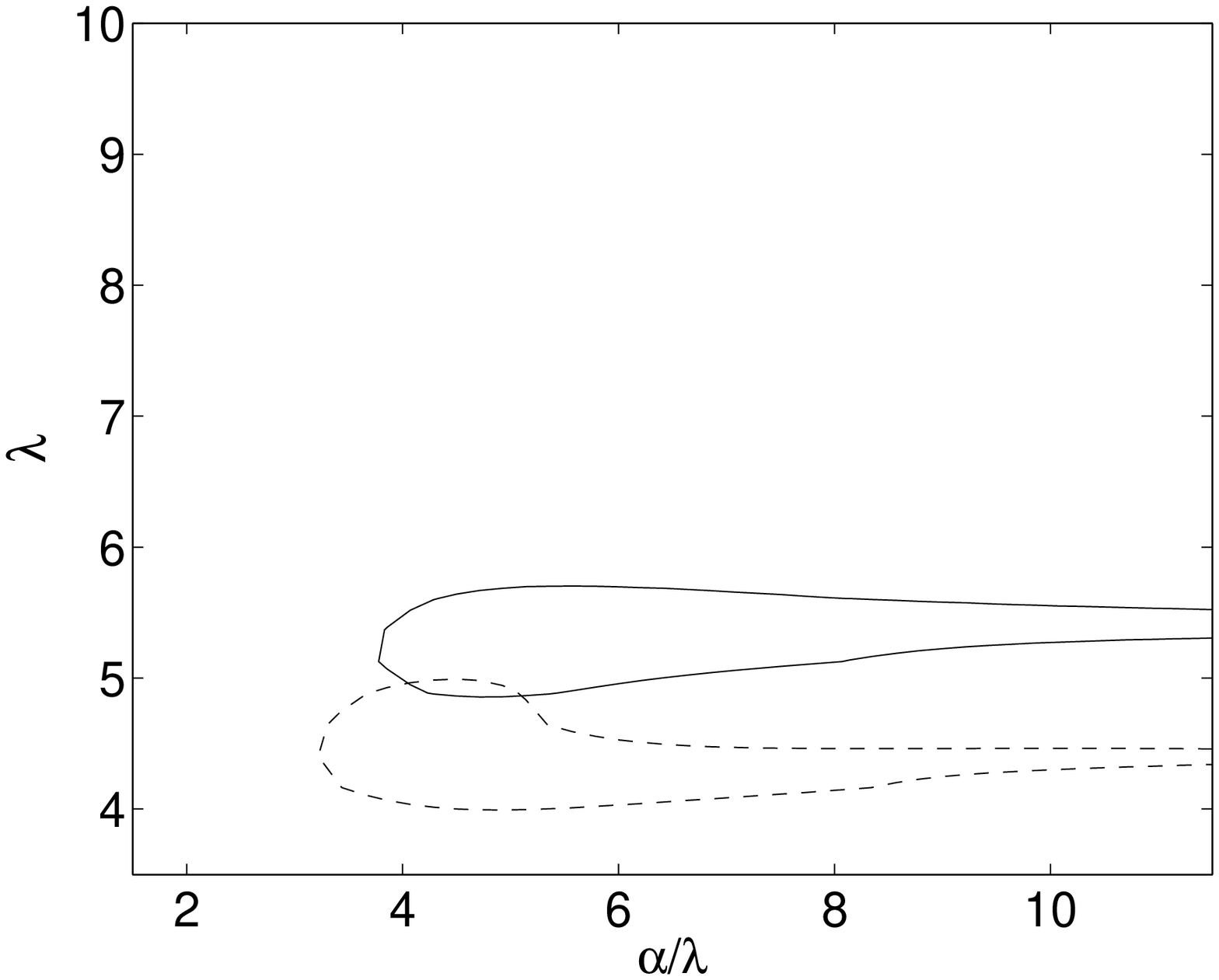}
\includegraphics[width=0.23\textwidth]{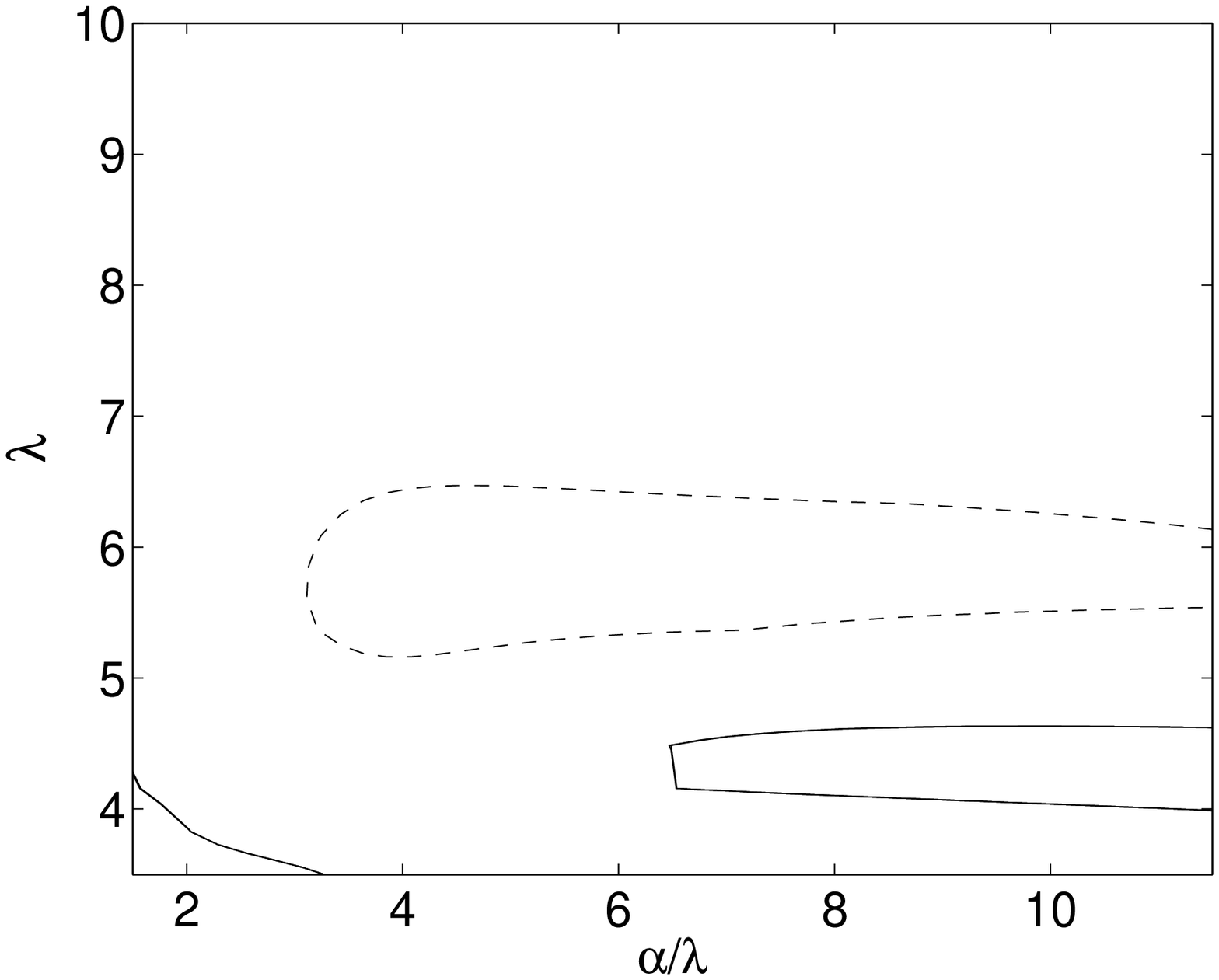}
\includegraphics[width=0.23\textwidth]{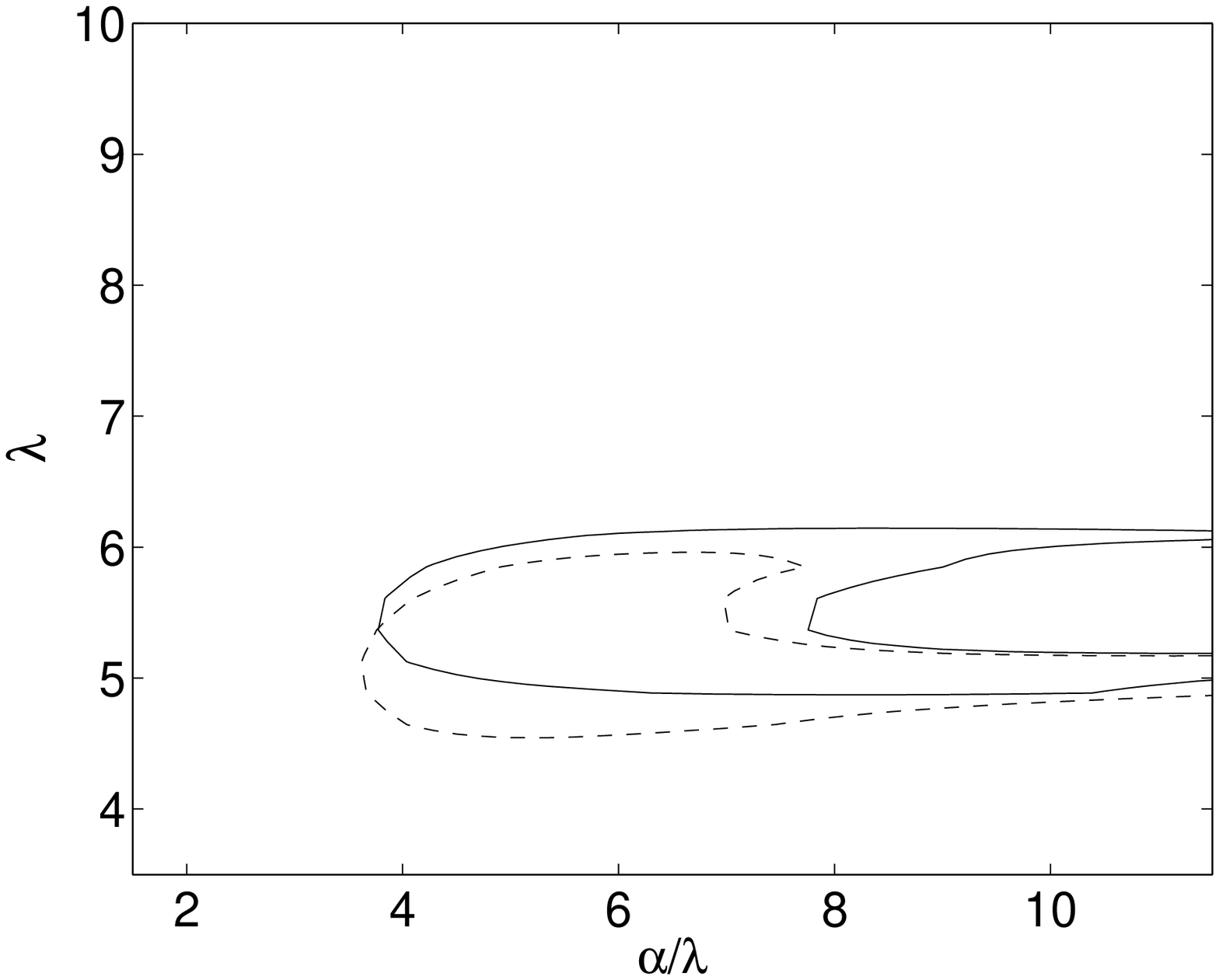}
\includegraphics[width=0.23\textwidth]{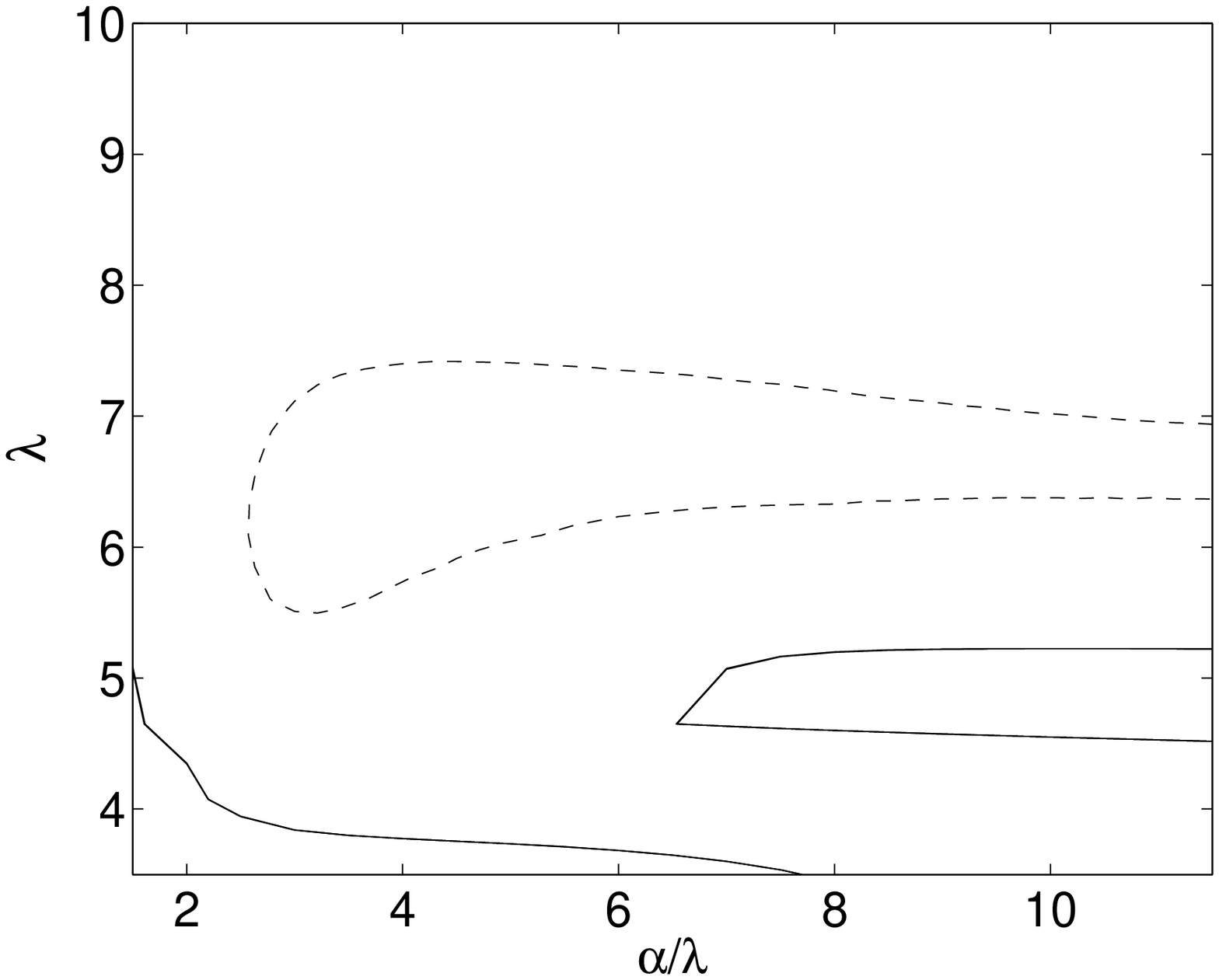}
\caption{\label{omplanes} Parameter constraints in the 
($\alpha/\lambda$, $\lambda$) plane. In the right panels, 
dashed contours (corresponding $\Delta\chi^2=2.3$) are 
constraints arising from SNeIa and the solid contours 
include also $\mathcal{R}$. In the left panels, we have 
in addition included the $\mathcal{A}$. There at the solid lines one has 
$G_*'=0$ at the present. In the upper pair of figures $\Omega_m=0.3$, in 
the middle panels $\Omega_m=0.35$ and in the lowest two figures 
$\Omega_m=0.4$. One notes that the background expansion does not very 
sensitively probe $\alpha$, and that requiring $G_*' = 0$ restricts very 
strictly the parameter combinations.}
\end{center}
\end{figure}

The combined constraints arising from the CMBR shift parameter, baryon 
oscillations and the SNeIa are presented in the shaded contours
in FIG. \ref{contours1}. There the dependence on the parameter $\alpha$ is 
not shown, but we have included also likelihood slices on three 
constant-$\Omega_m$ slices of the parameter space in FIG. 
(\ref{omplanes}), to explicitly show that the value of $\alpha$ is not so 
tightly constrained, as the contours on the $\alpha$-$\lambda$ planes have 
are nearly horisontal planes. The best-fit values for the $\chi^2$ are 
given in Table \ref{tabu_a}.
Note that we have reported the $\chi^2$ per effective degree of freedom, $\chi^2_{dof}$, thus 
taking into account that the coupled model has more free parameters than the 
$\Lambda$CDM model. The fit with CMBR and SNeIa is as good as with the concordance model,
but when the constraint from $\mathcal{A}$ is included the fit gets worse.
This conclusion is the same for both the Gold and the SNLS data. The latter data
gives slightly worse fit to the model than to $\Lambda$CDM; when $\mathcal{R}$-constraint
is included the goodness of fit is slightly better than for $\Lambda$CDM; and when the
$\mathcal{A}$-constraint is added the goodness of fit is significantly worse, difference
of $\chi^2$ being almost $\Delta\chi^2 = 9.5$. Taken at face value, the model is ruled out by the combining
the three kinematical tests. The reason why the baryon oscillation data seems to be in tension
with the other data for this model is that it fixes the present
matter density rather tightly to the $\Lambda$CDM region, $\Omega_m \approx 0.27$.
As a matter of fact in a flat universe this constraint is equivalent to 
requiring that the matter density
is $\Omega_m = 0.273 + 0.123(1+w_0) \pm 0.025$, where $w_0$ is an averaged
equation state between the present and $z_1$ \cite{Eisenstein:2005su}. On the other
hand, because the Gauss-Bonnet coupled field produces a very 
negative $w_{eff}$ today, the other data sets we included prefer higher matter 
densities than in the standard $\Lambda$CDM model. 

\begin{figure}
\begin{center}
\includegraphics[width=0.4\textwidth]{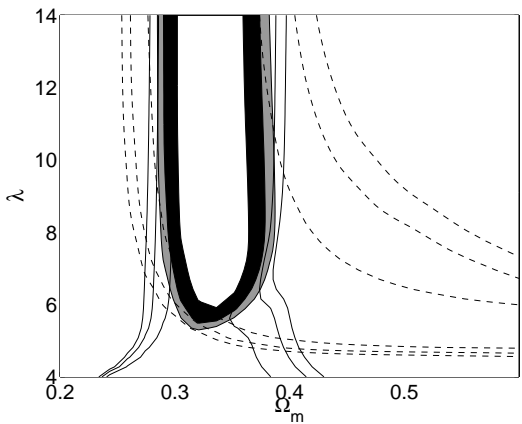}
\caption{\label{contours2} The confidence limits corresponding to $\Delta\chi^2=2.3$,
$4.61$ and $9.21$ for the model in the $\Omega_m$ - $\lambda$ plane. Dashed lines are constraints from 
the CMBR shift data alone, solid lines from the baryon oscillation scale alone, 
and shaded regions combine these data. The coupling $\alpha$ (marginalized over.
in the range $1.5\lambda < \alpha <10\lambda$.}
\end{center}
\end{figure}

However, on one hand it is not clear how model-independent probe of the background expansion
the baryon oscillation constraint in its present form is. Here the 
the non-negligible contribution of the scalar field to the energy density during 
matter domination probably affects the dependence of $\mathcal{A}$ on 
$\Omega_m$. In addition, the Gauss-Bonnet modification of gravity might 
have such effects on the nonlinear evolution of galaxy distributions 
that shift the scale at which the oscillation appear us today. Finally, 
in general in dark energy models with rapidly varying equations of 
state the validity of the approximation of lumping the observations 
to the single redhift $z_1$ has not been established\cite{Dick:2006ev}. 
For such reasons one has to be careful before ruling out alternative 
cosmologies based on the baryon oscillation constraint. 

On the other hand, as shown in Fig. \ref{contours2}, the baryon oscillations and the CMBR acoustic
scale alone or together do not limit the allowed parameter space very strictly. The best fit
to these observations is $\chi^2 \ll 1$, and they are compatible with 
lower matter densities than with the SNeIa data included. As mentioned previously, it is possible 
that the supernova constraints do not apply in their present form to the model, and hence
it is interesting to note that when they are discarded, the model matches excellently 
with the other observations. It is then also consistent with high values of 
$\lambda$, and the tension with nucleosynthesis bound, to be discussed next, disappears.

%Suspicions concerning the
%applicability of the baryon oscillation constraint for modified  
%gravity have been expressed also in Ref. 
%for other non-standard cosmologies in Ref. \cite{Alnes:2005rw}. 
%Nevertheless, we checked that our model produces the observed 
%angular scale of the radiation background sound horizon together with 
%the size of the oscillations detected in the structure at large scales. 
%When combined, these parameters probe very efficiently the properties of 
%the universe, whereas the SNeIa observations tell only about the $z<2$ 
%universe. Therefore the 
%fact that our model can predict at least the correct magnitude also for $\mathcal{R}$ 
%and $\mathcal{A}$ seem to indicate that the Gauss-Bonnet coupled quintessence is a promising 
%candidate for dark energy. 
%However, now the formally defined
%$\Omega_m$ presents the fractional energy density in the presence of 
%modified gravity, where the sum of these fractional densities is not
%equal to one even when the universe is flat. 
%
\begin{table} 
\begin{tabular}{|c||c|c||c|c|c|c|}
\hline
Data set  & \multicolumn{2}{|c||}{$\Lambda$CDM} & \multicolumn{3}{|c|}{$R^2_{GB}\phi$ model}  \\ 
\hline 
          & $\chi^2_{dof}$ & $\Omega_m$ & $\chi^2_{dof}$ & $\Omega_m$ & $\lambda$ \\   
\hline
 SNeIa     & $ 1.142 $  & $0.31^{\pm 0.04} $    &  1.140   & 
$0.22^{+0.32}_{-0.11}$      &  $3.3^{+4.3}_{-2.2}$  \\
\hline
 SNeIa+$\mathcal{R}$   & $ 1.144 $  & $0.28^{\pm 0.03}$  &  
$1.136$ & $0.45^{+0.04}_{-0.07}$ &  $5.4^{+1.0}_{-0.5}$  \\
\hline
 SNeIa+$\mathcal{R}$+$\mathcal{A}$ & $ 1.137 $  & $ 0.28^{\pm 0.02} $ &  
1.205  & $0.34^{+0.03}_{-0.02}$  &  $5.7^{+0.6}_{-0.5}$ \\
\hline
\end{tabular}
\caption{\label{tabu_a} The best-fit values for
$\Lambda$CDM model compared with fits of the coupled scalar field model 
for some parameter values when $\alpha$ is marginalized over. The error limits correspond
to change of unity in the $\chi^2$. The degrees of freedom 
in the first row are $157-d$, in the second $158-d$ and in the third $159-d$, where 
$d=1$ for the $\Lambda$CDM and $d=2$ for the G$\phi$ models.}
\end{table}

\subsubsection{Nucleosynthesis}

The nucleosynthesis bound restricts the allowed amount of dark
energy in the early universe. If the expansion rate is not due
to radiation alone, the prediction for the abundance of the light
elements produced during nucleosynthesis will be modified. This
places constraints on  $\Omega_\phi$ at $a \sim 10^{-10}$. As tight
constraints as $\Omega_\phi<0.045$ have been reported \cite{Bean:2001wt},
but more a conservative limit is $\Omega_\phi<0.2$ \cite{Copeland:2006wr}. 
Here one should keep in mind that the observationally 
preferred value of $\omega_b h^2$ might turn out to be different than in 
the concordance models (as it seems to be with $\omega_m h^2$. 
Nevertheless, if the scalar field is tracking during 
nucleosynthesis, the more conservative bound translates into 
$\lambda>6.3$. While with potential slopes corresponding to such values,
a good fit can still be achieved, the best fit values are typically at 
smaller $\lambda$ if the SNeIa data is taken into account. Therefore one might want 
to consider relaxing the nucleosynthesis constraint. 

There are several ways to do that. The simplest option is to assume that the 
scalar field has not reached the tracker solution yet, but begins to approach it 
during or after the nucleosynthesis. Then one just considers large 
enough initial values for the field, so that the potential term in Eq.(\ref{klein-gordon}) 
is small compared to the Hubble drag at very early times. This could be the natural
outcome of inflation \cite{Malquarti:2002bh}. If one however insists that 
the field must have entered the tracking phase well before nucleosynthesis, it 
would be necessary to modify the early dynamics of the field. If the slope of
the potential is steeper for small values of the field, the energy 
density residing in the quintessence field is subdominant until it reaches the 
shallower region, and choosing the parameters suitably, this could happen 
after a sufficient amount of light elements is produced. It might be 
reasonable to consider $\lambda$ as a function of the scale 
factor\cite{Neupane:2006ip}. Alternatively, an additional 
field dependent features in the coupling $f$ could be used to hold the 
field subdominant enough in very early universe. 
%We might return to these issues when considering 
%the viability of the action (\ref{action}) in the context of the 
%early inflation of the universe.  {\bf{(If we cannot do this do not forget to remove this sentence.)}}

\subsection{Solar System Constraints}
\label{solar}

Time variation of the gravitational constant is tightly constrained by
observations. These observations include various tests of the force
of gravity within the Solar system and laboratories, and indicate that
$(dG_*/dt)/G_*$ is less than about $10^{-12}$ per year, where $G_*$ is
the effective gravitational constant \cite{uzan}. This bounds 
translates into
\be \label{bound}
|\frac{G_*'}{G_*}| \lesssim 0.01.
\ee
To derive the variation of this constant for the coupled Gauss-Bonnet
gravity, we follow the approach of Ref.\cite{Amendola:2005cr} where
cosmological perturbation equations were considered at their Newtonian
limit, i.e. assuming small scales and small velocities
the Poisson equation was derived, and the effective strength of gravitational 
coupling was read from the resulting expression, which relates the gradient of the 
gravitational potential to the perturbations in matter density. We will first compute
results ensuing from this approach and then discuss the Post-Newtonian 
parameterization.

\subsubsection{Newtonian limit}

Consider subhorizon scales at a limit where the time derivates of the perturbations 
are much smaller than their gradients. In the Newtonian gauge we have 
$\varphi = -\Phi$ and $\varphi_A=\Psi$. Then $\kappa = 3H\Psi$. This gauge is also called 
zero-shear or longitudinal for the reason that there $\chi=0$.
The ADM energy constraint Eq.(\ref{admenergy}) is then, at this limit
\be
\frac{a^2\rho\delta}{k^2} = -2(1-8H\dot{f})\Phi - 8H^2\delta f.
\ee
The shear equation (\ref{propagation}) reveals there is effective anisotropic stress in
the sense that $\Phi \neq -\Psi$:
\be
(1-8H\dot{f})\Psi + (1-8\ddot{f})\Phi = -8(\dot{H}+H^2)\delta f.
\ee
To obtain the effective Poisson equation here, we need to eliminate
the scalar field perturbation from the above two equations. The fluctuation
$\delta\phi$ is given by the Klein-Gordon equation (\ref{klein}),
\be
\delta \phi = 8f'[2(\dot{H}+H^2)\Phi-H^2\Psi].
\ee
From the previous three equations, it is then straightforward to obtain 
the Poisson equation,
\be
\frac{k^2H^2}{a^2}\Psi = 8\pi G_*\rho\delta.
\ee
We find that the effective gravitational constant here
coincides with the expression (\ref{g_eff}). This is in spite that the
Newtonian limit is considered at a static configuration of the gravitational 
sources. This takes the approximation we used in Section (\ref{perturbations}) 
further, since there we derived the small scale limit of the dynamical
equations. We can undertand that the two $G_*$ agree since the time derivatives 
of say $\Psi$ are expected to be proportional to $\dot{\Psi} \sim H\Psi$. The
gradient terms are more important at small scales and hence
determine there the main contribution to the gravitational 
coupling of evolving structures.

Though not obvious from the formula (\ref{g_eff}), it equals one when the coupling 
goes to zero. Generally, when the coupling is significant, i.e. $\mu$ is 
of order one, then one expects the $G_*'/G_*$ to be of roughly 
of order one as well. This is confirmed by numerical evaluation of the time 
variation of $G_*$ for the model considered here\footnote{It is possible
to find an expression for the $G_* '/G_*$ in terms of $x$, $y$ and $\mu$, but
it is somewhat lenghty.}. As claimed in 
Ref. \cite{Amendola:2005cr}, one has to assume ''an accidental 
cancellation'' to satisfy the bound Eq. (\ref{bound}) in the presence of 
significant Gauss-Bonnet contribution to the energy density.
This means that we have to fine-tune the coupling $\alpha$ in order to eliminate 
the time variation of the effective gravitational coupling. Then the Newtonian
limit in general exhibits time-varying $G_*$, but at the present
this $G_*$ appears to us to be a constant. 
\begin{figure}
\begin{center}
\includegraphics[width=0.4\textwidth]{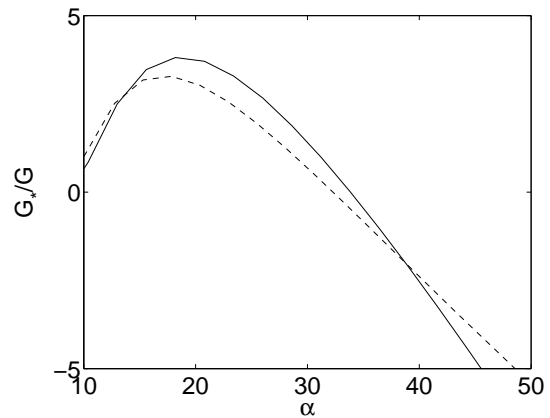}
\caption{\label{gvar} Variation of the gravitational constant, $G_*'/G_*$ at the 
present as a function of the coupling $\alpha$. The solid line is for the
first model in Table \ref{tabu} and the dashed for the second one.}
\end{center}
\end{figure}
In practice, the requirement of well-behaved Newtonian limit 
sets the magnitude of the coupling $\alpha$. Such a limit does not exist 
for all $\lambda$ and $\Omega_m$. When it is found and it requires 
$\alpha$ to be fixed with the accuracy of about $0.01$, see FIGs. 
\ref{omplanes} for actual constraints and \ref{gvar} for the variation as 
function of $\alpha$.
%Even this is not severe fine-tuning, given in comparison the ''hyperfine'' tuning 
%that is necessary when the dynamics of cosmological acceleration are 
%solely due to the coupling in the absence of a potential for the scalar 
%field  \cite{Esposito-Farese:2004cc}. Hence, in fact, our model improves this 
%fine-tuning situation.

In Table \ref{tabu} we report the $\chi^2_{dof}$, computed under the condition $G_* = 0$, 
for some parameter choices for our model.
%\footnote{The reason we do not plot confidence contours
%this time is the non-linearity of Eq.(\ref{g_eff}) results in untidy contours.}. 
Even with the coupling fixed in to yield $\dot{G}_*=0$, the coupled Gauss-Bonnet 
quintessence fits both the SNeIa and the CMBR shift parameter data as 
well as, or slightly better than the $\Lambda$CDM concordance model. However, again this 
is only when the baryon oscillation constraint is not taken into account: 
with the $\mathcal{A}$ combined with all the other data sets 
including the Solar system constraints the model is in blatant 
contradiction with the data.

\begin{table} 
\begin{tabular}{|c||c|c||c|c|c|c|}
\hline
Data set  & \multicolumn{2}{|c||}{$\Lambda$CDM} & \multicolumn{4}{|c|}{$R^2_{GB}\phi$ model}  \\ 
\hline 
          & $\chi^2_{dof}$ & $\Omega_m$ & $\chi^2_{dof}$ & $\Omega_m$ & $\lambda$ & $\alpha$ \\   
\hline
 SNeIa     & $ 1.142 $  & $ 0.314 $    &  1.146   & 0.42      & 5.1        
&   32.3     \\
\hline
 SNeIa+$\mathcal{R}$   & $ 1.144 $  & $ 0.277 $    &  1.141   & 0.44       
&  5.2     
&   33.8     \\
%\hline
% SNeIa+$\mathcal{R}$+$\mathcal{A}$ & $ 1.137 $  & $ 0.279 $    & 1.282   
%& 0.32       & 6.2        & 170     \\
\hline
\end{tabular}
\caption{\label{tabu} The best-fit values for
$\Lambda$CDM model compared with fits of the coupled scalar field model 
for some parameter values when the coupling $\alpha$ is set in order
to fix the present time variation of $G_*$ to zero. The degrees of freedom 
in the first row are $157-d$, in the second $158-d$ and in the third $159-d$, where 
$d=1$ for the $\Lambda$CDM and $d=3$ for the $R^2_{GB}\phi$ models.}
\end{table}

\subsubsection{Schwarzchild metric}
\label{sm}

The rather tight bounds we get could be loosened if one takes into account that cosmological 
variations of $G$ and other gauge-couplings might be different from the ones we 
measure on Earth or within our Solar system \citep{clifton,barrow1,barrow2,Barrow:2002ed}.  
The model is set up in such a way that the corrections
to the Einstein gravity will affect the overall expansion of the Universe,
while the Solar system is clearly subcosmological in scale and nonlinear in
nature. Thus the local values of the scalar field and the potentials could be 
something else than at cosmological scales where the linear 
perturbation equations of Appendix Eq.(\ref{peqs}) are expected to apply. If 
for example, the scalar field happened to be nearly frozen at our neighbourhood we would observe 
$G_*=G$ though at larger scales the linear perturbation would
evolve according to Eq.(\ref{d_evol}) where possibly $G_*' \neq 0$. 

%It would probably be worthwhile to consider the implications of these 
%theories for the Solar system experiments more carefully. 
Conventionally the constraints ensuing from Solar system experiments are 
imposed on the deviations from the Schwarzchild solution written in form of the 
Post-Newtonian parameters (PPN). Here we computed the Newtonian 
limit of the cosmological perturbation equations in the
FRW background to derive the constraints, and there is reason to doubt that
results ensuing within the PPN formalism would not be equivalent. Therefore
our considerations of the Newtonian limit in the previous subsection 
must be regarded as preliminary. More detailed study of the Solar system 
constraints is left for for future. Since the $R^2_{GB}$ 
is quadratic, its value for the Schwarzchild metric goes like $1/r^6$. However,  
due to the dynamics of the field, this does not necessarily mean that the 
curvature correction would dominate at small scales. Previously local constraints 
for a scalar-Gauss Bonnet coupling have been investigated in the case that the 
potential can be neglected\cite{Esposito-Farese:2003ze,Esposito-Farese:2004cc}. 
Even then all the effects of the scalar field can be negligible for any
Solar system experiment provided $f''(\phi_0)$ is negative. One also notes that
due to specific properties of $R^2_{GB}$, if the coupling $f(\phi_0)$ happens
to be in its minimum at present all the corrections trivially disappear. With these 
remarks in mind, it seems easy to adjust the function $f(\phi)$ in such 
a way that the local constraints would be satisfied (even if our preliminary
analysis would turn out to be inadequate and setting $\alpha$ to a certain value
would not guarantee the viability of the model). Such an adjustment would 
apparently ruin the potential of the model to alleviate coincidence problem, since the
slope of the coupling function $f(\phi)$ would change just today. However, it might
be possible to associate the change of the slope with the triggering of the acceleration,
since it is then that the curvature corrections become dynamically important and 
one could expect higher order modifications to enter in the play. We hope to address 
these issues more quantitatively elsewhere.

\section{Conclusions}
\label{conclusions}

We studied cosmological phenomenology of dark energy based upon a low-energy effective string
theory action featuring a compactification modulus and taking into account
the leading order curvature corrections. One is then lead to consider a generalized
scalar-tensor theory which includes a coupling to the Gauss-Bonnet invariant.
Specifically, we investigated the evolution of perturbations. There the simple closed form
equation (\ref{d_evol}) for the linear matter inhomogeneities, together with its effective gravitional 
constant (\ref{g_eff}) could be highlighted as one of the main results of 
the paper, since it enabled 
to find several new {\it model-independent results of perturbation evolution} in Gauss-Bonnet 
dark energy, these including the following.
\begin{itemize}
\item The evolution of matter perturbations is scale-invariant at small
scales in the presence of the Gauss-Bonnet term, and thus the shape of the matter
power spectrum is retained.
\item The growth rate of matter perturbations, which is easily extracted from
Eq.(\ref{d_evol}), can be compatible with observations even in the presence of
significant contribution from the Gauss-Bonnet interaction.  
\item The stability of perturbations can be read off from the expression for the effective  
gravitational constant. A divergence is possible.   
\end{itemize}
This gravitational constant cannot be deduced by matching Schwarzchild and FRW 
-type metrics
for subhorizon evolution of spherical matter overdensities, as the Jebsen-Birkhoff theorem
is not in general respected in these models. The Newtonian limit, as discussed in Section \ref{solar}, 
does feature this effective strength of gravity, which might be used to efficiently
constrain the coupling. However, the relevance of this limit to the experiments within
Solar system is an open question to be addressed more carefully in forthcoming studies.

In addition to such general considerations mentioned above, we chose a particular model and 
scrutinized its predictions for cosmology. Specifically, we characterized the potential and the 
coupling by exponential functions, as with the Nojiri-Odintsov-Sasaki modulus \cite{Nojiri:2005vv}.
This parameterization is appealing in its minimality, as within it one can describe Gauss-Bonnet dark
energy with just two extra degrees of freedom compared to the $\Lambda$CDM model. Moreover, the exponential
forms are well motivated on fundamental grounds, and allow dark energy solutions without
introducing unnatural scales into the Lagrangian. Many of the previous cosmological studies  
of this type of low-energy string theory have been concerned with asymptotic solutions and confined to cases where 
the Gauss-Bonnet term is subdominant, and thus only introducing small deviations to standard quintessence 
scenarios. Here allowed a crucial role for the Gauss-Bonnet term and examined its consequences in detail.
Such approach also proved to be very useful, as it revealed several {\it possibilities of cosmological evolution 
present already in this simple model}, these including the following.
\begin{itemize}
\item The Gauss-Bonnet coupling can act to switch the decelerating expansion into acceleration
      after a scaling matter era, thus perhaps alleviating the coincidence 
      problem.
\item The curvature interaction may momentarily push the universe into a phantom era,
      which will then not lead into a Big Rip singularity in the future. Hence the
      model can explain superacceleration.
\item However, the linear perturbations might diverge in the future. This might terminate
      the de Sitter phase, possibly helping to make contact with string theory.   
\end{itemize} 
Let us briefly comment each of these possibilities.

%We adopted an exponential potential for our quintessence field and allowed 
%the scalar field to evolve on the tracking solution, which is an attractor
%when the coupling is negligible. 
%Therefore the 
%coupling must grow exponentially with the field, and its slope must be 
%steeper than that of the coupling. In terms of the parameterization 
%(\ref{exps}), we hence required $\alpha>\lambda$. 
%For any other potential, 
%the coupling should also grow rapidly lest {\bf{(?!?)}} it would be completely negligible 
%today compared to other contributors to the energy densities. This is because
%the Gauss-Bonnet term $R^2_{GB}$ is proportional to a higher power of the Hubble 
%parameter than the standard energy densities.

The field is drawn into the scaling attractor virtually regardless of initial conditions,
and at late times acceleration can be onset provided simply $\alpha>\lambda$. 
This can be seen as a possible way towards solution of the coincidence 
problem,
though it is occasionally considered that in a preferable solution 
the future evolution of the Universe would be characterized by a constant ratio 
of $\Omega_\phi$ and $\Omega_m$ different from 1. However, a transition 
from a scaling matter era to an accelerating scaling era has been 
shown to be impossible for a very general class of scalar field models \cite{Amendola:2006qi}.
Another remark is that the question of why dark energy is beginning its 
domination today is now translated into the question that why the coupling 
comes into play just recently. The same holds for any other 
coupling mechanism proposed so far, but it is still fair to say that such mechanism 
can provide an approach to tackle the coincidence problem. 
The relevant dynamics are not always obvious when only the 
asymptotic behaviour is considered, and therefore it was only very 
recently \cite{Koivisto:2006xf, Tsujikawa:2006ph} found that 
the simple model considered here can (without flipping the field to a 
phantom, i.e. putting $\gamma=-1$ in the action (\ref{action})) feature 
a transition from a matter dominated to an accelerating phase. 

We also showed that during this transition, the 
dynamics of the coupling can cause a phantom expansion. Again, this phenomenon is 
unaccessible when only asymptotic behaviour or vacuum is taken into account. 
Several authors have shown that phantom dark energy does not necessarily lead a Big Rip 
singularity in the presence of the Gauss-Bonnet term. Here we found that 
this term can in fact ally with a canonical scalar field into an effective 
phantom energy. This is clearly a more appealing way to produce 
$w_{eff}<-1$, since thereby the phantom era can take place without 
the introduction strong energy condition violating components, while the Big Rip
singularity is still avoided.
% (though an explosion of linear perturbations 
%occurs possibly in the future cosmological evolution of this model). 

However, one of the peculiarities of this class of models is that their linear
perturbations can diverge, even while the background solution is not singular.
We argued that such a possibility does not necessarily indicate an inconsistency 
of the model. 
%In our case the divergence would occur just when the scalar mode would turn into 
%a ghost. This shows that, already at the classical level, one finds a theory with such a ghost
%insensible. 
For the specific model considered here, the divergence typically 
awaits in the future when the evolution would begin to asymptote to the de Sitter expansion. The dynamics is 
such that the (linearized) inhomogeneities grow explosively as the $s_{SC}<0$ limit is approached. 
Hence the linear approximation perhaps together with the
whole FRW description breaks down. 
%Thus the appearance of the ghost might be avoided.
It is tempting to believe that the de Sitter phase will not then be reached but
the inhomogeneities take over. This would enable to reconcile the accelerating universe
with string theory in these models in yet another way. Since if 
the acceleration is transient one can consistently define the set of 
observables in string theories\cite{Hellerman:2001yi}. Their present S-matrix 
formulations seem to be in odds with an eternally accelerating universe, 
like in the concordance model with the cosmological constant. Would the universe
however somehow pass through $c^2_{S}=0$, one would have to invoke higher-order 
curvature corrections or modifications of the exponential parameterization 
to escape the possible singularity.
%and eternity in de Sitter space. 

We calculated the implications of Gauss-Bonnet dark energy to    
various different cosmological and astrophysical observations and matched
them with several data sets. The general conclusion we arrive at is that 
the quadratic curvature coupling has interesting and non-trivial signatures while 
being in a good agreement with a large amount of data. We checked how
the Gauss-Bonnet interaction affects the evolution of scalar 
field fluctuations and how these in turn impact the large scale structure and
CMBR observations. Using combined datasets related to the CMBR 
shift parameter, baryon oscillation scale, supernovae Ia luminosity distance
and variations of the effective gravitational constant in the Solar system, we
derived confidence limits on the parameters of the model. The study presented not only
shows the requirements to be imposed on models of Gauss-Bonnet dark energy for them 
to exhibit completely realistic cosmology, but also quantifies the degree to
which these requirements can be within the most minimalistic set-ups.
Due to its simplicity, matching with observations, capability to shed light on 
the coincidence problem, possibility incorporate phantom (and perhaps also 
transient) acceleration, and not least its theoretical motivation, the 
Gauss-Bonnet quintessence seems to be a promising alternative to a cosmological
constant.

We end by mentioning some issues yet to be pursued. Among these are the nature of the divergence 
and the Solar System limit in these models. Also, we found that there is some
contrast with the amount of early quintessence and the nucleosynthesis bound and with the
present matter density and the baryon oscillation scale. Whether these enforce one to resort to more 
elaborate models depends on several assumptions, like the validity of the baryon oscillation constraint
and the initial conditions of the scalar field after inflation. One caveat is also that these
contrasts can disappear if the Gauss-Bonnet modification affects the intrinsic evolution of supernovae
and thus changes the effective luminosity-distance relationships of these objects. This does render 
some of our conclusions indecisive, but in an interesting way. Suppose that the coupling Gauss-Bonnet
modification does influence small scale physics considerably. Then one has to consider $f(\phi)$
such that it evolves to a constant as dark energy domination begins, so as to comply with the Solar system
experiments (which is very much the approach in standard scalar-tensor theories). Then one could
match all the observations, while the expansion of the universe at small $z$ could be something
considerably different than commonly established from the SNeIa luminosity-redshift curves. 
 
%Finally, let us mention that the early inflation of the universe 
%might also be incorporated within the framework presented here. {\bf{(We should develop this!)}}     

\acknowledgments
We thank G. Calcagni, A. De Felice, I. Neupane and S. Odintsov for useful discussions and clarifications of the 
ghost constraints. DFM acknowledges support from the Humboldt Foundation, from the Research Council of Norway 
through project number 159637/V30 and of the
Perimeter Institute for Theoretical Physics. TK is
supported by the Magnus Ehrnrooth Foundation.

\appendix

\section{Background} 

%\subsection{Equations of motion}
\label{eqs}

In the appendices we will consider some technical details relevant in solving 
the cosmological equations. The first appendix concerns only the kinematic
evolution model specified by Eq.(\ref{exps}). The equations in 
the second appendix are for perturbations in general for 
an action (\ref{action}). In fact, when one sets $\gamma=0$, they 
describe also modified Gauss-Bonnet models \cite{Cognola:2006eg} featuring a
function $f(R^2_{GB})$ in their action, since there exists a simple mapping between the 
theories\cite{Nojiri:2006je}.  

Employing the dimensionless variables 
(\ref{variables}) we can write the system (\ref{system}) to derive 
the explicit evolution equations for each individual variable. It would 
be possible to algebraically solve all but two of them, but 
we rather use the three variables related to the fractional energy 
densities of the kinetic term of the scalar field $x$, its 
potential $y$ and the coupling $\mu$. The coupled ordinary 
differential equation system for these three variables is then  
\begin{widetext}
\be \label{system2a}
x'  =  -\frac{2\epsilon(-3 + x^2 + 6\mu) + 3x^2(1 + w) +
       3(1 + w)(-3 + 3\mu + y) + x(3\alpha \mu - \lambda y)}{2x^2 + 3\mu}x,
\ee
\be \label{system2b}
y' =  -(2\epsilon + x\lambda)y,
\ee
\be \label{system2c}
\mu' =  \frac{2\alpha x^3 + 2\epsilon(3 + x^2 - 3\mu) - 3x^2(1 + w) + x\lambda y -
      3(1 + w)(-3 + 3\mu + y)}{2x^2 + 3\mu}\mu,
\ee
where the slow-roll parameter $\epsilon$ is given by
\be \label{eps}
\epsilon = -\frac{2\alpha x^3\mu - 3\mu^2 + 
           2x^4(-1 + w) +
           2x^2[\mu(-1 + 3w) + (1 + w)(-3 + y)] + \lambda x\mu y}{
           4x^2(-1 + \mu) - 3\mu^2}.
\ee
\end{widetext}
Here $w$ is given by $w=\Omega^0_r/(3(\Omega^0_r+a\Omega^0_m))$.
Since the field is attracted to 
the scaling solution E regardless of the initial values for $x$ and $y$, 
the only initial condition that makes difference is for $\mu$. However, 
given the exponents of the coupling and the potential, the initial 
condition for $\mu$ is determined by the required amount of 
matter today $\Omega^0_m$. The fractional matter density $\Omega_m$ as a 
function of $\log(a)$ is given by $\Omega_m = (3-x^2-3\mu-y)/3$. Since on the other hand 
$\Omega_m = \Omega^0_m a^{-3(1+w)}/H^2$, we can readily obtain the Hubble 
parameter as function of redshift from the solution for $x$, $y$ and 
$\mu$. One consistency check that we have made is the comparison of the 
numerical derivative of the Hubble parameter with the 
algebraic solution for $\epsilon$, Eq.(\ref{eps}). One might be concerned that the 
system (\ref{system2a})-(\ref{eps}) is not well behaved in the limit that 
the denominators in the equation go to zero. The denominator vanishes when the field is constant. However, 
then also the numerators go to zero. We have found no numerical 
problems in the situations when the sign of the field velocity happens 
to get flipped. On the other hand, the denominator of Eq.(\ref{eps}) would vanish also
when $\mu=\frac{2}{3}x^2[1\pm\sqrt{x^2-3}]$. This would occur only when 
the field would be decreasing at enormous speed, and such a configuration
is indeed impossible to reach.

For completeness, we also mention that for the fixed point G we have
\be
x = 10 + \frac{100-9\alpha^2}{1000-54\alpha^2+9r^{1/3}} +
      \frac{1}{3\alpha}k,
\ee
and
\be
\mu & = & -\Big[27\lambda^4 +3rk + 100k^2 + 20(500+3r+50k)
\nonumber \\
& + & 3\alpha^2(320+26k + 5k^2)\Big]/(9k^2)
\ee  
where $r=a(2000-264\alpha^2+9\alpha^4)^{1/2}$ and
$k=(1000-54\alpha^2+9r)^{1/3}$.

%\subsection{Attractors}
%\label{attr}
%
%The other fixed points could be found easily, but due to non-linearity
%of the system we did not exact conditions for the stability of the fixed
%point $C$ in the presence of the coupling. However, the general conditions
%are not relevant to our particular case, since in the early universe the
%coupling is negligible, and thus the scalar field is attracted to the fixed
%point $E$. From this point the coupling always carries the field to
%the fixed point $C$. This behaviour does not depend on the parameters
%$\alpha$ or $\lambda$. We show a phase portrait of the model in FIG. (\ref{pp}).

\section{Linearly Perturbed Equations} \label{peqs}

In this appendix we consider the scalar perturbation equations for coupled 
Gauss-Bonnet gravity in a general gauge. These equations are consistent with the ones presented in 
Refs.\cite{Cartier:2001is,Hwang:2005hb}. 

The metric is defined as in Eq.(\ref{metric}). 
Recall that the variables satisfy the constraints 
  \be 
 \chi \equiv a\beta+a^2\dot{\psi} \,, \quad
 \kappa \equiv 3(H\varphi_A -
 \dot{\varphi})+\frac{k^2}{a^2}\chi\,.
 \ee
Due to general structure of covariantly modified gravity \cite{tomimod1}, the 
last three conservation equations are deducible from the set of 
field equations (\ref{admenergy})-(\ref{ray}) when the matter content is known. 

\begin{widetext}
The energy constraint ($G^0_0$ component of the field equation) is
 \be \label{admenergy}
   \left(1-8H\dot{f}\right)\frac{k^2}{a^2}\varphi
  + \left(\frac{1}{2}\gamma\dot{\phi}^2 + 12H^3\dot{f}\right)\varphi_A
 - \left(H - 12H^2\dot{f}\right)\kappa = \frac{1}{2}\left(\delta \rho_m +
   \gamma\dot{\phi}\dot{\delta\phi} + V'\delta\phi\right) +
  4H^2\left( 3H\dot{\delta f} +\frac{k^2}{a^2}\delta f\right),
 \ee
and the momentum constraint ($G^0_i$ component) is
 \be \label{mc}
 \left(1-8H\dot{f}\right)\left(\kappa - \frac{k^2}{a^2}\chi\right)
 - 12H^2\dot{f}\varphi_A = \frac{3}{2}\left[a\left(\rho_m+p_m\right)v_m + \gamma\dot{\phi}\delta\phi\right]
 - 12H^2\left(\dot{\delta f} - H\delta f\right).
 \ee
The shear propagation equation ($G^i_j-\frac{1}{3}\delta^i_jG^k_k$
component) reads
 \be
 \label{propagation}
 \left(1-8H\dot{f}\right)\dot{\chi} +
\left[H-8\left(H\ddot{f}+\dot{H}\dot{f}
 + H^2\dot{f}\right)\right]\chi - \left(1-8H\dot{f}\right)\varphi_A
 - \left(1-8\ddot{f}\right)\varphi = \Pi_m - 8\left(
    \dot{H}+H^2\right)\delta f.
  \ee
The Raychaudhuri equation ($G^k_k-G^0_0$ component) is now given by
 \be\label{ray}
 \left(1-8H\dot{f}\right)\dot{\kappa} & + &
 \left(2H-8H\ddot{f}-8\dot{H}\dot{f}-12H^2\dot{f}\right)\kappa
 - 12H^2\dot{f}\dot{\varphi_A}
 \nonumber \\
 & + & \left(3\dot{H}-\frac{k^2}{a^2} + 2\gamma\dot{\phi}^2 - 24H^2\ddot{f}-48\dot{H}H\dot{f}-12H^3\dot{f} +
        8H\dot{f}\frac{k^2}{a^2}\right)\varphi_A
  +  8\left(\ddot{f}-H\dot{f}\right)\frac{k^2}{a^2}\varphi
 \nonumber \\
 & = &
 \frac{1}{2}\left(\delta\rho_m+3\delta p_m\right) + 2\gamma\dot{\phi}\dot{\delta\phi} - V'\delta\phi -
 4\left[3H^2\ddot{\delta f} + \left(2\dot{H}+H^2\right)\left(3H\dot{\delta f} + \frac{k^2}{a^2}\delta f\right)\right].
 \ee
The previous three equations can be considered as evolution equations for $\varphi$, $\chi$ and $\kappa$, 
respectively. For instance, the Raychaudhuri equation might be written as $\dot{\kappa} = $ 
sources$/(1-8H\dot{f})$. Also the other equations imply a divergence of perturbations in the
case that $8H\dot{f}=1$. In the models considered here, $8H\dot{f}<1$ is always satisfied; this condition
in fact is necessary for absence of tensor ghost modes \cite{Calcagni:2006ye}. Similar considerations 
apply for the factor $(1-8\ddot{f})$ appearing in some of the equations. One might understand this by noting
that the first condition for stability of scalar modes is automatically satisfied given the
conditions for stability of the tensor modes \cite{Calcagni:2006ye}. The second condition for stability of scalar 
modes, which is not always guaranteed in models considered here, is discussed extensively in Section \ref{stability}.
%Due to general structure of covariantly modified gravity\cite{tomimod1}, the conservation equations for the fluid 
%perturbations follow from the above four field equations. 
The equation of motion for the scalar field is
 \be \label{klein}
 \gamma\left(\ddot{\delta \phi} +  3H\dot{\delta \phi}\right)  + \left(\frac{k^2}{a^2} +V''\right)\delta\phi
 =  \gamma\dot{\phi}\left( \dot{\varphi_A} + \kappa\right)
 + \gamma\left(2\ddot{\phi}+3H\dot{\phi}\right)\varphi_A
 - f''\bar{R}^2_{GB} \delta\phi - f'\delta R^2_{GB},
 \ee
where
\be
\delta R^2_{GB} = 4H^2\delta R-16\dot{H}\left(H\kappa-\frac{k^2}{a^2}\varphi\right), \quad\delta R =
2\left[-\dot{\kappa}-4H\kappa+\left(\frac{k^2}{a^2}-3\dot{H}\right)\varphi_A+2\frac{k^2}{a^2}\varphi\right].
\ee
The continuity equations for matter are
\be \label{dd}
\dot{\delta}_m = 3H\left(w_m-c_m^2\right)\delta_m + \left(1+w_m\right)\left(-\frac{k^2}{a}v_m+\kappa-3H\varphi_A\right),
\ee
\be \label{dv}
\dot{v}_m=\left(3c_m^2-1\right)Hv_m+\frac{1}{a}\left[\varphi_A + \frac{c_m^2\delta_m}{1+w_m} 
           -\frac{2k^2}{3(1+w_m)a^2\rho_m}\Pi_m\right].
\ee
To implement the model in the CAMB code, we have included there an integration of the background
equations described in the Appendix \ref{eqs}. The solution is then interpolated to obtain it at
arbitrary points. The perturbation equations are modified to be consistent with the set of equations  
listed above, as written in the synchronous gauge (with the CAMB variables slightly different from those $h$, 
$\eta$ and $\chi$ we discussed in \ref{perturbations}). Modifications have to be taken into account  
in the evolution equations for perturbations to obtain the matter power spectrum, and in the evaluation of 
their line-of-sight integral to obtain in addition the CMBR spectrum. 

\end{widetext}

\bibliography{refs}

\begin{thebibliography}{93}
\expandafter\ifx\csname natexlab\endcsname\relax\def\natexlab#1{#1}\fi
\expandafter\ifx\csname bibnamefont\endcsname\relax
  \def\bibnamefont#1{#1}\fi
\expandafter\ifx\csname bibfnamefont\endcsname\relax
  \def\bibfnamefont#1{#1}\fi
\expandafter\ifx\csname citenamefont\endcsname\relax
  \def\citenamefont#1{#1}\fi
\expandafter\ifx\csname url\endcsname\relax
  \def\url#1{\texttt{#1}}\fi
\expandafter\ifx\csname urlprefix\endcsname\relax\def\urlprefix{URL }\fi
\providecommand{\bibinfo}[2]{#2}
\providecommand{\eprint}[2][]{\url{#2}}

\bibitem[{\citenamefont{Riess et~al.}(2004)}]{Riess:2004nr}
\bibinfo{author}{\bibfnamefont{A.~G.} \bibnamefont{Riess}} \bibnamefont{et~al.}
  (\bibinfo{collaboration}{Supernova Search Team}),
  \bibinfo{journal}{Astrophys. J.} \textbf{\bibinfo{volume}{607}},
  \bibinfo{pages}{665} (\bibinfo{year}{2004}), \eprint{astro-ph/0402512}.

\bibitem[{\citenamefont{Astier et~al.}(2006)}]{Astier:2005qq}
\bibinfo{author}{\bibfnamefont{P.}~\bibnamefont{Astier}} \bibnamefont{et~al.},
  \bibinfo{journal}{Astron. Astrophys.} \textbf{\bibinfo{volume}{447}},
  \bibinfo{pages}{31} (\bibinfo{year}{2006}), \eprint{astro-ph/0510447}.

\bibitem[{\citenamefont{Spergel et~al.}(2006)}]{Spergel:2006hy}
\bibinfo{author}{\bibfnamefont{D.~N.} \bibnamefont{Spergel}}
  \bibnamefont{et~al.} (\bibinfo{year}{2006}), \eprint{astro-ph/0603449}.

\bibitem[{\citenamefont{Tegmark et~al.}(2004)}]{sloan}
\bibinfo{author}{\bibfnamefont{M.}~\bibnamefont{Tegmark}} \bibnamefont{et~al.}
  (\bibinfo{collaboration}{SDSS}), \bibinfo{journal}{Phys. Rev.}
  \textbf{\bibinfo{volume}{D69}}, \bibinfo{pages}{103501}
  (\bibinfo{year}{2004}), \eprint{astro-ph/0310723}.

\bibitem[{\citenamefont{Copeland et~al.}(2006)\citenamefont{Copeland, Sami, and
  Tsujikawa}}]{Copeland:2006wr}
\bibinfo{author}{\bibfnamefont{E.~J.} \bibnamefont{Copeland}},
  \bibinfo{author}{\bibfnamefont{M.}~\bibnamefont{Sami}}, \bibnamefont{and}
  \bibinfo{author}{\bibfnamefont{S.}~\bibnamefont{Tsujikawa}}
  (\bibinfo{year}{2006}), \eprint{hep-th/0603057}.

\bibitem[{\citenamefont{Bertone et~al.}(2005)\citenamefont{Bertone, Hooper, and
  Silk}}]{bertone}
\bibinfo{author}{\bibfnamefont{G.}~\bibnamefont{Bertone}},
  \bibinfo{author}{\bibfnamefont{D.}~\bibnamefont{Hooper}}, \bibnamefont{and}
  \bibinfo{author}{\bibfnamefont{J.}~\bibnamefont{Silk}},
  \bibinfo{journal}{Phys. Rept.} \textbf{\bibinfo{volume}{405}},
  \bibinfo{pages}{279} (\bibinfo{year}{2005}), \eprint{hep-ph/0404175}.

\bibitem[{\citenamefont{Caldwell}(2002)}]{Caldwella}
\bibinfo{author}{\bibfnamefont{R.~R.} \bibnamefont{Caldwell}},
  \bibinfo{journal}{Phys. Lett.} \textbf{\bibinfo{volume}{B545}},
  \bibinfo{pages}{23} (\bibinfo{year}{2002}), \eprint{astro-ph/9908168}.

\bibitem[{\citenamefont{Brookfield
  et~al.}(2006{\natexlab{a}})\citenamefont{Brookfield, van~de Bruck, Mota, and
  Tocchini-Valentini}}]{brookfield2}
\bibinfo{author}{\bibfnamefont{A.~W.} \bibnamefont{Brookfield}},
  \bibinfo{author}{\bibfnamefont{C.}~\bibnamefont{van~de Bruck}},
  \bibinfo{author}{\bibfnamefont{D.~F.} \bibnamefont{Mota}}, \bibnamefont{and}
  \bibinfo{author}{\bibfnamefont{D.}~\bibnamefont{Tocchini-Valentini}},
  \bibinfo{journal}{Phys. Rev.} \textbf{\bibinfo{volume}{D73}},
  \bibinfo{pages}{083515} (\bibinfo{year}{2006}{\natexlab{a}}),
  \eprint{astro-ph/0512367}.

\bibitem[{\citenamefont{Bertolami and Mota}(1999)}]{bertolami}
\bibinfo{author}{\bibfnamefont{O.}~\bibnamefont{Bertolami}} \bibnamefont{and}
  \bibinfo{author}{\bibfnamefont{D.~F.} \bibnamefont{Mota}},
  \bibinfo{journal}{Phys. Lett.} \textbf{\bibinfo{volume}{B455}},
  \bibinfo{pages}{96} (\bibinfo{year}{1999}), \eprint{gr-qc/9811087}.

\bibitem[{\citenamefont{Mota and van~de Bruck}(2004)}]{carsten}
\bibinfo{author}{\bibfnamefont{D.~F.} \bibnamefont{Mota}} \bibnamefont{and}
  \bibinfo{author}{\bibfnamefont{C.}~\bibnamefont{van~de Bruck}},
  \bibinfo{journal}{Astron. Astrophys.} \textbf{\bibinfo{volume}{421}},
  \bibinfo{pages}{71} (\bibinfo{year}{2004}), \eprint{astro-ph/0401504}.

\bibitem[{\citenamefont{Koivisto et~al.}(2005)\citenamefont{Koivisto,
  Kurki-Suonio, and Ravndal}}]{tomicard}
\bibinfo{author}{\bibfnamefont{T.}~\bibnamefont{Koivisto}},
  \bibinfo{author}{\bibfnamefont{H.}~\bibnamefont{Kurki-Suonio}},
  \bibnamefont{and} \bibinfo{author}{\bibfnamefont{F.}~\bibnamefont{Ravndal}},
  \bibinfo{journal}{Phys. Rev.} \textbf{\bibinfo{volume}{D71}},
  \bibinfo{pages}{064027} (\bibinfo{year}{2005}), \eprint{astro-ph/0409163}.

\bibitem[{\citenamefont{Koivisto and Mota}(2006{\natexlab{a}})}]{koivisto}
\bibinfo{author}{\bibfnamefont{T.}~\bibnamefont{Koivisto}} \bibnamefont{and}
  \bibinfo{author}{\bibfnamefont{D.~F.} \bibnamefont{Mota}},
  \bibinfo{journal}{Phys. Rev.} \textbf{\bibinfo{volume}{D73}},
  \bibinfo{pages}{083502} (\bibinfo{year}{2006}{\natexlab{a}}),
  \eprint{astro-ph/0512135}.

\bibitem[{\citenamefont{Capozziello}(2002)}]{Capozziello02}
\bibinfo{author}{\bibfnamefont{S.}~\bibnamefont{Capozziello}},
  \bibinfo{journal}{Int. J. Mod. Phys.} \textbf{\bibinfo{volume}{D11}},
  \bibinfo{pages}{483} (\bibinfo{year}{2002}), \eprint{gr-qc/0201033}.

\bibitem[{\citenamefont{Nojiri and
  Odintsov}(2006{\natexlab{a}})}]{Nojiri:2006ri}
\bibinfo{author}{\bibfnamefont{S.}~\bibnamefont{Nojiri}} \bibnamefont{and}
  \bibinfo{author}{\bibfnamefont{S.~D.} \bibnamefont{Odintsov}}
  (\bibinfo{year}{2006}{\natexlab{a}}), \eprint{hep-th/0601213}.

\bibitem[{\citenamefont{Straumann}(2006)}]{Straumann:2006tv}
\bibinfo{author}{\bibfnamefont{N.}~\bibnamefont{Straumann}},
  \bibinfo{journal}{Mod. Phys. Lett.} \textbf{\bibinfo{volume}{A21}},
  \bibinfo{pages}{1083} (\bibinfo{year}{2006}), \eprint{hep-ph/0604231}.

\bibitem[{\citenamefont{Borowiec et~al.}(2006)\citenamefont{Borowiec,
  Godlowski, and Szydlowski}}]{Borowiec:2006qr}
\bibinfo{author}{\bibfnamefont{A.}~\bibnamefont{Borowiec}},
  \bibinfo{author}{\bibfnamefont{W.}~\bibnamefont{Godlowski}},
  \bibnamefont{and}
  \bibinfo{author}{\bibfnamefont{M.}~\bibnamefont{Szydlowski}}
  (\bibinfo{year}{2006}), \eprint{astro-ph/0607639}.

\bibitem[{\citenamefont{Woodard}(2006)}]{Woodard:2006nt}
\bibinfo{author}{\bibfnamefont{R.~P.} \bibnamefont{Woodard}}
  (\bibinfo{year}{2006}), \eprint{astro-ph/0601672}.

\bibitem[{\citenamefont{Faraoni}(2006)}]{Faraoni:2006sy}
\bibinfo{author}{\bibfnamefont{V.}~\bibnamefont{Faraoni}}
  (\bibinfo{year}{2006}), \eprint{astro-ph/0610734}.

\bibitem[{\citenamefont{Nojiri and
  Odintsov}(2006{\natexlab{b}})}]{Nojiri:2006gh}
\bibinfo{author}{\bibfnamefont{S.}~\bibnamefont{Nojiri}} \bibnamefont{and}
  \bibinfo{author}{\bibfnamefont{S.~D.} \bibnamefont{Odintsov}},
  \bibinfo{journal}{Phys. Rev.} \textbf{\bibinfo{volume}{D74}},
  \bibinfo{pages}{086005} (\bibinfo{year}{2006}{\natexlab{b}}),
  \eprint{hep-th/0608008}.

\bibitem[{\citenamefont{Nojiri and
  Odintsov}(2006{\natexlab{c}})}]{Nojiri:2006su}
\bibinfo{author}{\bibfnamefont{S.}~\bibnamefont{Nojiri}} \bibnamefont{and}
  \bibinfo{author}{\bibfnamefont{S.~D.} \bibnamefont{Odintsov}}
  (\bibinfo{year}{2006}{\natexlab{c}}), \eprint{hep-th/0610164}.

\bibitem[{\citenamefont{Amendola
  et~al.}(2006{\natexlab{a}})\citenamefont{Amendola, Polarski, and
  Tsujikawa}}]{Amendola:2006kh}
\bibinfo{author}{\bibfnamefont{L.}~\bibnamefont{Amendola}},
  \bibinfo{author}{\bibfnamefont{D.}~\bibnamefont{Polarski}}, \bibnamefont{and}
  \bibinfo{author}{\bibfnamefont{S.}~\bibnamefont{Tsujikawa}}
  (\bibinfo{year}{2006}{\natexlab{a}}), \eprint{astro-ph/0603703}.

\bibitem[{\citenamefont{Koivisto and Kurki-Suonio}(2006)}]{tomimod2}
\bibinfo{author}{\bibfnamefont{T.}~\bibnamefont{Koivisto}} \bibnamefont{and}
  \bibinfo{author}{\bibfnamefont{H.}~\bibnamefont{Kurki-Suonio}},
  \bibinfo{journal}{Class. Quant. Grav.} \textbf{\bibinfo{volume}{23}},
  \bibinfo{pages}{2355} (\bibinfo{year}{2006}), \eprint{astro-ph/0509422}.

\bibitem[{\citenamefont{Amarzguioui et~al.}(2005)\citenamefont{Amarzguioui,
  Elgaroy, Mota, and Multamaki}}]{morad}
\bibinfo{author}{\bibfnamefont{M.}~\bibnamefont{Amarzguioui}},
  \bibinfo{author}{\bibfnamefont{O.}~\bibnamefont{Elgaroy}},
  \bibinfo{author}{\bibfnamefont{D.~F.} \bibnamefont{Mota}}, \bibnamefont{and}
  \bibinfo{author}{\bibfnamefont{T.}~\bibnamefont{Multamaki}}
  (\bibinfo{year}{2005}), \eprint{astro-ph/0510519}.

\bibitem[{\citenamefont{Koivisto}(2006{\natexlab{a}})}]{tomimod3}
\bibinfo{author}{\bibfnamefont{T.}~\bibnamefont{Koivisto}},
  \bibinfo{journal}{Phys. Rev.} \textbf{\bibinfo{volume}{D73}},
  \bibinfo{pages}{083517} (\bibinfo{year}{2006}{\natexlab{a}}),
  \eprint{astro-ph/0602031}.

\bibitem[{\citenamefont{Li et~al.}(2006)\citenamefont{Li, Chan, and
  Chu}}]{Li:2006ag}
\bibinfo{author}{\bibfnamefont{B.}~\bibnamefont{Li}},
  \bibinfo{author}{\bibfnamefont{K.~C.} \bibnamefont{Chan}}, \bibnamefont{and}
  \bibinfo{author}{\bibfnamefont{M.~C.} \bibnamefont{Chu}}
  (\bibinfo{year}{2006}), \eprint{astro-ph/0610794}.

\bibitem[{\citenamefont{Elizalde et~al.}(2004)\citenamefont{Elizalde, Nojiri,
  and Odintsov}}]{st1}
\bibinfo{author}{\bibfnamefont{E.}~\bibnamefont{Elizalde}},
  \bibinfo{author}{\bibfnamefont{S.}~\bibnamefont{Nojiri}}, \bibnamefont{and}
  \bibinfo{author}{\bibfnamefont{S.~D.} \bibnamefont{Odintsov}},
  \bibinfo{journal}{Phys. Rev.} \textbf{\bibinfo{volume}{D70}},
  \bibinfo{pages}{043539} (\bibinfo{year}{2004}), \eprint{hep-th/0405034}.

\bibitem[{\citenamefont{Boisseau et~al.}(2000)\citenamefont{Boisseau,
  Esposito-Farese, Polarski, and Starobinsky}}]{st2}
\bibinfo{author}{\bibfnamefont{B.}~\bibnamefont{Boisseau}},
  \bibinfo{author}{\bibfnamefont{G.}~\bibnamefont{Esposito-Farese}},
  \bibinfo{author}{\bibfnamefont{D.}~\bibnamefont{Polarski}}, \bibnamefont{and}
  \bibinfo{author}{\bibfnamefont{A.~A.} \bibnamefont{Starobinsky}},
  \bibinfo{journal}{Phys. Rev. Lett.} \textbf{\bibinfo{volume}{85}},
  \bibinfo{pages}{2236} (\bibinfo{year}{2000}), \eprint{gr-qc/0001066}.

\bibitem[{\citenamefont{Albrecht et~al.}(2002)\citenamefont{Albrecht, Burgess,
  Ravndal, and Skordis}}]{Albrecht:2001xt}
\bibinfo{author}{\bibfnamefont{A.}~\bibnamefont{Albrecht}},
  \bibinfo{author}{\bibfnamefont{C.~P.} \bibnamefont{Burgess}},
  \bibinfo{author}{\bibfnamefont{F.}~\bibnamefont{Ravndal}}, \bibnamefont{and}
  \bibinfo{author}{\bibfnamefont{C.}~\bibnamefont{Skordis}},
  \bibinfo{journal}{Phys. Rev.} \textbf{\bibinfo{volume}{D65}},
  \bibinfo{pages}{123507} (\bibinfo{year}{2002}), \eprint{astro-ph/0107573}.

\bibitem[{\citenamefont{Kainulainen and Sunhede}(2006)}]{Kainulainen:2004vk}
\bibinfo{author}{\bibfnamefont{K.}~\bibnamefont{Kainulainen}} \bibnamefont{and}
  \bibinfo{author}{\bibfnamefont{D.}~\bibnamefont{Sunhede}},
  \bibinfo{journal}{Phys. Rev.} \textbf{\bibinfo{volume}{D73}},
  \bibinfo{pages}{083510} (\bibinfo{year}{2006}), \eprint{astro-ph/0412609}.

\bibitem[{\citenamefont{Callan et~al.}(1985)\citenamefont{Callan, Martinec,
  Perry, and Friedan}}]{Callan}
\bibinfo{author}{\bibfnamefont{J.}~\bibnamefont{Callan},
  \bibfnamefont{Curtis~G.}}, \bibinfo{author}{\bibfnamefont{E.~J.}
  \bibnamefont{Martinec}}, \bibinfo{author}{\bibfnamefont{M.~J.}
  \bibnamefont{Perry}}, \bibnamefont{and}
  \bibinfo{author}{\bibfnamefont{D.}~\bibnamefont{Friedan}},
  \bibinfo{journal}{Nucl. Phys.} \textbf{\bibinfo{volume}{B262}},
  \bibinfo{pages}{593} (\bibinfo{year}{1985}).

\bibitem[{\citenamefont{Gross and Sloan}(1987)}]{gross}
\bibinfo{author}{\bibfnamefont{D.~J.} \bibnamefont{Gross}} \bibnamefont{and}
  \bibinfo{author}{\bibfnamefont{J.~H.} \bibnamefont{Sloan}},
  \bibinfo{journal}{Nucl. Phys.} \textbf{\bibinfo{volume}{B291}},
  \bibinfo{pages}{41} (\bibinfo{year}{1987}).

\bibitem[{\citenamefont{Aros et~al.}(2000)\citenamefont{Aros, Contreras, Olea,
  Troncoso, and Zanelli}}]{Aros:1999id}
\bibinfo{author}{\bibfnamefont{R.}~\bibnamefont{Aros}},
  \bibinfo{author}{\bibfnamefont{M.}~\bibnamefont{Contreras}},
  \bibinfo{author}{\bibfnamefont{R.}~\bibnamefont{Olea}},
  \bibinfo{author}{\bibfnamefont{R.}~\bibnamefont{Troncoso}}, \bibnamefont{and}
  \bibinfo{author}{\bibfnamefont{J.}~\bibnamefont{Zanelli}},
  \bibinfo{journal}{Phys. Rev. Lett.} \textbf{\bibinfo{volume}{84}},
  \bibinfo{pages}{1647} (\bibinfo{year}{2000}), \eprint{gr-qc/9909015}.

\bibitem[{\citenamefont{Olea}(2005)}]{Olea:2005gb}
\bibinfo{author}{\bibfnamefont{R.}~\bibnamefont{Olea}}, \bibinfo{journal}{JHEP}
  \textbf{\bibinfo{volume}{06}}, \bibinfo{pages}{023} (\bibinfo{year}{2005}),
  \eprint{hep-th/0504233}.

\bibitem[{\citenamefont{Carter and
  Neupane}(2005{\natexlab{a}})}]{Carter:2005fu}
\bibinfo{author}{\bibfnamefont{B.~M.~N.} \bibnamefont{Carter}}
  \bibnamefont{and} \bibinfo{author}{\bibfnamefont{I.~P.}
  \bibnamefont{Neupane}} (\bibinfo{year}{2005}{\natexlab{a}}),
  \eprint{hep-th/0512262}.

\bibitem[{\citenamefont{Neupane}(2006{\natexlab{a}})}]{Neupane:2006dp}
\bibinfo{author}{\bibfnamefont{I.~P.} \bibnamefont{Neupane}}
  (\bibinfo{year}{2006}{\natexlab{a}}), \eprint{hep-th/0602097}.

\bibitem[{\citenamefont{Tsujikawa}(2006)}]{Tsujikawa:2006is}
\bibinfo{author}{\bibfnamefont{S.}~\bibnamefont{Tsujikawa}},
  \bibinfo{journal}{Annalen Phys.} \textbf{\bibinfo{volume}{15}},
  \bibinfo{pages}{302} (\bibinfo{year}{2006}), \eprint{hep-th/0606040}.

\bibitem[{\citenamefont{Carter and
  Neupane}(2005{\natexlab{b}})}]{Carter:2005sd}
\bibinfo{author}{\bibfnamefont{B.~M.~N.} \bibnamefont{Carter}}
  \bibnamefont{and} \bibinfo{author}{\bibfnamefont{I.~P.}
  \bibnamefont{Neupane}} (\bibinfo{year}{2005}{\natexlab{b}}),
  \eprint{hep-th/0510109}.

\bibitem[{\citenamefont{Neupane}(2006{\natexlab{b}})}]{Neupane:2006ip}
\bibinfo{author}{\bibfnamefont{I.~P.} \bibnamefont{Neupane}}
  (\bibinfo{year}{2006}{\natexlab{b}}), \eprint{hep-th/0605265}.

\bibitem[{\citenamefont{Koivisto and
  Mota}(2006{\natexlab{b}})}]{Koivisto:2006xf}
\bibinfo{author}{\bibfnamefont{T.}~\bibnamefont{Koivisto}} \bibnamefont{and}
  \bibinfo{author}{\bibfnamefont{D.~F.} \bibnamefont{Mota}}
  (\bibinfo{year}{2006}{\natexlab{b}}), \eprint{astro-ph/0606078}.

\bibitem[{\citenamefont{Nojiri et~al.}(2005)\citenamefont{Nojiri, Odintsov, and
  Sasaki}}]{Nojiri:2005vv}
\bibinfo{author}{\bibfnamefont{S.}~\bibnamefont{Nojiri}},
  \bibinfo{author}{\bibfnamefont{S.~D.} \bibnamefont{Odintsov}},
  \bibnamefont{and} \bibinfo{author}{\bibfnamefont{M.}~\bibnamefont{Sasaki}},
  \bibinfo{journal}{Phys. Rev.} \textbf{\bibinfo{volume}{D71}},
  \bibinfo{pages}{123509} (\bibinfo{year}{2005}), \eprint{hep-th/0504052}.

\bibitem[{\citenamefont{Tsujikawa and Sami}(2006)}]{Tsujikawa:2006ph}
\bibinfo{author}{\bibfnamefont{S.}~\bibnamefont{Tsujikawa}} \bibnamefont{and}
  \bibinfo{author}{\bibfnamefont{M.}~\bibnamefont{Sami}}
  (\bibinfo{year}{2006}), \eprint{hep-th/0608178}.

\bibitem[{\citenamefont{Sami et~al.}(2005)\citenamefont{Sami, Toporensky,
  Tretjakov, and Tsujikawa}}]{Sami:2005zc}
\bibinfo{author}{\bibfnamefont{M.}~\bibnamefont{Sami}},
  \bibinfo{author}{\bibfnamefont{A.}~\bibnamefont{Toporensky}},
  \bibinfo{author}{\bibfnamefont{P.~V.} \bibnamefont{Tretjakov}},
  \bibnamefont{and}
  \bibinfo{author}{\bibfnamefont{S.}~\bibnamefont{Tsujikawa}},
  \bibinfo{journal}{Phys. Lett.} \textbf{\bibinfo{volume}{B619}},
  \bibinfo{pages}{193} (\bibinfo{year}{2005}), \eprint{hep-th/0504154}.

\bibitem[{\citenamefont{Calcagni et~al.}(2005)\citenamefont{Calcagni,
  Tsujikawa, and Sami}}]{Calcagni:2005im}
\bibinfo{author}{\bibfnamefont{G.}~\bibnamefont{Calcagni}},
  \bibinfo{author}{\bibfnamefont{S.}~\bibnamefont{Tsujikawa}},
  \bibnamefont{and} \bibinfo{author}{\bibfnamefont{M.}~\bibnamefont{Sami}},
  \bibinfo{journal}{Class. Quant. Grav.} \textbf{\bibinfo{volume}{22}},
  \bibinfo{pages}{3977} (\bibinfo{year}{2005}), \eprint{hep-th/0505193}.

\bibitem[{\citenamefont{Nojiri et~al.}(2006{\natexlab{a}})\citenamefont{Nojiri,
  Odintsov, and Sami}}]{Nojiri:2006je}
\bibinfo{author}{\bibfnamefont{S.}~\bibnamefont{Nojiri}},
  \bibinfo{author}{\bibfnamefont{S.~D.} \bibnamefont{Odintsov}},
  \bibnamefont{and} \bibinfo{author}{\bibfnamefont{M.}~\bibnamefont{Sami}}
  (\bibinfo{year}{2006}{\natexlab{a}}), \eprint{hep-th/0605039}.

\bibitem[{\citenamefont{Cognola et~al.}(2006)\citenamefont{Cognola, Elizalde,
  Nojiri, Odintsov, and Zerbini}}]{Cognola:2006eg}
\bibinfo{author}{\bibfnamefont{G.}~\bibnamefont{Cognola}},
  \bibinfo{author}{\bibfnamefont{E.}~\bibnamefont{Elizalde}},
  \bibinfo{author}{\bibfnamefont{S.}~\bibnamefont{Nojiri}},
  \bibinfo{author}{\bibfnamefont{S.~D.} \bibnamefont{Odintsov}},
  \bibnamefont{and} \bibinfo{author}{\bibfnamefont{S.}~\bibnamefont{Zerbini}}
  (\bibinfo{year}{2006}), \eprint{hep-th/0601008}.

\bibitem[{\citenamefont{Nojiri et~al.}(2006{\natexlab{b}})\citenamefont{Nojiri,
  Odintsov, and Gorbunova}}]{Nojiri:2005am}
\bibinfo{author}{\bibfnamefont{S.}~\bibnamefont{Nojiri}},
  \bibinfo{author}{\bibfnamefont{S.~D.} \bibnamefont{Odintsov}},
  \bibnamefont{and} \bibinfo{author}{\bibfnamefont{O.~G.}
  \bibnamefont{Gorbunova}}, \bibinfo{journal}{J. Phys.}
  \textbf{\bibinfo{volume}{A39}}, \bibinfo{pages}{6627}
  (\bibinfo{year}{2006}{\natexlab{b}}), \eprint{hep-th/0510183}.

\bibitem[{\citenamefont{Aref'eva et~al.}(2005)\citenamefont{Aref'eva, Koshelev,
  and Vernov}}]{Aref'eva:2005fu}
\bibinfo{author}{\bibfnamefont{I.~Y.} \bibnamefont{Aref'eva}},
  \bibinfo{author}{\bibfnamefont{A.~S.} \bibnamefont{Koshelev}},
  \bibnamefont{and} \bibinfo{author}{\bibfnamefont{S.~Y.}
  \bibnamefont{Vernov}}, \bibinfo{journal}{Phys. Rev.}
  \textbf{\bibinfo{volume}{D72}}, \bibinfo{pages}{064017}
  (\bibinfo{year}{2005}), \eprint{astro-ph/0507067}.

\bibitem[{\citenamefont{Aref'eva and Koshelev}(2006)}]{Aref'eva:2006et}
\bibinfo{author}{\bibfnamefont{I.~Y.} \bibnamefont{Aref'eva}} \bibnamefont{and}
  \bibinfo{author}{\bibfnamefont{A.~S.} \bibnamefont{Koshelev}}
  (\bibinfo{year}{2006}), \eprint{hep-th/0605085}.

\bibitem[{\citenamefont{Baccigalupi et~al.}(2000)\citenamefont{Baccigalupi,
  Matarrese, and Perrotta}}]{Baccigalupi:2000je}
\bibinfo{author}{\bibfnamefont{C.}~\bibnamefont{Baccigalupi}},
  \bibinfo{author}{\bibfnamefont{S.}~\bibnamefont{Matarrese}},
  \bibnamefont{and} \bibinfo{author}{\bibfnamefont{F.}~\bibnamefont{Perrotta}},
  \bibinfo{journal}{Phys. Rev.} \textbf{\bibinfo{volume}{D62}},
  \bibinfo{pages}{123510} (\bibinfo{year}{2000}), \eprint{astro-ph/0005543}.

\bibitem[{\citenamefont{Pettorino et~al.}(2005)\citenamefont{Pettorino,
  Baccigalupi, and Perrotta}}]{Pettorino:2005pv}
\bibinfo{author}{\bibfnamefont{V.}~\bibnamefont{Pettorino}},
  \bibinfo{author}{\bibfnamefont{C.}~\bibnamefont{Baccigalupi}},
  \bibnamefont{and} \bibinfo{author}{\bibfnamefont{F.}~\bibnamefont{Perrotta}},
  \bibinfo{journal}{JCAP} \textbf{\bibinfo{volume}{0512}}, \bibinfo{pages}{003}
  (\bibinfo{year}{2005}), \eprint{astro-ph/0508586}.

\bibitem[{\citenamefont{Amendola}(2000)}]{Amendola:1999er}
\bibinfo{author}{\bibfnamefont{L.}~\bibnamefont{Amendola}},
  \bibinfo{journal}{Phys. Rev.} \textbf{\bibinfo{volume}{D62}},
  \bibinfo{pages}{043511} (\bibinfo{year}{2000}), \eprint{astro-ph/9908023}.

\bibitem[{\citenamefont{Manera and Mota}(2005)}]{manera}
\bibinfo{author}{\bibfnamefont{M.}~\bibnamefont{Manera}} \bibnamefont{and}
  \bibinfo{author}{\bibfnamefont{D.~F.} \bibnamefont{Mota}}
  (\bibinfo{year}{2005}), \eprint{astro-ph/0504519}.

\bibitem[{\citenamefont{Brookfield
  et~al.}(2006{\natexlab{b}})\citenamefont{Brookfield, van~de Bruck, Mota, and
  Tocchini-Valentini}}]{brookfield1}
\bibinfo{author}{\bibfnamefont{A.~W.} \bibnamefont{Brookfield}},
  \bibinfo{author}{\bibfnamefont{C.}~\bibnamefont{van~de Bruck}},
  \bibinfo{author}{\bibfnamefont{D.~F.} \bibnamefont{Mota}}, \bibnamefont{and}
  \bibinfo{author}{\bibfnamefont{D.}~\bibnamefont{Tocchini-Valentini}},
  \bibinfo{journal}{Phys. Rev. Lett.} \textbf{\bibinfo{volume}{96}},
  \bibinfo{pages}{061301} (\bibinfo{year}{2006}{\natexlab{b}}),
  \eprint{astro-ph/0503349}.

\bibitem[{\citenamefont{Bean and Magueijo}(2001)}]{Bean:2000zm}
\bibinfo{author}{\bibfnamefont{R.}~\bibnamefont{Bean}} \bibnamefont{and}
  \bibinfo{author}{\bibfnamefont{J.}~\bibnamefont{Magueijo}},
  \bibinfo{journal}{Phys. Lett.} \textbf{\bibinfo{volume}{B517}},
  \bibinfo{pages}{177} (\bibinfo{year}{2001}), \eprint{astro-ph/0007199}.

\bibitem[{\citenamefont{Huey and Wandelt}(2004)}]{Huey:2004qv}
\bibinfo{author}{\bibfnamefont{G.}~\bibnamefont{Huey}} \bibnamefont{and}
  \bibinfo{author}{\bibfnamefont{B.~D.} \bibnamefont{Wandelt}}
  (\bibinfo{year}{2004}), \eprint{astro-ph/0407196}.

\bibitem[{\citenamefont{Koivisto}(2005)}]{Koivisto:2005nr}
\bibinfo{author}{\bibfnamefont{T.}~\bibnamefont{Koivisto}},
  \bibinfo{journal}{Phys. Rev.} \textbf{\bibinfo{volume}{D72}},
  \bibinfo{pages}{043516} (\bibinfo{year}{2005}), \eprint{astro-ph/0504571}.

\bibitem[{\citenamefont{Nunes and Mota}(2004)}]{nunes}
\bibinfo{author}{\bibfnamefont{N.~J.} \bibnamefont{Nunes}} \bibnamefont{and}
  \bibinfo{author}{\bibfnamefont{D.~F.} \bibnamefont{Mota}}
  (\bibinfo{year}{2004}), \eprint{astro-ph/0409481}.

\bibitem[{\citenamefont{Uzan}(2003)}]{uzan}
\bibinfo{author}{\bibfnamefont{J.-P.} \bibnamefont{Uzan}},
  \bibinfo{journal}{Rev. Mod. Phys.} \textbf{\bibinfo{volume}{75}},
  \bibinfo{pages}{403} (\bibinfo{year}{2003}), \eprint{hep-ph/0205340}.

\bibitem[{\citenamefont{Mota and Shaw}(2006{\natexlab{a}})}]{doug1}
\bibinfo{author}{\bibfnamefont{D.~F.} \bibnamefont{Mota}} \bibnamefont{and}
  \bibinfo{author}{\bibfnamefont{D.~J.} \bibnamefont{Shaw}}
  (\bibinfo{year}{2006}{\natexlab{a}}), \eprint{hep-ph/0606204}.

\bibitem[{\citenamefont{Mota and Shaw}(2006{\natexlab{b}})}]{doug2}
\bibinfo{author}{\bibfnamefont{D.~F.} \bibnamefont{Mota}} \bibnamefont{and}
  \bibinfo{author}{\bibfnamefont{D.~J.} \bibnamefont{Shaw}}
  (\bibinfo{year}{2006}{\natexlab{b}}), \eprint{hep-ph/0608078}.

\bibitem[{\citenamefont{Antoniadis
  et~al.}(1992{\natexlab{a}})\citenamefont{Antoniadis, Gava, and
  Narain}}]{Antoniadis2}
\bibinfo{author}{\bibfnamefont{I.}~\bibnamefont{Antoniadis}},
  \bibinfo{author}{\bibfnamefont{E.}~\bibnamefont{Gava}}, \bibnamefont{and}
  \bibinfo{author}{\bibfnamefont{K.~S.} \bibnamefont{Narain}},
  \bibinfo{journal}{Nucl. Phys.} \textbf{\bibinfo{volume}{B383}},
  \bibinfo{pages}{93} (\bibinfo{year}{1992}{\natexlab{a}}),
  \eprint{hep-th/9204030}.

\bibitem[{\citenamefont{Damour and Polyakov}(1994)}]{Damour:1994zq}
\bibinfo{author}{\bibfnamefont{T.}~\bibnamefont{Damour}} \bibnamefont{and}
  \bibinfo{author}{\bibfnamefont{A.~M.} \bibnamefont{Polyakov}},
  \bibinfo{journal}{Nucl. Phys.} \textbf{\bibinfo{volume}{B423}},
  \bibinfo{pages}{532} (\bibinfo{year}{1994}), \eprint{hep-th/9401069}.

\bibitem[{\citenamefont{Antoniadis
  et~al.}(1992{\natexlab{b}})\citenamefont{Antoniadis, Gava, and
  Narain}}]{Antoniadis3}
\bibinfo{author}{\bibfnamefont{I.}~\bibnamefont{Antoniadis}},
  \bibinfo{author}{\bibfnamefont{E.}~\bibnamefont{Gava}}, \bibnamefont{and}
  \bibinfo{author}{\bibfnamefont{K.~S.} \bibnamefont{Narain}},
  \bibinfo{journal}{Phys. Lett.} \textbf{\bibinfo{volume}{B283}},
  \bibinfo{pages}{209} (\bibinfo{year}{1992}{\natexlab{b}}),
  \eprint{hep-th/9203071}.

\bibitem[{\citenamefont{Antoniadis et~al.}(1994)\citenamefont{Antoniadis,
  Rizos, and Tamvakis}}]{Antoniadis}
\bibinfo{author}{\bibfnamefont{I.}~\bibnamefont{Antoniadis}},
  \bibinfo{author}{\bibfnamefont{J.}~\bibnamefont{Rizos}}, \bibnamefont{and}
  \bibinfo{author}{\bibfnamefont{K.}~\bibnamefont{Tamvakis}},
  \bibinfo{journal}{Nucl. Phys.} \textbf{\bibinfo{volume}{B415}},
  \bibinfo{pages}{497} (\bibinfo{year}{1994}), \eprint{hep-th/9305025}.

\bibitem[{\citenamefont{Gasperini and Veneziano}(2003)}]{Gasperini:2002bn}
\bibinfo{author}{\bibfnamefont{M.}~\bibnamefont{Gasperini}} \bibnamefont{and}
  \bibinfo{author}{\bibfnamefont{G.}~\bibnamefont{Veneziano}},
  \bibinfo{journal}{Phys. Rept.} \textbf{\bibinfo{volume}{373}},
  \bibinfo{pages}{1} (\bibinfo{year}{2003}), \eprint{hep-th/0207130}.

\bibitem[{\citenamefont{Koivisto}(2006{\natexlab{b}})}]{tomimod1}
\bibinfo{author}{\bibfnamefont{T.}~\bibnamefont{Koivisto}},
  \bibinfo{journal}{Class. Quant. Grav.} \textbf{\bibinfo{volume}{23}},
  \bibinfo{pages}{4289} (\bibinfo{year}{2006}{\natexlab{b}}),
  \eprint{gr-qc/0505128}.

\bibitem[{\citenamefont{Cartier et~al.}(2001)\citenamefont{Cartier, Hwang, and
  Copeland}}]{Cartier:2001is}
\bibinfo{author}{\bibfnamefont{C.}~\bibnamefont{Cartier}},
  \bibinfo{author}{\bibfnamefont{J.-c.} \bibnamefont{Hwang}}, \bibnamefont{and}
  \bibinfo{author}{\bibfnamefont{E.~J.} \bibnamefont{Copeland}},
  \bibinfo{journal}{Phys. Rev.} \textbf{\bibinfo{volume}{D64}},
  \bibinfo{pages}{103504} (\bibinfo{year}{2001}), \eprint{astro-ph/0106197}.

\bibitem[{\citenamefont{Bardeen}(1980)}]{Bardeen:1980kt}
\bibinfo{author}{\bibfnamefont{J.~M.} \bibnamefont{Bardeen}},
  \bibinfo{journal}{Phys. Rev.} \textbf{\bibinfo{volume}{D22}},
  \bibinfo{pages}{1882} (\bibinfo{year}{1980}).

\bibitem[{\citenamefont{Hwang and Noh}(2005)}]{Hwang:2005hb}
\bibinfo{author}{\bibfnamefont{J.-c.} \bibnamefont{Hwang}} \bibnamefont{and}
  \bibinfo{author}{\bibfnamefont{H.}~\bibnamefont{Noh}},
  \bibinfo{journal}{Phys. Rev.} \textbf{\bibinfo{volume}{D71}},
  \bibinfo{pages}{063536} (\bibinfo{year}{2005}), \eprint{gr-qc/0412126}.

\bibitem[{\citenamefont{Ma and Bertschinger}(1995)}]{Ma:1995ey}
\bibinfo{author}{\bibfnamefont{C.-P.} \bibnamefont{Ma}} \bibnamefont{and}
  \bibinfo{author}{\bibfnamefont{E.}~\bibnamefont{Bertschinger}},
  \bibinfo{journal}{Astrophys. J.} \textbf{\bibinfo{volume}{455}},
  \bibinfo{pages}{7} (\bibinfo{year}{1995}), \eprint{astro-ph/9506072}.

\bibitem[{\citenamefont{Lue et~al.}(2004)\citenamefont{Lue, Scoccimarro, and
  Starkman}}]{Lue:2003ky}
\bibinfo{author}{\bibfnamefont{A.}~\bibnamefont{Lue}},
  \bibinfo{author}{\bibfnamefont{R.}~\bibnamefont{Scoccimarro}},
  \bibnamefont{and} \bibinfo{author}{\bibfnamefont{G.}~\bibnamefont{Starkman}},
  \bibinfo{journal}{Phys. Rev.} \textbf{\bibinfo{volume}{D69}},
  \bibinfo{pages}{044005} (\bibinfo{year}{2004}), \eprint{astro-ph/0307034}.

\bibitem[{\citenamefont{Lewis et~al.}(2000)\citenamefont{Lewis, Challinor, and
  Lasenby}}]{Lewis:1999bs}
\bibinfo{author}{\bibfnamefont{A.}~\bibnamefont{Lewis}},
  \bibinfo{author}{\bibfnamefont{A.}~\bibnamefont{Challinor}},
  \bibnamefont{and} \bibinfo{author}{\bibfnamefont{A.}~\bibnamefont{Lasenby}},
  \bibinfo{journal}{Astrophys. J.} \textbf{\bibinfo{volume}{538}},
  \bibinfo{pages}{473} (\bibinfo{year}{2000}), \eprint{astro-ph/9911177}.

\bibitem[{\citenamefont{Ferreira and Joyce}(1998)}]{Ferreira:1997hj}
\bibinfo{author}{\bibfnamefont{P.~G.} \bibnamefont{Ferreira}} \bibnamefont{and}
  \bibinfo{author}{\bibfnamefont{M.}~\bibnamefont{Joyce}},
  \bibinfo{journal}{Phys. Rev.} \textbf{\bibinfo{volume}{D58}},
  \bibinfo{pages}{023503} (\bibinfo{year}{1998}), \eprint{astro-ph/9711102}.

\bibitem[{\citenamefont{Hellerman et~al.}(2001)\citenamefont{Hellerman,
  Kaloper, and Susskind}}]{Hellerman:2001yi}
\bibinfo{author}{\bibfnamefont{S.}~\bibnamefont{Hellerman}},
  \bibinfo{author}{\bibfnamefont{N.}~\bibnamefont{Kaloper}}, \bibnamefont{and}
  \bibinfo{author}{\bibfnamefont{L.}~\bibnamefont{Susskind}},
  \bibinfo{journal}{JHEP} \textbf{\bibinfo{volume}{06}}, \bibinfo{pages}{003}
  (\bibinfo{year}{2001}), \eprint{hep-th/0104180}.

\bibitem[{\citenamefont{Kawai et~al.}(1998)\citenamefont{Kawai, Sakagami, and
  Soda}}]{Kawai:1998ab}
\bibinfo{author}{\bibfnamefont{S.}~\bibnamefont{Kawai}},
  \bibinfo{author}{\bibfnamefont{M.-a.} \bibnamefont{Sakagami}},
  \bibnamefont{and} \bibinfo{author}{\bibfnamefont{J.}~\bibnamefont{Soda}},
  \bibinfo{journal}{Phys. Lett.} \textbf{\bibinfo{volume}{B437}},
  \bibinfo{pages}{284} (\bibinfo{year}{1998}), \eprint{gr-qc/9802033}.

\bibitem[{\citenamefont{Kawai and Soda}(1999)}]{Kawai:1999pw}
\bibinfo{author}{\bibfnamefont{S.}~\bibnamefont{Kawai}} \bibnamefont{and}
  \bibinfo{author}{\bibfnamefont{J.}~\bibnamefont{Soda}},
  \bibinfo{journal}{Phys. Lett.} \textbf{\bibinfo{volume}{B460}},
  \bibinfo{pages}{41} (\bibinfo{year}{1999}), \eprint{gr-qc/9903017}.

\bibitem[{\citenamefont{De~Felice et~al.}(2006)\citenamefont{De~Felice,
  Hindmarsh, and Trodden}}]{DeFelice:2006pg}
\bibinfo{author}{\bibfnamefont{A.}~\bibnamefont{De~Felice}},
  \bibinfo{author}{\bibfnamefont{M.}~\bibnamefont{Hindmarsh}},
  \bibnamefont{and} \bibinfo{author}{\bibfnamefont{M.}~\bibnamefont{Trodden}}
  (\bibinfo{year}{2006}), \eprint{astro-ph/0604154}.

\bibitem[{\citenamefont{Calcagni et~al.}(2006)\citenamefont{Calcagni,
  de~Carlos, and De~Felice}}]{Calcagni:2006ye}
\bibinfo{author}{\bibfnamefont{G.}~\bibnamefont{Calcagni}},
  \bibinfo{author}{\bibfnamefont{B.}~\bibnamefont{de~Carlos}},
  \bibnamefont{and} \bibinfo{author}{\bibfnamefont{A.}~\bibnamefont{De~Felice}}
  (\bibinfo{year}{2006}), \eprint{hep-th/0604201}.

\bibitem[{\citenamefont{Dick et~al.}(2006)\citenamefont{Dick, Knox, and
  Chu}}]{Dick:2006ev}
\bibinfo{author}{\bibfnamefont{J.}~\bibnamefont{Dick}},
  \bibinfo{author}{\bibfnamefont{L.}~\bibnamefont{Knox}}, \bibnamefont{and}
  \bibinfo{author}{\bibfnamefont{M.}~\bibnamefont{Chu}},
  \bibinfo{journal}{JCAP} \textbf{\bibinfo{volume}{0607}}, \bibinfo{pages}{001}
  (\bibinfo{year}{2006}), \eprint{astro-ph/0603247}.

\bibitem[{\citenamefont{Nesseris and Perivolaropoulos}(2006)}]{Nesseris:2006jc}
\bibinfo{author}{\bibfnamefont{S.}~\bibnamefont{Nesseris}} \bibnamefont{and}
  \bibinfo{author}{\bibfnamefont{L.}~\bibnamefont{Perivolaropoulos}},
  \bibinfo{journal}{Phys. Rev.} \textbf{\bibinfo{volume}{D73}},
  \bibinfo{pages}{103511} (\bibinfo{year}{2006}), \eprint{astro-ph/0602053}.

\bibitem[{\citenamefont{Odman et~al.}(2003)\citenamefont{Odman, Melchiorri,
  Hobson, and Lasenby}}]{Odman:2002wj}
\bibinfo{author}{\bibfnamefont{C.~J.} \bibnamefont{Odman}},
  \bibinfo{author}{\bibfnamefont{A.}~\bibnamefont{Melchiorri}},
  \bibinfo{author}{\bibfnamefont{M.~P.} \bibnamefont{Hobson}},
  \bibnamefont{and} \bibinfo{author}{\bibfnamefont{A.~N.}
  \bibnamefont{Lasenby}}, \bibinfo{journal}{Phys. Rev.}
  \textbf{\bibinfo{volume}{D67}}, \bibinfo{pages}{083511}
  (\bibinfo{year}{2003}), \eprint{astro-ph/0207286}.

\bibitem[{\citenamefont{Wang and Mukherjee}(2006)}]{Wang:2006ts}
\bibinfo{author}{\bibfnamefont{Y.}~\bibnamefont{Wang}} \bibnamefont{and}
  \bibinfo{author}{\bibfnamefont{P.}~\bibnamefont{Mukherjee}}
  (\bibinfo{year}{2006}), \eprint{astro-ph/0604051}.

\bibitem[{\citenamefont{Eisenstein et~al.}(2005)}]{Eisenstein:2005su}
\bibinfo{author}{\bibfnamefont{D.~J.} \bibnamefont{Eisenstein}}
  \bibnamefont{et~al.}, \bibinfo{journal}{Astrophys. J.}
  \textbf{\bibinfo{volume}{633}}, \bibinfo{pages}{560} (\bibinfo{year}{2005}),
  \eprint{astro-ph/0501171}.

\bibitem[{\citenamefont{Bean et~al.}(2001)\citenamefont{Bean, Hansen, and
  Melchiorri}}]{Bean:2001wt}
\bibinfo{author}{\bibfnamefont{R.}~\bibnamefont{Bean}},
  \bibinfo{author}{\bibfnamefont{S.~H.} \bibnamefont{Hansen}},
  \bibnamefont{and}
  \bibinfo{author}{\bibfnamefont{A.}~\bibnamefont{Melchiorri}},
  \bibinfo{journal}{Phys. Rev.} \textbf{\bibinfo{volume}{D64}},
  \bibinfo{pages}{103508} (\bibinfo{year}{2001}), \eprint{astro-ph/0104162}.

\bibitem[{\citenamefont{Malquarti and Liddle}(2002)}]{Malquarti:2002bh}
\bibinfo{author}{\bibfnamefont{M.}~\bibnamefont{Malquarti}} \bibnamefont{and}
  \bibinfo{author}{\bibfnamefont{A.~R.} \bibnamefont{Liddle}},
  \bibinfo{journal}{Phys. Rev.} \textbf{\bibinfo{volume}{D66}},
  \bibinfo{pages}{023524} (\bibinfo{year}{2002}), \eprint{astro-ph/0203232}.

\bibitem[{\citenamefont{Amendola et~al.}(2005)\citenamefont{Amendola,
  Charmousis, and Davis}}]{Amendola:2005cr}
\bibinfo{author}{\bibfnamefont{L.}~\bibnamefont{Amendola}},
  \bibinfo{author}{\bibfnamefont{C.}~\bibnamefont{Charmousis}},
  \bibnamefont{and} \bibinfo{author}{\bibfnamefont{S.~C.} \bibnamefont{Davis}}
  (\bibinfo{year}{2005}), \eprint{hep-th/0506137}.

\bibitem[{\citenamefont{Clifton et~al.}(2005)\citenamefont{Clifton, Mota, and
  Barrow}}]{clifton}
\bibinfo{author}{\bibfnamefont{T.}~\bibnamefont{Clifton}},
  \bibinfo{author}{\bibfnamefont{D.~F.} \bibnamefont{Mota}}, \bibnamefont{and}
  \bibinfo{author}{\bibfnamefont{J.~D.} \bibnamefont{Barrow}},
  \bibinfo{journal}{Mon. Not. Roy. Astron. Soc.}
  \textbf{\bibinfo{volume}{358}}, \bibinfo{pages}{601} (\bibinfo{year}{2005}),
  \eprint{gr-qc/0406001}.

\bibitem[{\citenamefont{Mota and Barrow}(2004{\natexlab{a}})}]{barrow1}
\bibinfo{author}{\bibfnamefont{D.~F.} \bibnamefont{Mota}} \bibnamefont{and}
  \bibinfo{author}{\bibfnamefont{J.~D.} \bibnamefont{Barrow}},
  \bibinfo{journal}{Mon. Not. Roy. Astron. Soc.}
  \textbf{\bibinfo{volume}{349}}, \bibinfo{pages}{291}
  (\bibinfo{year}{2004}{\natexlab{a}}), \eprint{astro-ph/0309273}.

\bibitem[{\citenamefont{Mota and Barrow}(2004{\natexlab{b}})}]{barrow2}
\bibinfo{author}{\bibfnamefont{D.~F.} \bibnamefont{Mota}} \bibnamefont{and}
  \bibinfo{author}{\bibfnamefont{J.~D.} \bibnamefont{Barrow}},
  \bibinfo{journal}{Phys. Lett.} \textbf{\bibinfo{volume}{B581}},
  \bibinfo{pages}{141} (\bibinfo{year}{2004}{\natexlab{b}}),
  \eprint{astro-ph/0306047}.

\bibitem[{\citenamefont{Barrow and Mota}(2002)}]{Barrow:2002ed}
\bibinfo{author}{\bibfnamefont{J.~D.} \bibnamefont{Barrow}} \bibnamefont{and}
  \bibinfo{author}{\bibfnamefont{D.~F.} \bibnamefont{Mota}},
  \bibinfo{journal}{Class. Quant. Grav.} \textbf{\bibinfo{volume}{19}},
  \bibinfo{pages}{6197} (\bibinfo{year}{2002}), \eprint{gr-qc/0207012}.

\bibitem[{\citenamefont{Esposito-Farese}(2003)}]{Esposito-Farese:2003ze}
\bibinfo{author}{\bibfnamefont{G.}~\bibnamefont{Esposito-Farese}}
  (\bibinfo{year}{2003}), \eprint{gr-qc/0306018}.

\bibitem[{\citenamefont{Esposito-Farese}(2004)}]{Esposito-Farese:2004cc}
\bibinfo{author}{\bibfnamefont{G.}~\bibnamefont{Esposito-Farese}},
  \bibinfo{journal}{AIP Conf. Proc.} \textbf{\bibinfo{volume}{736}},
  \bibinfo{pages}{35} (\bibinfo{year}{2004}), \eprint{gr-qc/0409081}.

\bibitem[{\citenamefont{Amendola
  et~al.}(2006{\natexlab{b}})\citenamefont{Amendola, Quartin, Tsujikawa, and
  Waga}}]{Amendola:2006qi}
\bibinfo{author}{\bibfnamefont{L.}~\bibnamefont{Amendola}},
  \bibinfo{author}{\bibfnamefont{M.}~\bibnamefont{Quartin}},
  \bibinfo{author}{\bibfnamefont{S.}~\bibnamefont{Tsujikawa}},
  \bibnamefont{and} \bibinfo{author}{\bibfnamefont{I.}~\bibnamefont{Waga}}
  (\bibinfo{year}{2006}{\natexlab{b}}), \eprint{astro-ph/0605488}.

\end{thebibliography}

\end{document}